\documentclass[fleqn,usenatbib]{mnras}

\usepackage{newtxtext,newtxmath}

\usepackage[T1]{fontenc}

\DeclareRobustCommand{\VAN}[3]{#2}
\let\VANthebibliography\thebibliography
\def\thebibliography{\DeclareRobustCommand{\VAN}[3]{##3}\VANthebibliography}

\usepackage{graphicx}	
\usepackage{amsmath}	
\usepackage{supertabular}

\usepackage{pdflscape}	

\title[Galaxy -- halo connection from $z=0.25 \rightarrow 4$]{The connection between massive galaxies and their dark matter haloes from $z=0.25$ to $z \sim 4$ from the combined COSMOS, XMM-LSS and ECDF-S fields}

\author[N. Stylianou et al.]{N. Stylianou,$^{1}$\thanks{E-mail: natalia.stylianou@physics.ox.ac.uk}
Matt J. Jarvis,$^{1,2}$
R. G. Varadaraj,$^{1}$ 
P. W. Hatfield,$^{1}$
J. Hamlett,$^{1}$ 
C. L. Hale,$^{3, 1}$
T. Yasin,$^{1}$ 
\and
\\
$^{1}$Astrophysics, Department of Physics, University of Oxford, Keble Road, Oxford OX1 3RH, UK\\
$^{2}$Department of Physics, University of the Western Cape, Bellville 7535, Cape Town, South Africa \\
$^{3}$Institute for Astronomy, University of 
Edinburgh, Royal Observatory Edinburgh, Blackford Hill, Edinburgh, EH9 3HJ, UK
}

\date{Accepted XXX. Received YYY; in original form ZZZ}

\pubyear{\the\year{}}

\begin{document}
\label{firstpage}
\pagerange{\pageref{firstpage}--\pageref{lastpage}}
\maketitle

\begin{abstract}
Understanding how galaxies populate and grow within dark matter haloes is essential for linking the observed evolution of galaxies to the hierarchical growth of structure. We present a galaxy clustering analysis using optical and near-infrared data from the UltraVISTA and VIDEO surveys covering $9.3$ deg$^2$ across the COSMOS, XMM-LSS and ECDF-S fields, $\sim 5\times$ larger than previous comparable studies, providing sufficient volume to determine the halo occupation of the most massive galaxies to $z\sim 4$. We measure the angular two-point correlation function for stellar-mass-selected galaxies over $0.25<z<4$ and by combining the three fields, we reduce the cosmic variance by a factor of $\sim2$ compared to the individual fields. The stellar-to-halo mass ratio peaks at a halo mass of $M_{h}\approx10^{12}$\,M$_{\odot}$ and is a maximum at $z\sim 1$. For the most massive galaxies, the peak halo mass ($M_{h}^{\rm peak}$) increases from $z=1\rightarrow 0$, while the peak stellar mass in such haloes remains approximately constant, indicating that at $z<1$ massive galaxy formation becomes less efficient. We also find a steady increase in the satellite fraction since $z\sim 1$, from which we infer that massive halos continue to grow hierarchically while their central galaxies experience suppressed stellar-mass growth. We suggest this is due to the central massive galaxies having their growth controlled by feedback from active galactic nuclei feedback, likely from low-power stochastic radio emission.

\end{abstract}




\section{Introduction} 
\label{sec:Intro}
A central prediction of the $\Lambda$CDM framework is that structure grows hierarchically from primordial density fluctuations into a cosmic web dominated by dark matter, within which self-gravitating haloes provide the sites where galaxies form, accrete gas, and build up their stellar content (\citealt{1985ApJ...292..371D}; \citealt{1978MNRAS.183..341W}; \citealt{1991ApJ...379...52W}). 

However, we cannot directly observe the evolution of the dark matter haloes, instead we are forced to use the galaxies that reside in these haloes as biased tracers \citep{1984ApJ...284L...9K} of the underlying dark matter distribution. Galaxies of different types and mass trace this underlying dark matter distribution in different ways, often described by their bias; the ratio between the galaxy clustering and dark matter clustering amplitude. The differences in the bias for different galaxies are due to the strong link between the build-up of baryonic mass and dark-matter halo mass (\citealt{2010ApJ...717..379B}; \citealt{2013ApJ...770...57B}; \citealt{2018ARA&A..56..435W}), alongside a plethora of baryonic effects that may alter the observed properties of galaxies. Such effects can be broadly split into internal physical mechanisms within the galaxies themselves, e.g. feedback from supernovae (\citealt{1974MNRAS.169..229L}; \citealt{1986ApJ...303...39D}; \citealt{2000MNRAS.317..697E}) or the supermassive black holes residing at the centres of massive galaxies (\citealt{2012ARA&A..50..455F}; \citealt{Croton2006}; \citealt{Bower2006}), or environmental effects, e.g. ram-pressure stripping (\citealt{1972ApJ...176....1G}; \citealt{2008MNRAS.383..593M}; \citealt{2014ApJ...781L..40E}), strangulation (\citealt{1980ApJ...237..692L}; \citealt{2008ApJ...672L.103K}; \citealt{2015MNRAS.446.1939F}) and galaxy harassment (\citealt{1981ApJ...243...32F}; \citealt{1996Natur.379..613M}; \citealt{1998ApJ...495..139M}). Furthermore, assembly bias (\citealt{2007MNRAS.374.1303C}; \citealt{2007MNRAS.377L...5G}; \citealt{2014MNRAS.443.3044Z}), also adds additional complexity to a simple relationship between galaxy properties and halo properties by introducing the concept that haloes that form earlier often have more evolved galaxies within them, than later-formed haloes of a similar mass.
Therefore investigations of the galaxy-halo connection link the growth of stellar mass to the evolution of halo mass, thereby capturing the integrated effects of star formation, quenching, and feedback over the history of the Universe (\citealt{2015ARA&A..53...51S}; \citealt{2013ApJ...770...57B}; \citealt{2013MNRAS.428.3121M}). Determining this connection observationally is therefore fundamental to any empirical description of galaxy evolution in the standard cosmological model.

A key observational consequence of the galaxy-halo connection is that the galaxies residing in different halo environments cluster differently, making galaxy clustering a sensitive probe of halo occupation. Since the galaxy clustering signal depends on the distribution of dark matter as well as on the masses and abundances of the these haloes, measurements of galaxy clustering can be used to constrain the halo populations associated with observed galaxy samples. Within this context, halo occupation distribution (HOD) models provide a particularly useful framework by statistically describing how galaxies populate haloes of different masses (\citealt{PeacockSmith2000}; \citealt{2002PhR...372....1C}; \citealt{2005ApJ...630....1Z}). 

HOD modelling starts from the assumption that a dark matter halo develops a central galaxy of a given mass at some time in its evolution, and as this halo grows through gravitational processes, it then develops a population of satellite galaxies once it reaches a given total dark matter halo mass. The simplest HOD models often prescribe that there is a minimum halo mass that is required to form the first central galaxy of a given stellar mass and then a minimum halo mass, traced by the central galaxy, at which the first satellite of a given stellar mass is formed. The growth in the number of satellites galaxies is then often described by a power-law increase with halo mass. Variations from this basic model introduce a typical halo mass in addition to the minimum halos mass for a satellite to form and the addition of a smoothing term that allows there to be scatter in the halo mass and the central galaxy stellar mass. 
This type of HOD is similar to abundance matching \citep{2004ApJ...609...35K, 2004MNRAS.353..189V, 2006ApJ...647..201C, 2010ApJ...717..379B}, where galaxies and dark matter haloes are essentially ordered by mass and the highest stellar mass galaxies (often using a galaxy stellar mass function) are matched to the highest mass haloes (from $N-$body simulations) at a given epoch, again with some scatter in this relation. More complicated abundance matching techniques specifically address sub-haloes \citep[e.g.][]{Angulo2009}, assembly bias \citep[e.g.][]{Contreras2021} and aim to overcome mass-resolution effects in the $N-$body simulations \citep{2016MNRAS.459.3040G}. A related approach was adopted by \cite{2016MNRAS.460.1173R}, who used halo abundance matching applied to the BigMultiDark (N)-body simulation to reproduce the clustering of  SDSS-III/BOSS-CMASS (Sloan Digital Sky Survey III; \cite{2011AJ....142...72E},  Baryon Oscillation Spectroscopic Survey; \cite{2013AJ....145...10D}) galaxies at $0.43<z<0.7$. Further extensions of the standard HOD formalism also relax the assumption that halo mass alone determines galaxy occupation. For example, \cite{2016MNRAS.460.2552H} introduced the decorated HOD formalism, in which the occupation of galaxies may depend on secondary halo properties at fixed halo mass, thereby allowing galaxy assembly bias to be included within the HOD framework. \cite{2017MNRAS.472.3570H} introduced an analytic quenching term that allowed constraints on the impact of the central galaxy on the star-forming properties of satellite galaxies.

More complicated forms of HOD model aim to start from a functional form for the stellar-to-halo mass relation and its relation to the galaxy stellar mass function \citep{2011ApJ...738...45L}. This model with many more free parameters ensures that the galaxy stellar mass function and the stellar-to-halo mass relation, determined from the clustering of galaxies, are internally consistent, at the expense of complexity and potentially bedding-in underlying assumptions that are difficult to extract in the absence of other data such as gravitational lensing.
HOD-based analyses have therefore become a key tool for interpreting galaxy clustering measurements and linking them to theoretical models of structure formation \citep{2005ApJ...633..791Z}.

Clustering-based HOD studies have traced the galaxy–halo connection over an increasingly broad redshift range. At low redshift, SDSS analyses established that more luminous galaxies occupy more massive haloes, that red galaxies cluster more strongly than blue galaxies at fixed luminosity, and that luminous red galaxies predominantly reside as centrals in massive haloes \citep{2011ApJ...736...59Z,2009ApJ...707..554Z}. Comparisons between DEEP2 galaxies at $z\sim1$ and SDSS galaxies at $z\sim0$ further showed evolution in the central galaxy luminosity--halo mass relation, with brighter central galaxies at fixed halo mass at earlier times \citep{2007ApJ...667..760Z}.

Larger photometric surveys subsequently extended these measurements to intermediate redshifts. CFHTLS-Wide measurements over $0.2<z<1.2$ showed evolution in the luminosity-to-halo mass relation with redshift and galaxy type \citep{2012A&A...542A...5C}, while Hyper Suprime-Cam Subaru (HSC) measurements over $0.3<z<1.4$ found stellar-to-halo mass ratios broadly consistent with empirical expectations \citep{2020ApJ...904..128I,2013ApJ...770...57B,2019MNRAS.488.3143B}. Near-infrared surveys have extended such analyses to higher redshift: UltraVISTA and VIDEO studies found that more massive galaxies inhabit more massive haloes and that galaxy bias increases with stellar mass and redshift \citep{2015MNRAS.449..901M,2016MNRAS.459.2618H}. SMUVS further probed the galaxy--halo connection to $z\sim5$ \citep{2018ApJ...853...69C}, while cross-correlations of red and blue populations have provided evidence for galaxy conformity and constraints on halo quenching \citep{2017MNRAS.472.3570H,Paquereau2026}.

At $z\gtrsim2$, spectroscopic surveys have begun to extend HOD constraints on the galaxy--halo connection into the early Universe. Using VUDS galaxies over $2<z<5$, \citet{2015A&A...583A.128D} measured the evolution of galaxy clustering and inferred the characteristic halo masses and stellar-to-halo mass ratios of high-redshift star-forming galaxies.
At $z\gtrsim3$, Lyman-break galaxy clustering has enabled constraints on the galaxy--halo connection into the early Universe. These studies generally find significant evolution in star-formation efficiency and the stellar-halo mass relation (SHMR), although the inferred trends depend on redshift and galaxy selection \citep{2017ApJ...841....8I,2016ApJ...821..123H,Dalmassco2024,Ye2025}. More recently, joint clustering and abundance analyses spanning $0.2<z<5.5$ have found that the halo mass associated with peak star-formation efficiency evolves with redshift \citep{2022A&A...664A..61S}. Combining multiple fields has also demonstrated the importance of cosmic variance: \cite{2024arXiv241104256Z} found that a COSMOS-only analysis can shift the inferred $M_{\rm h}^{\rm peak}$ by up to $\sim0.15$ dex at $z<1.5$. COSMOS-Web measurements are now extending constraints on star-formation efficiency to still earlier epochs \citep{2025A&A...695A..20S}.

Despite this progress, the galaxy--halo connection remains poorly constrained at high stellar masses and high redshift, where galaxies are rare and field-to-field variance becomes increasingly important. Many previous studies have relied on the relatively small COSMOS field, making both sample size and cosmic variance important limitations. Here, we combine the depth of UltraVISTA \citep{2012A&A...544A.156M} and VIDEO \citep{2013MNRAS.428.1281J} with coverage across COSMOS, XMM-LSS, and ECDF-S to study massive galaxies out to $z=4$. The resulting larger effective volume increases the high-mass galaxy sample, reduces sensitivity to cosmic variance, and improves clustering constraints, particularly on the two-halo term, allowing halo occupation to be probed in a regime that remains difficult to constrain with smaller-area surveys.

This paper is organised as follows. In Section \ref{sec:Data}, we provide a detailed overview of the optical and near-infrared galaxy catalogues we created and used in this study, as well as the data processing and sample selection process adopted to obtain a final sample. 
In Section \ref{sec:TPCF}, we discuss the angular two-point correlation function and its application in measuring the clustering of galaxies. 
The HOD model and its associated parameters, as well as the fitting process, are then explained in Section \ref{sec:HOD}. 
The observed angular correlation functions for galaxies in different redshift and stellar mass bins, along with the results of the HOD model fitting are presented in Section \ref{sec:Results}, followed by a discussion of the implications for the galaxy-halo connection. 
Finally, we conclude and summarise our findings in Section \ref{sec:Conclusions}.

Throughout this work, all magnitudes are given in the AB magnitude system \citep{1983ApJ...266..713O} and a standard $\Lambda$CDM cosmology is assumed $H_{0} = 70$\,km s$^{-1}$\,Mpc$^{-1}$, $\Omega_{\rm M} = 0.3$, $\Omega_{\Lambda} = 0.7$.

\section{Data} 
\label{sec:Data}

\subsection{Data Acquisition and Analysis} 
\label{sec:Data Acquisition and Analysis}
This section presents an overview of the optical and near-infrared data utilized in this study, encompassing three different fields: COSMOS, XMM Large Scale Structure (XMM-LSS), and Extended Chandra Deep Field South (ECDF-S).

The near-infrared data for the COSMOS field \citep{2007ApJS..172..150S} were sourced from the sixth and last data release (DR6) of the UltraVISTA Survey \citep{2012A&A...544A.156M}. The DR6 data provide a uniform depth over the full COSMOS field for the broad-band $Y J H Ks$ filters, as opposed to the `deep' and `ultra-deep' stripes from the previous data releases. 
Specifically, the $5\sigma$ depth in the deep stripes of the $Y$, $J$ and $H$ bands has increased substantially by $1.1$ mag and the $Ks$ band depth has improved by $0.3$ mag compared to the preceding DR4 and DR5 data. 
For the XMM-LSS and the ECDF-S fields the near-infrared data were obtained from the final release of the VISTA Deep Extragalactic Observations (VIDEO) survey \citep{2013MNRAS.428.1281J}.

The VISTA data are complemented by overlapping optical data in the $G R I Z y$ broad-bands from Data Release 3 of the Hyper Suprime-Cam Subaru Strategic Program (HSC-SSP, \citealt{2022PASJ...74..247A}), for the COSMOS and XMM-LSS fields. The HSC-SSP DR3 data measurements have an increase in $5\sigma$ depth of $\sim0.1-0.7$ mag compared to the previously released DR2 data \citep{2019PASJ...71..114A}.
For the ECDF-S field, we used HSC optical data by \cite{2019RNAAS...3....5N}, in the $G$, $I$, $Z$ bands. We excluded the single central HSC $R$-band pointing from our analysis due to its poor coverage. The HSC $I$ and $Z$ bands suffer from poor seeing as a result of varied observational conditions. 
Therefore, additional optical data were used for the ECDF-S field from the VST Optical Imaging of the CDFS and ES1 Fields (VOICE, \citealt{2016heas.confE..26V}) Survey, in the $u g r $ bands. 
We omitted the VST $i$-band due to significantly poorer sensitivity than the HSC data in this field.
The XMM-LSS and EDCF-F fields are both split into three VISTA tiles of $1.5$ deg$^2$ each, denoted XMM1, XMM2, XMM3 and CDFS1, CDFS2, CDFS3, respectively. The $5\sigma$ depths for each optical and near-IR photometric band used in the COSMOS, XMM-LSS and ECDF-S fields are shown in Table~\ref{tab:5s_depths}, and the vast majority of the data used have similar seeing-limited resolution of $\sim 0.8$\,arcsec, apart from some regions and filters in CDFS, which we compensate for as described below. We note that even with this resolution, some galaxies will be blended, particularly at low redshift. However, this will only effect the very small scale ($<10$ arcsec) clustering signal and does not impact on our analysis.

\begin{table*}
    \centering
    \caption{The $5\sigma$ detection depths for the COSMOS field and the XMM-LSS and ECDF-S tiles. The depths are computed using 2 arcsec diameter circular apertures positioned on empty regions of the image. The telescope associated with each filter band (i.e., VISTA, HSC, VST) is specified in the final column.}
    \begin{tabular}{ccccccccc}
    \hline \hline
         Filters &  COSMOS & XMM1 & XMM2 & XMM3 & CDFS1 & CDFS2 & CDFS3 & Telescope \\
         \hline
         $Y$&   25.9  & 25.2 & 25.1 & 25.2 & 24.6 & 24.9 & 24.6  &VISTA\\
         $J$&   25.8  & 24.7 & 24.7 & 24.7 & 25.3 & 25.3 & 25.2  &VISTA\\
         $H$&   25.5  & 24.2 & 24.3 & 24.3 & 24.1 & 24.3 & 24.3  &VISTA\\
         $Ks$&  25.1  & 23.8 & 23.9 & 23.9 & 23.8 & 23.7 & 23.7  &VISTA\\
         \hline
         $G$&  27.4   & 27.6 & 27.1 & 27.1 & 26.5 & 26.2 & 26.4 &HSC\\
         $R$&  27.1   & 27.1 & 26.6 & 26.5 & - & - &  - &HSC\\
         $I$&  26.9   & 26.9 & 26.2 & 26.2 & 25.6 & 25.5 & 25.5 &HSC\\
         $Z$&  26.3   & 26.5 & 25.9 & 25.9 & 25.1 & 24.8 & 24.8 &HSC\\
         $y$&  25.3   & 25.6 & 24.7 & 24.7 & - & - &  - &HSC\\
         \hline
         $u$&  -   & - & - & - & 25.5 & 25.6 & 25.4  &VST\\
         $g$&  -   & - & - & - & 26.0 & 25.9 & 26.0  &VST\\
         $r$&  -   & - & - & - & 26.0 & 26.0 & 26.0  &VST\\
         \hline
    \end{tabular}
    \label{tab:5s_depths}
\end{table*}

Furthermore, for XMM-LSS and ECDF-S, we selected objects with $Ks$ magnitudes brighter than $23.8$, since the VIDEO survey achieves a completeness of 90 per cent at this magnitude limit \citep{2013MNRAS.428.1281J}. For the COSMOS sample, we applied $Ks$ magnitude cut-off depth of $Ks<25$. This selection allows us to focus on the brightest galaxies for subsequent parameter calculations.

\begin{figure*}
    \centering
    \includegraphics[scale=0.46]{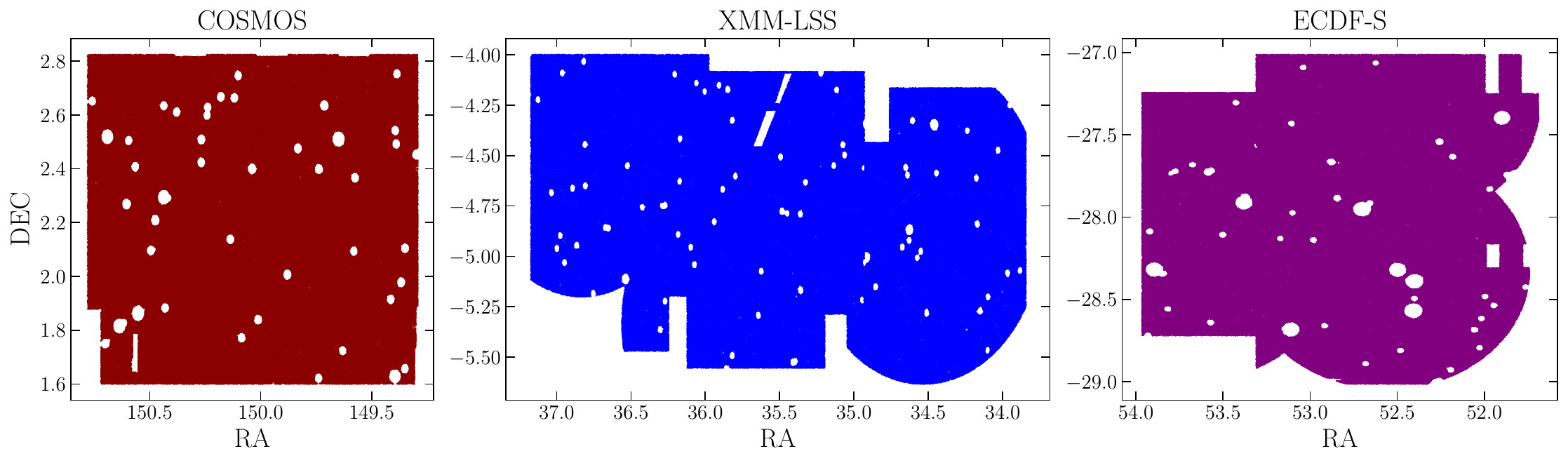}
    \caption{Spatial distribution of galaxies in each field after masking. It is important to note that the three fields are not shown to scale with each other, since the two VIDEO fields (XMM-LSS and ECDF-S) are about three times larger than the COSMOS field.}
    \label{fig:masking_plots}
\end{figure*}

The source detection in all three fields was conducted in the $Ks$ band image using SExtractor \citep{1996A&AS..117..393B}, and subsequent forced photometry was applied to capture data from all other bands. A circular diameter of 2 arcsec was used for aperture photometry. 
To account for additional noise and variations in the HSC and VST data due to the variable seeing in the ECDF-S field, certain adjustments and corrections were required. We adjusted the background mesh size to enable SExtractor to capture and remove small-scale variations in the background on scales down to 32 pixels, thus reducing residual noise structures across the field and leading to more uniform source detection and photometry. Atmospheric seeing variations cause the Point Spread Function (hereafter PSF) to vary across the ECDF-S field, scattering light unevenly and biasing aperture photometry. To account for this, we applied a local aperture flux PSF correction to each object using a PSF grid. This grid quantifies the amount of light enclosed within a specific aperture size for a given PSF across each filter. This approach scales the aperture fluxes in each band as if they were observed under a common PSF, thereby homogenising photometry and reducing colour biases and subsequent incorrect photometric redshift estimates.

A mask was further applied to each catalogue to exclude any faulty detector regions with dead pixels and nearby bright stars. It is essential to note that while the COSMOS field is entirely covered by the HSC pointings, in the case of the XMM-LSS and ECDF-S fields, additional masking was implemented to eliminate any regions without HSC coverage. Similarly, extra masking was applied to condense the ECDF-S field into the VOICE coverage region. The plots in Figure \ref{fig:masking_plots} illustrate the results of the masking process for each field. Therefore, the total combined area utilised in this study encompasses $\sim9.3$ deg$^2$, comprising $1.7$ deg$^2$ for the COSMOS field, $4.3$ deg$^2$ for the XMM-LSS field and $3.3$ deg$^2$ for the ECDF-S field.

\subsection{Photometric Redshift Estimates} 
\label{sec:photo-zs}

The photometric redshifts for the COSMOS and XMM-LSS sources were mainly obtained from \cite{Hatfield2022}, which used the same data. \cite{Hatfield2022} used a combination of template fitting and machine learning techniques in 10 filters (4 VISTA filters, 5 HSC filters and the u filter from the Canada-France-Hawaii Telescope Large Area U-band Deep Survey, CLAUDS; \citealt{2019MNRAS.489.5202S}) to obtain accurate estimates of photometric redshifts. For template fitting estimates, spectral energy distribution (SED) fitting was performed using LEPHARE (\citealt{1999MNRAS.310..540A}; \citealt{2006A&A...457..841I}) using galaxy and active galactic nuclei templates following the approach outlined in \cite{2020MNRAS.494.1771A} and for the machine learning-based predictions the algorithm GPz was employed (see \citealt{2016MNRAS.462..726A}; \citealt{2016MNRAS.462..726A}). Additionally, a Hierarchical Bayesian model was utilised to combine these two estimates (the method for which is further discussed in \citealt{2018MNRAS.477.5177D}), resulting in enhanced photo-$z$ prediction accuracy compared to the utilisation of each method separately. The photometric redshift estimates from \cite{Hatfield2022} were found to be of comparable quality to those from COSMOS2020 (\citealt{Weaver2022}), despite the latter having a larger number of bands in their analysis. We therefore use the \cite{Hatfield2022} photometric redshifts for consistency across the three fields.
In addition to obtaining photometric redshift measurements, we also used the probability density functions (PDFs) for each redshift estimate from \cite{Hatfield2022}. These PDFs provide a measurement of the uncertainty associated with the photo-$z$ estimates.

For the ECDF-S field, we do not apply the same hybrid method with the machine learning algorithm GPz, since the filter set for ECDF-S is very different (as discussed in Section \ref{sec:Data Acquisition and Analysis}). Consequently, a model trained with COSMOS and XMM-LSS filters cannot easily be used to infer redshifts in ECDF-S.
Therefore, for the ECDF-S field and any unmatched COSMOS and XMM-LSS sources, we used LePHARE to get photometric redshift measurements, which uses SED fitting techniques and $\chi^2$ minimisation to best match the observed photometry with a library of redshifted galaxy templates. We utilise the  COSMOS SED input template provided by LePHARE, which consists of 32 spectral energy distributions used in the \cite{2009ApJ...690.1236I} photo-$z$ study for the COSMOS field. These templates have been adjusted to account for dust attenuation using the extinction law of \cite{2000ApJ...533..682C}, with $E(B - V)$ values from $0.0$ to $1.5$.
The photometric flux errors were set to be a minimum of 5 per cent and only bands with available data for each individual galaxy were included in the template fitting process. The result of LePHARE's SED fitting is a best-fit photometric redshift measurement, and a corresponding Probability Density Function (PDF) for each redshift estimate.

\subsection{Spectroscopic Redshifts} \label{sec: Spectroscopic Redshifts}
Additionally, we have used spectroscopic redshifts (spec-z) where available for all three fields from The Spitzer Spectroscopic Data Fusion, \cite{vaccari_2025_15176245}, which merges miscellaneous spectroscopic information available within different extragalactic survey fields, including the recent data compilation of \cite{Khostovan2026} in the COSMOS field. We cross-matched the galaxy and spectroscopic catalogues using a maximum separation of 1 arcsec, with most of the resulting matches having angular separations well below this limit.

The number of galaxies with spec-$z$ used in this paper is $34,105$ for COSMOS, $49,053$ for XMM-LSS, and $16,942$ for ECDF-S.
For these galaxies, we adjust their PDFs to reflect their spec-$z$ values, with the probability set to one where each spec-$z$ value is located and zero elsewhere, hence reflecting the precision of \textbf{spectroscopic} redshifts compared to photometric redshifts.

\subsection{Sample Selection} \label{sec: Sample Selection}

\subsubsection{Galaxy -- Star Separation} \label{sec: Star Elimination}

\begin{figure*}
    \centering
    \includegraphics[scale=0.34]{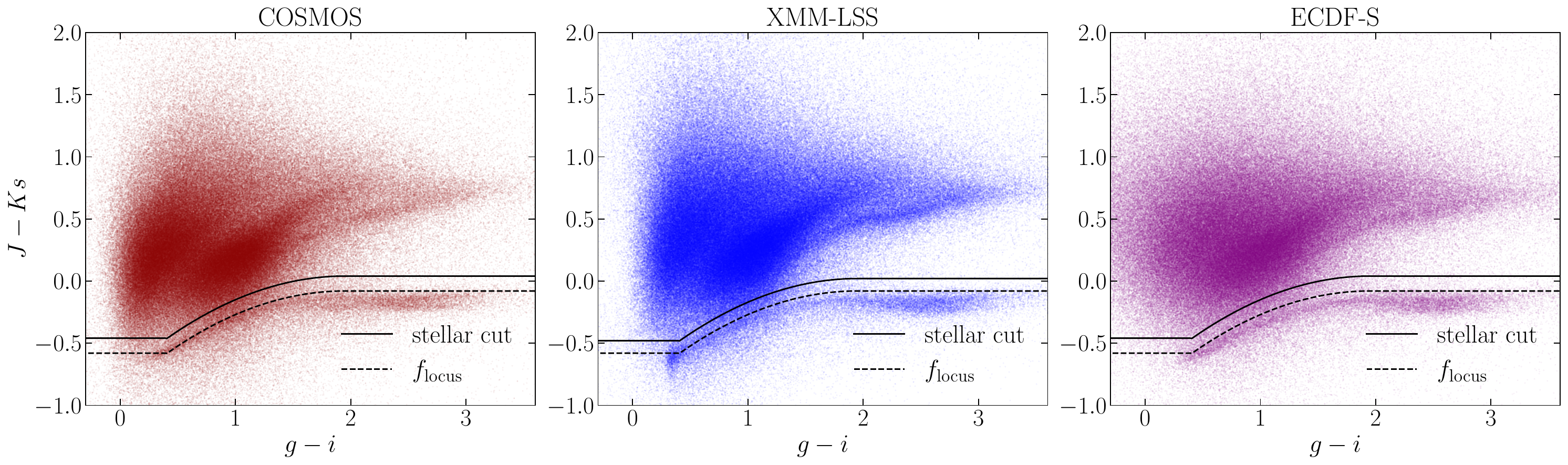}
    \caption{Colour-colour diagrams on all three fields with $G-I$ from HSC versus $J-Ks$ from VISTA. The plots depict sources with $Ks < 25$ in COSMOS and $Ks < 23.8$ in XMM-LSS and ECDF-S. The dashed line indicates the stellar locus fit in the optical/NIR data and the solid line denotes the stellar cut separating the stars at the bottom from the galaxies at the top.}
    \label{fig:colour-colour_diagrams}
\end{figure*}

To exclude stars from our galaxy catalogue, we adopt a colour-based criterion as described by \cite{2013MNRAS.428.1281J} that follows the methodology of \cite{2010MNRAS.404...86B}. We define a stellar locus ($f_{\mathrm{locus}}$) as previously used by \cite{2016MNRAS.459.2618H} and \cite{2012MNRAS.423..132M} in a colour-colour diagram of $J - Ks$ versus $G - I$. The $f_{\mathrm{locus}}$ is expressed by the following quadratic equation:
\begin{align}
                    &-0.58                       & x \leq 0.4 \nonumber \\
    f_{\mathrm{locus}}(x)=   &-0.88 + 0.82x - 0.21x^2     & 0.4<x\ \leq 1.9 \\
                    &-0.58                       & x > 1.9 \nonumber
\end{align}
where $x = G - I$. Any sources whose colours fall below this criterion 
$$J-Ks < 0.12 + f_{\mathrm{locus}}(G-I)$$ 
are then removed from our sample (see Figure \ref{fig:colour-colour_diagrams}).

\subsubsection{Point Source Removal} \label{sec: point-source removal}
We anticipate the majority of quasars to be considerably brighter than their associated host galaxies in the $Ks$ band. According to \cite{2015MNRAS.448.2665W}, in the VIDEO-depth data, a morphology cut based on the SExtractor K\_CLASS\_STAR parameter can be used to distinguish quasars. This parameter quantifies the degree of compactness of a source on a scale of 0 to 1,  where a value of 1 indicates a point-like appearance, which is the expected morphology for a quasar. Specifically, \cite{2015MNRAS.448.2665W} found that below a $Ks$-band magnitude of $\sim22.4$, the K\_CLASS\_STAR parameter can reliably identify a point-like source (see Figure 2 in \citealt{2015MNRAS.448.2665W}). Therefore, we consider an object to be point-like and remove it from our sample if $Ks < 18$ and $\text{K\_CLASS\_STAR} > 0.95$, or if $18\leq Ks \leq 22$ and $\text{K\_CLASS\_STAR} > 0.98$.

\subsubsection{Stellar Mass Estimation} \label{sec:Stellar Mass Estimation}
To estimate the stellar masses of the galaxies, we use LePHARE's spectral energy distribution fitting method by fixing the redshift of each galaxy to either the spectroscopic redshift where available, or the best-fit photometric redshift otherwise, during the template fitting process. Additionally, we utilise the total fluxes instead of aperture flux measurements as they provide a more accurate representation of a galaxy's total luminosity and stellar mass content and hence ensure more precise estimates during the SED fitting.
For the stellar mass estimation, we used a set of templates based on \cite{2003MNRAS.344.1000B}. These SED templates include two metallicities (one $Z_\odot$ and one 20 per cent $Z_\odot$) and for each there are $10$ star formation histories available; one with constant star formation rate and $9$ different exponentially decaying star formation rates of time-scales, $\tau = 0.1, 0.3, 1, 2, 3, 5, 10, 15, 30$ Gyr and finally they offer 58 ages for the stellar population spanning from $0.01$ Gyr to $13.5$ Gyr.

We note that AGN may also contribute to the emission, and artificially enhance the flux and thus we would overestimate the stellar mass. The AGN-dominated sources, i.e. those with bright nuclear emission, should be removed by our point source removal discussed in Section~\ref{sec: point-source removal}. Low-level AGN contamination could still be an issue, however, this is not likely to have a large affect on the near-infrared emission which provides the strongest constraints on the stellar masses as any obscured AGN would still need to be highly accreting to heat the dust to very high temperatures that increased the near-infrared emission. Furthermore, the main source of uncertainty on stellar mass for many of our objects arises the photometric redshifts, rather than the specific spectral energy distribution fit given a redshift.

\subsubsection{Stellar Mass Completeness} \label{sec: Stellar Mass Completeness}
\begin{figure*}
    \centering
    \includegraphics[scale=0.39]{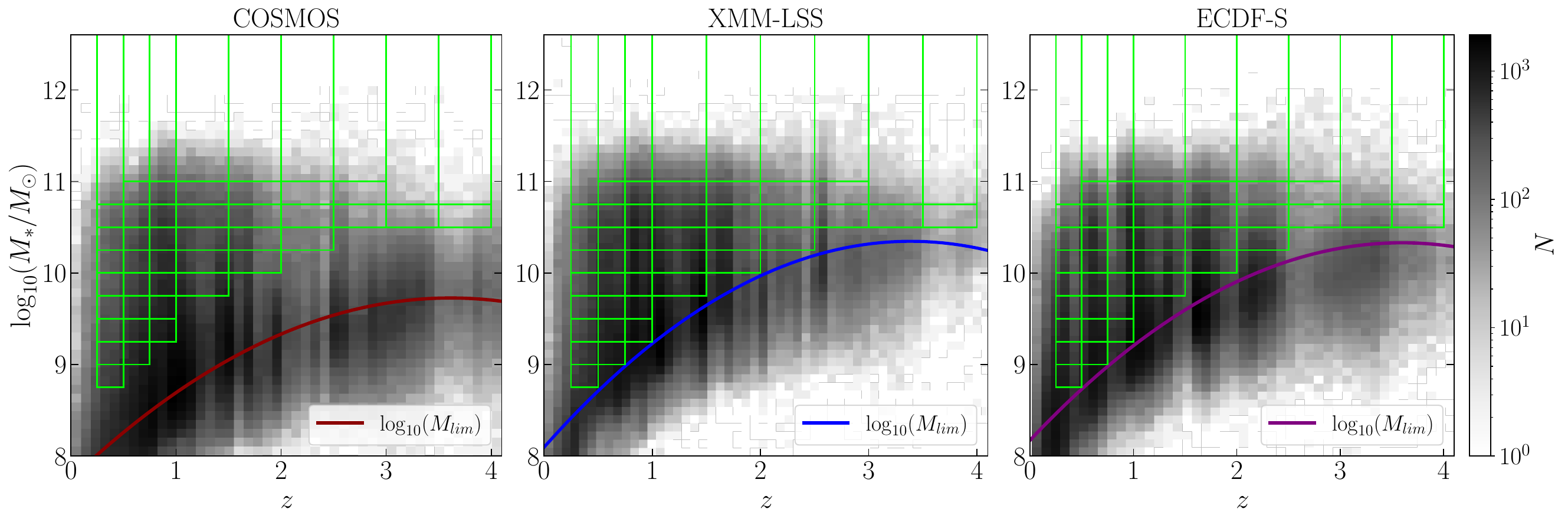}
    \caption{Stellar mass against the redshift of galaxies in COSMOS, XMM-LSS and ECDF-S after applying masks, a $Ks$ magnitude cut, star and point source elimination. The plots are coloured by the number of galaxies in each field, indicated in the colour-bar. The $\log_{10}(M_{\rm lim})$ lines represent the stellar mass completeness limits for each field as described in Section \ref{sec: Stellar Mass Completeness}. The green boxes depict the stellar mass and redshift sample selections used in subsequent analysis. These plots also show how galaxies from each field contribute to different parts of the stellar mass-redshift parameter space.}
    \label{fig:stellar_mass_vs_redshift}
\end{figure*}

We determine the stellar mass completeness limit using the methodology outlined by \cite{2010A&A...523A..13P}. The stellar mass completeness limit ($M_{\rm lim}$) is described as the estimated mass each galaxy would have assuming its observed aparent $Ks$-band magnitude is equal to the adopted magnitude limit of the sample ($Ks_{\rm lim}$). In this case, the magnitude limit as assigned for the COSMOS field is $Ks_{\rm lim}=25$ and for the XMM-LSS and ECDF-S fields is $Ks_{\rm lim}=23.8$. Hence, we first calculate an individual limiting stellar mass for every galaxy using:
\begin{equation}
\log_{10}(M_{\rm lim})=\log_{10}(M_*) +0.4(Ks-Ks_{\rm lim})
\label{eq2: M_lim}
\end{equation}
To establish a representative boundary, the galaxies are divided into redshift bins and within each bin, we select only the faintest 20 per cent of galaxies in $Ks$. These galaxies lie closest to the detection limit and therefore provide the most relevant estimate of the stellar masses that could be missed at that redshift. Including brighter galaxies in this estimate would instead bias and artificially inflate the estimated mass completeness limit unnecessarily. Therefore, the stellar mass completeness limit in each redshift bin is taken to be the 90th percentile of the $M_{\rm lim}$ distribution for this faint subsample, meaning that 90 per cent of these near-limit galaxies would still be included above this stellar mass \citep[in accordance with][]{2016A&A...586A..23D,2021A&A...646A.151P,2024A&A...687A.117F}. The resulting redshift-dependent limits are then fitted with a quadratic function in $\log_{10}(M_\star)$, providing the smooth completeness curves shown in Figure \ref{fig:stellar_mass_vs_redshift} for each field.

The determination of the minimum stellar mass limit helps guide the selection of appropriate stellar mass thresholds for each redshift sample. This ensures that the samples are complete in terms of stellar mass and allows for accurate calculation of angular correlation functions in subsequent analyses.

\subsection{Final Samples} \label{sec: Final Samples}
\subsubsection{Redshift Bin Selection} \label{sec: Redshift Bin Selection}
After the selection process described in the previous sections, the final full catalogues consist of a total of $1,381,779$ galaxies ($459,278$ for COSMOS, $492,317$ for XMM-LSS and $428, 215$ for ECDF-S). 
We adopt a redshift range spanning $z=0.25$ to $z=4$ for the three fields because that is the limit to which we can reasonably accurately probe stellar mass, given the majority of stellar mass is contained in low-mass red stars, which emit most of their emission longward of 4000\,\AA\ in the rest-frame. Beyond $z=4$, the 4000\,\AA\ break is redshifted out of the K-band.
Our redshift bins have a separation of $0.25$ between $z=0.25$ and $z=1$ and of $0.5$ between $z=1$ and $z=4$ to retain a reasonable number of source per redshift and mass bin (Figure~\ref{fig:stellar_mass_vs_redshift}).

\subsubsection{Stellar Mass Threshold Selection} \label{sec: Stellar Mass Threshold Selection}
In addition to selecting samples based on redshift, we also choose samples based on stellar mass thresholds, which are determined starting from the stellar mass completeness limit calculated in Section \ref{sec: Stellar Mass Completeness}. We recognise that there are additional mass samples that could be selected and modelled individually for the COSMOS field since the mass completeness limits are lower than those of the XMM-LSS field. However, for this combined analysis of the three fields, we set the stellar mass thresholds to follow the higher mass completeness limit defined by the shallower data in the XMM-LSS and ECDF-S fields, as indicated by the blue and purple $M_{\rm lim}$ lines in Figure \ref{fig:stellar_mass_vs_redshift}. This is appropriate for our purposes, as our analysis is specifically focused on the high-mass galaxy regime. The stellar mass thresholds chosen for each redshift bin are indicated by the horizontal green lines in Figure \ref{fig:stellar_mass_vs_redshift}.

\subsubsection{Redshift Bin Probability Weighting} \label{sec: Redshift Bin Probability Weighting}
To determine the probability of a galaxy residing in a specific redshift bin, we divide the photo-$z$ PDFs of each galaxy into 9 corresponding redshift bins. As a result, during the clustering measurement, galaxies with a higher likelihood of existing in a particular redshift bin are assigned a greater weight in the calculation compared to those with lower probabilities.

When conducting a clustering analysis, various methods have been used to allocate galaxies in redshift bins. This can be done by selecting galaxies based on whether their peak redshift in the photo-$z$ PDF falls within the $z$-bin or if their lower and upper $1\sigma$ values are within the $z$-bin limits \citep{2024arXiv241104256Z} or within $\pm 0.5$ of the $z$-bin limits \citep{2022A&A...664A..61S}. Alternatively, one can assign weights to every galaxy according to how much of the probability density of the photometric redshift lies within the given redshift bin, in this case, all galaxies contribute to all bins if their redshift probability is non-zero in the given bin \citep[e.g.][]{2016MNRAS.459.2618H}. These approaches help to mitigate the impact of galaxies with highly uncertain redshifts, which could otherwise introduce errors into the analysis.

In this work, we chose to include the weighted contributions, based on the PDF of the photometric redshifts, from all the galaxies in the three available fields. We do, however, make some adjustments. To account for the effect of unreliable photo-$z$s, especially ones with long PDF tails systematically extending to lower or higher redshifts, we modify our photo-$z$ PDFs by essentially adopting a galaxy mass function prior, whereby we perturb the photometric redshift PDF by the expected number density in the given redshift bin, accounting for the fact that the co-moving volume in each redshift bin varies and that the expected number of massive galaxies is less than the expected number of low mass galaxies according to the shape of the mass function.

To weight the PDFs, we use the GSMF ($\phi$) from \cite{2021MNRAS.506.4933A} (for $0.1<z<2$) and from \cite{2021MNRAS.503.4413M} (for $2<z<4$), using the stellar mass of each galaxy in each $z$-bin. Incorporating both volume and galaxy stellar mass function weighting helps refine the redshift distributions and makes the weighted photo-$z$PDFs better reflect the true redshift likelihood, leading to more reliable clustering estimates. We note this has a small overall effect on our results, but smooths out the relations we find.

Obviously, to do this we also need to determine the stellar mass for a given observed galaxy across all redshift bins. However, the stellar masses for each galaxy were initially calculated based on the peak redshift ($z_{\rm peak}$) in the SED fitting (see Section \ref{sec:Stellar Mass Estimation}). 
To properly account for the contribution of all galaxies in each $z$-bin, we therefore correct the stellar masses to match each corresponding $z$-bin median. Assuming the SED fitting remains the same, then the stellar mass, to first order, scales with the near-infrared luminosity, which in turn scales with the luminosity distance squared. To address this, we scaled the stellar masses using the squared ratio of the luminosity distances between the $z_{\rm peak}$ and the target $z$-bin as shown by:
\begin{equation}
    M^{\ast}_{z-\rm bin} = M^{*}_{z_{\rm peak}} (D_{z-{
    \rm bin}} / D_{z_{\rm peak}})^{2}
\end{equation}
where $M^{\ast}_{z-bin}$ is the stellar mass of the galaxy at the median of the $z$-bin, $D_{z-bin}$ is the luminosity distance at that $z$-bin, $M^{\ast}_{z_{\rm peak}}$ is the stellar mass of the galaxy at the peak redshift 
and $D_{z_{\rm peak}}$ is the luminosity distance at the peak redshift. We note that this does not account for changes in mass-to-light ratios that may lead to slightly different stellar masses for different "best-fit" SEDs at different redshifts, but this effect is relatively small compared to the uncertainty in the photometric redshifts.

Table \ref{tab:hod_params_combined_table} presents a detailed breakdown of the mean and PDF-weighted stellar masses, as well as the number of observed galaxies, for the selected sub-samples, as determined by the modified PDFs.


\section{The angular two-point correlation function}
\label{sec:TPCF}

An accurate measurement of the clustering of galaxies in three-dimensional space is achieved through the spatial correlation function. However, it requires exact distance measurements between galaxies, and this can be challenging as they are affected by their peculiar velocities, which are not directly measured, and photometric redshifts do not provide sufficient accuracy for this purpose.
One way to overcome this limitation is by using the angular two-point correlation function denoted as $\omega(\theta)$, which utilizes the angular distance in the sky as a measure of separation between galaxies, see \cite{Peebles1980}.

\subsection{Measuring the Two-Point Correlation Function} \label{sec:Measuring the TPCF}
The Landy \& Szalay estimator is a commonly used method to calculate the angular correlation function, see \cite{1993ApJ...412...64L}.  
This estimator uses both the observed galaxy distribution and a random distribution to measure the excess probability of finding galaxies at different angular separations than would be expected for a random sample.
For a given angular separation $\theta$, the two-point correlation function, $\omega(\theta)$, is calculated as:
\begin{equation}
\omega(\theta) = \frac{DD(\theta) - 2DR(\theta) + RR(\theta)}{RR(\theta)}
\end{equation}
where $\omega(\theta)$ represents the angular two-point correlation function, $DD(\theta)$ is the normalised number of galaxy-galaxy pairs, $DR(\theta)$ is the normalised number of galaxy-random pairs, and $RR(\theta)$ is the normalised number of random-random pairs, at a specific angular separation $\theta$.

We produce a number of random galaxies ($3,000,000$ for COSMOS, $7,000,000$ for XMM-LSS and $6,000,000$ for ECDF-S) that cover the same region in RA and DEC as the real data, with identical masking regions applied. Using a substantially larger random galaxy sample than the observed galaxy sample ensures that the uncertainty in $\omega(\theta)$ is dominated by the data rather than by Poisson noise in the random pair counts.

The angular two-point correlation function (TPCF) calculation is weighted using the fraction of the PDF of the photometric redshift for each galaxy residing in the specified redshift bin, so galaxies with a higher likelihood of existing in a particular redshift-bin are given more weight.
The $DD(\theta)$, $DR(\theta)$ and $RR(\theta)$ pairs are normalised based on their weight, ensuring that each pair of galaxies in the data or random catalogues is appropriately accounted for. In the random catalogue, where there is no weight assigned, the normalisation factor corresponds to the number of random-random galaxy pairs, $RR(\theta)$, which can be calculated using $N_{R}(N_{R}-1)/2$, where $N_{R}$ represents the number of random galaxies. However, since weighted factors exist in our data sample due to the uncertainty in the photo-$z$s, each galaxy pair in the real data catalogue is normalised by a different factor. The normalisation factor for data - data pairs, $DD(\theta)$, is determined by $N_{d}(N_{d}-1)/2$, where $N_{d}$ is the sum of weights for real galaxies. Lastly, dividing by $N_{d} N_{R}$ normalises the contributions from data-random pairs, $DR(\theta)$.
We compute the TPCF measurements using the TreeCorr software package \citep{2004MNRAS.352..338J}\footnote{The relevant TreeCorr code can be found in this GitHub repository: \url{https://github.com/rmjarvis/TreeCorr}}.

\subsection{Covariance Matrix} \label{sec:Covariance Matrix}
The covariance matrix of the two-point correlation function accounts for correlations between different angular separation bins. This is important because each galaxy can contribute to multiple bins depending on its angular position and the chosen binning scheme. The full covariance matrix is calculated with the jackknife method:
\begin{equation}
    C_{i,j} = \frac{N_{patch} - 1}{N_{patch}}\sum_{k=1}^{N_{patch}} \left(\omega_k(\theta_i) - \bar{\omega}(\theta_i) \right) \times \left(\omega_k(\theta_j) - \bar{\omega}(\theta_j)\right)
\end{equation}
where $\omega_k(\theta_i)$ refers to the angular two-point correlation function at $i$th angular separation for the $k$th jackknife sample, and $\bar{\omega}(\theta_i)$ is the average angular correlation function from all the $N_{patch}$ jackknife samples, at the $i$th angular separation.
We implement this with TreeCorr by dividing the catalogues into patches, where $N_{patch}=24, 60, 46$ for COSMOS, XMM-LSS and ECDF-S, respectively, and recalculating the correlation function excluding one patch at a time. The number of patches in each field was chosen to give approximately the same area per patch across the three fields, while balancing the need for enough jackknife realisations to estimate the covariance matrix against maintaining patches that are sufficiently large and contain enough sources for each omission to produce a meaningful change in the measured correlation function.

\subsection{Integral Constraint} \label{sec:Integral Constraint}
The measured two-point correlation function is subject to underestimation at large angles due to the finite size of the individual fields. As we move towards the boundaries of the field, there are fewer galaxies and hence a decline in the observed correlation function measurements. The integral constraint ($K_{IC}$) aims to correct this negative offset \citep{1977ApJ...217..385G}, which generally begins to impact on the measured correlation function at 15-30 per cent of the largest angular scale probed, although it also depends on the amplitude of the signal and the field geometry.

The integral constraint is regarded as constant for a one-size field and can be estimated numerically \citep{roche1999angular} by:
\begin{equation}
    K_{IC} = \frac{\sum{RR(\theta)\omega_\mathrm{mod}(\theta)}}{\sum{RR(\theta)}}.
\end{equation}
In practice, the observed angular correlation ($\omega_{obs}$) is not a precise representation of the true angular correlation function due to the integral constraint. However, during modelling, the model represents the true angular correlation function ($\omega_{\rm mod}$), which allows us to use a corrected version of the model with the integral constraint included. Hence, instead of making upward corrections to the data as in \cite{2012A&A...542A...5C}, we can make downward corrections to the model to account for this discrepancy, similar to \cite[e.g.][]{2011MNRAS.416.3017B, 2016MNRAS.459.2618H,2022A&A...664A..61S}:

\begin{equation}\label{eq:K_ic_subtraction}
    \omega_{obs}(\theta) = \omega_\mathrm{mod}(\theta) - K_{IC} ,
\end{equation}
where $K_{IC}$ is re-evaluated for each $\omega_\mathrm{mod}$.

\section{Halo Occupation Distribution Modelling}
\label{sec:HOD}
\subsection{HOD Model Theory \& Parameters} \label{sec:HOD_model}
To model the two-point correlation function, we adopt the 5-parameter HOD model proposed by \cite{2005ApJ...633..791Z}. This model characterises the relationship between galaxies and their host dark matter halos, distinguishing between central and satellite galaxies within halos. 

The five parameters describe:
\begin{itemize}
    \item$M_{\rm min}$: the characteristic halo mass at which 50 per cent of haloes host a central galaxy
    
    \item$M_{1}$: the typical halo mass for a halo to host one satellite galaxy
    
    \item$M_{0}$: the halo mass below which no satellites are formed
    
    \item$\sigma_{\log_{10}M}$: the scatter in stellar mass at fixed halo mass  
        
    \item$\alpha$: the slope of satellites growing with halo mass 
\end{itemize}

For this work, we adopt an HOD model that uses the virial mass definition, the halo mass function by \cite{2016MNRAS.456.2486D}, and the halo bias model developed by \cite{2010ApJ...724..878T}. It also assumes a halo mass concentration relation based on  \cite{2008MNRAS.390L..64D}, and a Navarro-Frenk-White distribution profile for the halo and satellite over-density \citep{1997ApJ...490..493N}.

In the absence of galaxy–galaxy lensing information, the additional flexibility of more complex HOD prescriptions, such as the 11-parameter model \citep{2011ApJ...738...45L}, is difficult to constrain robustly from clustering alone and can introduce substantial parameter degeneracies. We therefore adopt the standard five-parameter HOD, which provides a sufficiently flexible description of the central and satellite occupations while limiting the number of poorly constrained degrees of freedom.

Starting from the halo model, we use the LSST Dark Energy Science Collaboration's Core Cosmology Library \citep[CCL; ][]{CCL}\footnote{The LSST Dark Energy Science Collaboration’s Core Cosmology Library (CCL) can be accessed from \url{https://github.com/LSSTDESC/CCL}} to compute the galaxy-galaxy power spectrum, $P_{gg}(k)$, which is divided into the commonly referred to 1-halo, $P_{gg}^{1h}(k)$, and 2-halo, $P_{gg}^{2h}(k)$, terms. The 1-halo term describes the clustering of galaxies within a halo, while the 2-halo term quantifies the clustering between separate haloes. These terms are given by:
\begin{equation}
P_{gg}(z,k) = P_{gg}^{1h}(z,k) + P_{gg}^{2h}(z,k) ,
\end{equation}
with
\begin{align}
P_{gg}^{1h}(k) &= \frac{1}{\langle n^{tot}_{g} \rangle^2} \int \mathrm{d}M_h \, \frac{\mathrm{d}n}{\mathrm{d}M_h} \, \langle N_{\mathrm{central}}|M_h \rangle \nonumber \\
&\quad \times \left[ \langle N_{\mathrm{sat}}|M_h \rangle^2 u_{\mathrm{sat}}^2(k) + 2 \langle N_{\mathrm{sat}}|M_h \rangle u_{\mathrm{sat}}(k) \right]
\end{align}
and
\begin{align}
    P_{gg}^{2h}(k) &= \left( \frac{1}{\langle n^{tot}_{g} \rangle} \int \mathrm{d}M_h \, \frac{\mathrm{d}n}{\mathrm{d}M_h} \, b_h(M_h) \, \langle N_{\mathrm{central}}|M_h \rangle \right. \nonumber \\
    &\quad \left. \times \left[ 1 + \langle N_{\mathrm{sat}}|M_h \rangle u_{\mathrm{sat}}(k) \right] \right)^2 P_{\mathrm{lin}}(k)
\end{align}
where $M_h$ is the halo mass, $\frac{\mathrm{d}n}{\mathrm{d}M_h}$ is the halo mass function, $b_h(M_h)$ is the halo bias, $u_{\mathrm{sat}}(k)$ is the Fourier transform of the normalised density profile of satellite galaxies and $P_{lin}(k)$ is the linear matter power spectrum. Additionally, $\langle N_{\mathrm{central}}|M_h \rangle$ and $\langle N_{\mathrm{sat}}|M_h \rangle$ are the mean number of central and satellite galaxies, respectively, while $\langle n^{tot}_{g} \rangle$ is the total mean galaxy density defined as:
\begin{equation}
    \langle n^{tot}_{g} \rangle = \int \mathrm{d}M_h \, \frac{\mathrm{d}n}{\mathrm{d}M_h} \, \langle N_{\mathrm{central}}|M_h \rangle \left[ 1 + \langle N_{\mathrm{sat}}|M_h \rangle \right] .
\end{equation}

\noindent The total number of galaxies in the model can therefore be determined by multiplying $\langle n^{tot}_{g} \rangle$ with the comoving volume.

The galaxy bias, $b_g$, can be viewed as the mean halo bias weighted by the number of galaxies in each halo, and it can be derived from the quantities mentioned above as such:
\begin{equation} \label{eq:galaxy_bias}
    b_{g} = \frac{1}{\langle n^{\mathrm{tot}}_{g} \rangle} \int \mathrm{d}M_h \, \frac{\mathrm{d}n}{\mathrm{d}M_h} \, b_h(M_h) \, \langle N_{\mathrm{central}}|M_h \rangle  \left[ 1 + \langle N_{\mathrm{sat}}|M_h \rangle \right]
\end{equation}

The central, satellite and total number of galaxies in a halo are parameterised by their halo occupation distributions in the following way:
\begin{equation} \label{eq:Ncent}
    \langle N_\mathrm{central}|M_{h} \rangle = \frac{1}{2} \left(1 + \text{erf} \left(\frac{\log_{10}M_{h} - \log_{10}M_{\rm min}}{\sigma_{\log_{10}M}} \right) \right) ,
\end{equation}
\begin{equation} \label{eq:Nsat}
    \langle N_\mathrm{sat}|M_{h}  \rangle = \Theta(M_{h} - M_{0}) \left(\frac{M_{h} - M_{0}}{M_{1}}\right)^{\alpha} ,
\end{equation}
\begin{equation} \label{eq:Ntotal}
    \langle N_\mathrm{total}|M_{h} \rangle = \langle N_\mathrm{central}|M_{h} \rangle \left[ 1 + \langle N_\mathrm{sat}|M_{h} \rangle \right] .
\end{equation}

\noindent We assume that satellite galaxies can only form in halos that already host a central galaxy. The term $\Theta(M_{h} - M_{0})$ is the Heaviside step function, which sets $\langle N_\mathrm{sat}|M_{h}  \rangle=0$ for $M_h < M_0$. Therefore, the satellite occupation function ensures that satellites reside only in halos with masses $M_h \geq M_0$.

Once the galaxy power spectrum, $P_{gg}(k,z)$, is computed, we project it into an angular power spectrum, $C_\ell$, using the Limber approximation. For a normalised galaxy redshift distribution $p(z)$, the angular power spectrum can be written as:
\begin{equation}
    C_\ell = \int {\rm d}z  \frac{H(z)}{c \chi^2(z)} p^2(z) 
    P_{gg}\left(k=\frac{\ell+1/2}{\chi(z)}, z\right),
\end{equation}
where $\chi(z)$ is the comoving distance, $H(z)$ is the Hubble parameter, and $p(z)$ describes the normalised redshift distribution of the galaxies (see Section~\ref{sec:RedshiftDistribution}). The factor $H(z)/c$ arises from converting between integration over comoving distance and integration over redshift.

We then obtain the modelled angular two-point correlation function, $\omega_{\rm mod}(\theta)$, from $C_\ell$ through:
\begin{equation}
    \omega_{\rm mod}(\theta) =
    \sum_\ell \frac{2\ell+1}{4\pi}  C_\ell 
    P_\ell(\cos\theta)
\end{equation}
where $P_\ell(\cos\theta)$ is the Legendre polynomial. This expression is for galaxy-galaxy clustering, where both density fields are scalar and have zero spin. This transforms the clustering signal from harmonic space into the real-space angular correlation function, giving the excess probability of finding galaxy pairs separated by an angle $\theta$.

\subsection{Cosmic Variance}
\label{sec: cosmic variance}
Accurately determining the number density of galaxies is fundamental for understanding the galaxy-halo relationship, and it directly informs the HOD modelling. However, number density measurements are subject to uncertainties arising from the statistical uncertainty that occurs when observations are limited to a finite volume of the universe. This uncertainty arises because the universe's matter distribution is not uniform; it contains regions of low and high density. Consequently, cosmic variance becomes particularly significant in surveys with limited volumes, where the sampled regions may not encompass the full diversity of cosmic environments. This limitation can result in systematic biases in estimating galaxy volume densities, as certain density fluctuations may be over- or under-represented. For instance, deep galaxy surveys covering small areas are especially susceptible to cosmic variance, which can dominate over Poisson errors and, for example, it could lead to galaxy stellar mass function estimates that do not accurately reflect the universal galaxy stellar mass function at a given redshift. Our multi-field approach provides an important advantage over single-field studies, because the larger effective survey volume and independent lines of sight reduce the impact of cosmic variance, allowing for the inferred HOD parameters in subsequent sections to be more representative of the wider galaxy population.

\cite{2004ApJ...600L.171S} offered predictions that can be used to measure the cosmic variance based on the mean redshift and survey volume, however, galaxy clustering and bias are strongly dependent on stellar mass, making it crucial to account for this mass dependence when estimating cosmic variance. Therefore, in this work, we calculate the uncertainty arising from cosmic variance, following the guidelines provided in the 'Cosmic variance cookbook' by \cite{2011ApJ...731..113M}.

We first measure the dark matter cosmic variance ($\sigma_{\mathrm{dm}}$) as a function of mean redshift ($\bar{z}$) and redshift bin size ($\Delta z$). \cite{2011ApJ...731..113M} estimated $\sigma_{\mathrm{dm}}$ by integrating the dark matter correlation function, derived from CDM theory, over narrow, deep sections of the sky (pencil beam geometry), at a given redshift $\bar{z}$, assuming a reference bin size of $\Delta z_{\rm ref} = 0.2$. The dark matter cosmic variance for different $\Delta z$ can be estimated as follows:
\begin{equation}
    \sigma_{\mathrm{dm}}(\Delta z, \bar{z}) = \sigma_{\mathrm{dm}}(\Delta z_{\rm ref} = 0.2, \bar{z})\sqrt{\frac{0.2}{\Delta z}}.
\end{equation}

\begin{figure}
    \centering
    \includegraphics[scale=0.42]{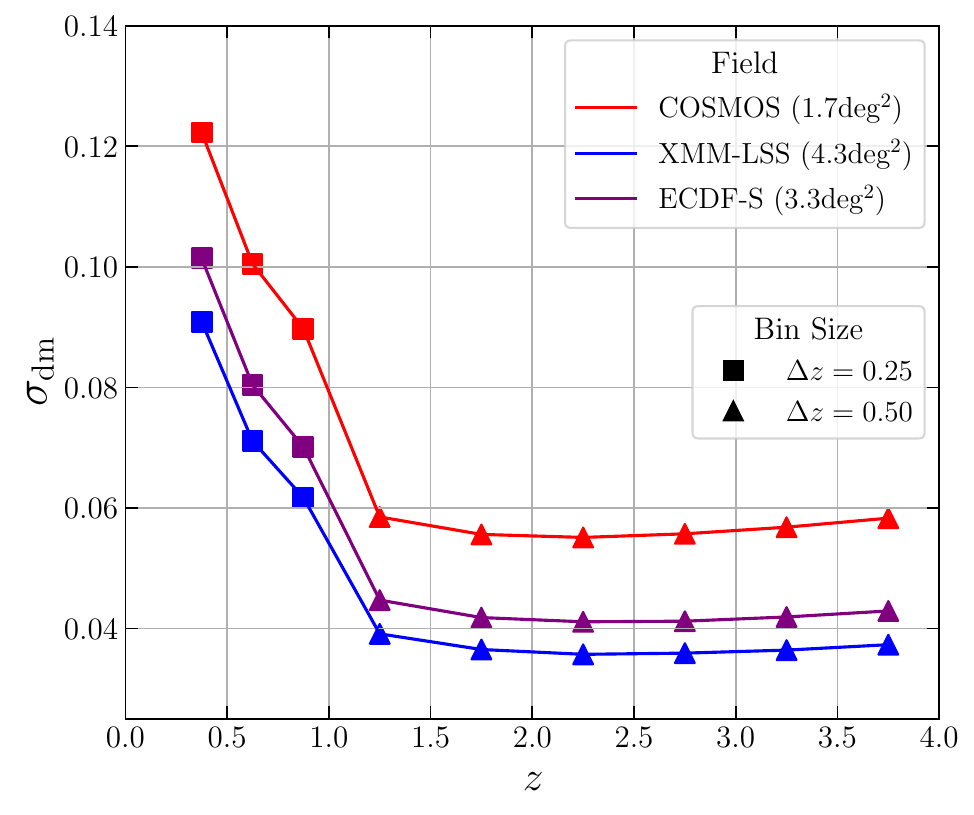}
    \caption{Dark matter cosmic variance, $\sigma_{\mathrm{dm}}$, as a function of mean redshift for the three fields used. The different markers represent the different redshift bin sizes $\Delta z$ that match the $z$-bins used in subsequent analyses.}  \label{fig:sigma_dm_plot}
\end{figure}

In Figure \ref{fig:sigma_dm_plot}, we show the dark matter cosmic variance across our three survey fields as a function of mean redshift, with the corresponding $\Delta z$ values indicated. As anticipated, $\sigma_{\mathrm{dm}}$ decreases with increasing field area and redshift bin size. This trend reflects the fact that larger volumes sample a more representative portion of the Universe, thereby reducing the influence of large-scale density fluctuations.

The cosmic variance $\sigma_{\mathrm{cv}}$, can then be calculated by multiplying the dark matter cosmic variance $\sigma_{\mathrm{dm}}$ with the galaxy bias $b(M_{\ast}, \bar{z})$ at a given redshift:
\begin{equation}
    \sigma_{\mathrm{cv}}(M_{\ast}, \bar{z}, \Delta z) = b(M_{\ast}, \bar{z}) \times \sigma_{\mathrm{dm}}(\Delta z, \bar{z}),
\end{equation}
where the $b(M_{\ast}, \bar{z})$ depends on the stellar mass range of the sample, and it is calculated during the HOD modelling for the specific stellar mass range defined by each sample. The resulting $\sigma_{\mathrm{cv}}$ is expressed as a fraction, and we then multiply it by the observed number of galaxies, as a proxy for the expected number of galaxies, to estimate the cosmic variance on the expected number of galaxies. We note that although this accounts for the cosmic variance on the total number of galaxies, it does not account for the variance on the measurement of the two-point correlation function. As pointed out by \cite{2018MNRAS.477.3760H}, two populations
can have the same average number counts, but different cosmic variance, i.e. they have the same 1-point statistics, but different 2-point statistics. 
Bearing in mind this caveat, we calculate $\sigma_{\mathrm{cv}}$ based on the number density for each redshift and stellar mass range in each field separately and then add it in quadrature to the Poisson uncertainties $\sigma_{\rm Poisson}$ (see Section~\ref{sec:Fitting_process}).

\subsection{Redshift Distribution} \label{sec:RedshiftDistribution}

A key element in converting the theoretical 3D galaxy power spectrum of a population of galaxies into the projected angular spectrum is the redshift distribution of galaxies within the given $z$-bin. This is non-trivial for samples dominated by photometric redshifts, where in essence every galaxy across all redshifts could contribute some weight to the given redshift bin, but the summation of all weights over all redshifts would obviously not reflect the true redshift distribution of galaxies from which the clustering is measured. Various methods have been used to determine the $N(z)$ for samples dominated by galaxies with photometric redshifts. For example, \cite{2015MNRAS.449..901M} adopts the best-fit photometric redshifts and checks the level of potential contamination using cross-correlations with different redshift bins,
whereas \cite{2016MNRAS.459.2618H} measure their TPCF using the weighted contributions from the photometric PDFs within the redshift bin of interest.  Both of these methods leads to sharp boundaries at the bin edges, which may not be truly representative of the effective $N(z)$.
\cite{2022A&A...664A..61S} adopt an alternative method of adopting an additional prior on the photometric redshifts based on the apparent magnitudes of the galaxies within the sample, coupled with softer edges in the binned $N(z)$s by requiring that each galaxy has its lower and
upper $1\sigma$ values on the photometric redshift within $\pm 0.5$ of the redshift-bin limits.
\cite{2024arXiv241104256Z} construct the redshift distributions by stacking the individual photometric-redshift probability distributions ($P(z)$) of galaxies in each redshift bin, while requiring the 16th and 84th percentiles of each galaxy’s redshift PDF to lie within the $z$-bin limits.

However, there is no perfect solution, as the photometric redshift probability distributions often do not adequately reflect the real probability distributions \citep[see e.g.][]{Schmidt2020}, but can be biased on a multitude of directions depending on the intrinsic galaxy colour, templates used, filter combinations and depth of the data.

\begin{figure*}
    \centering
    \includegraphics[scale=0.32]{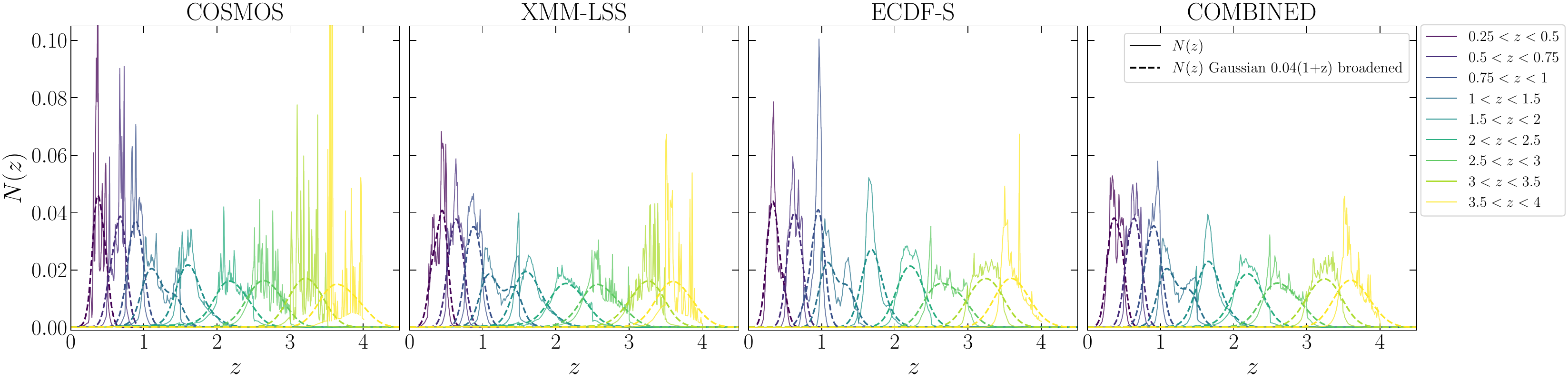}
    \caption{The normalised $N(z)$ redshift distributions (solid lines) of the COSMOS, XMM-LSS and ECDF-S fields and COMBINED are shown for each redshift bin used in this study. The $N(z)$ distributions broadened by a $0.04(1+z)$ Gaussian that are used during the modelling are also shown in dashed lines for each field.}
    \label{fig:N_z_plot}
\end{figure*}

Naively, a simple way to compute the redshift distribution $N(z)$ is to leverage the photometric redshift likelihoods provided by a photo-$z$ estimation code for each galaxy. In this framework, each galaxy is assigned a likelihood function, $L(\boldsymbol{o}|z)$, which represents the probability of observing its set of fluxes, $\boldsymbol{o}$, at a given redshift. By stacking these likelihoods for all galaxies in a selected sample, we get:
$$N(z) = \sum_{i}^{N_{\rm sample}} L_{i}(\boldsymbol{o}|z)$$
This approach naturally incorporates the uncertainty in photometric redshifts, acknowledging that a galaxy’s true redshift might lie outside the defined bin. 

However, \cite{2021A&A...647A.117E} have demonstrated that simply using the output likelihood $L(\boldsymbol{o}|z)$ from a photo-$z$ estimation code to build $N(z)$ can introduce biases in the mean redshift, reaching $\sim 0.01(1+z)$. Their findings compare this simply constructed $N(z)$ with both the actual redshift distribution and the probability integral transform (PIT) statistic, suggesting that the method can lead to mis-calibrated $N(z)$ distributions. 

Therefore, following \cite{2021A&A...647A.117E}, we adopt a more refined approach for computing $N(z)$ by incorporating a magnitude-dependent prior, $Pr(z|m_{0})$ (as described in \citealt{2006ApJS..162...20B}), which accounts for the magnitude-dependent nature of the redshift distribution and corrects for potential biases. The photo-$z$ prior $Pr(z|m_{0})$ is constructed for every $z$-bin by splitting the galaxies into magnitude bins and summing their likelihoods, as such:

$$Pr(z|m_{0}) = \sum_{i}^{N_{z-{\rm bin}}} L_{i}(\boldsymbol{o}|z) \Theta(m_{0,i} |m_{0})$$
where $\Theta(m_{0,i} |m_{0})$ is equal to 1 if the object’s magnitude $m_{0,i}$ falls within the magnitude bin centred at $m_{0}$ and 0 if it falls outside.
The redshift distribution can therefore be computed as follows:

$$N(z) = \sum_{i}^{N_{\rm sample}} L_{i}(\boldsymbol{o}|z) Pr(z|m_{0})$$

To obtain a representative $N(z)$, we first consider the probability contribution of each galaxy to the redshift bin. We include galaxies for which the combined weight within the $z$-bin, given by the sum of their $P(z)$ over that redshift interval, is above 0.2\footnote{We also test weight fractions of 0.3 and 0.5 and this does not change our results within the uncertainties.}. This means that galaxies are selected based on their probability of contributing sufficiently to each $z$-bin. By incorporating this and a magnitude prior suggested by \cite{2021A&A...647A.117E}, we construct a redshift distribution $N(z)$ that more accurately represents the true underlying distribution of the galaxy sample and thereby helps mitigate potential biases that could be introduced into the angular correlation function.

Moreover, \cite{2022A&A...664A..61S} shows that while mean redshift biases have only a modest impact (around a 3 per cent change) on the angular correlation function, as discussed by \cite{2021A&A...647A.117E}, misrepresentations in the shape of $N(z)$, especially the width of its wings, can result in significant discrepancies in $\omega(\theta)$ (up to 20–50 per cent differences at certain redshifts). 

We therefore convolve each $N(z)$ with a Gaussian with width $0.04(1+z)$, used here as a representative photometric-redshift uncertainty. The purpose of this step is to broaden the wings of the $N(z)$ redshift distribution, since the wings can have a strong effect on the predicted angular clustering. Physically, a broader redshift distribution projects the galaxy sample over a larger line-of-sight distance. As a result, some galaxy pairs that are close in angular separation are not close in three-dimensional space, and their contribution to the true clustering signal is weaker. This projection effect lowers the angular correlation amplitude, and therefore, a wider $N(z)$ can help avoid over-predicting the clustering amplitude. We test other smoothing  widths of $0.02, 0.03$ and $0.05(1+z)$ and find that our results are robust against this choice. We return to this in Section~\ref{sec:Fitting_process}, where we discuss the impact of this choice on the fitted HOD parameters. 


Figure \ref{fig:N_z_plot} shows the $N(z)$ redshift distribution and its applied Gaussian broadening for the three fields separately and combined, per redshift bin used in this paper. The observed spikes in the $N(z)$ distribution are representative of the certainty reflected in the likelihood function, $L(\boldsymbol{o}|z)$, of galaxies with spectroscopic redshifts.

\subsection{Fitting Process} \label{sec:Fitting_process}
We fit the angular two-point correlation functions and estimate the parameters of our model (described in Section \ref{sec:HOD_model}) using a Markov chain Monte Carlo (MCMC) approach, executed in the {\sc emcee} code \citep{2013PASP..125..306F}, by minimising the $\chi^2$ value.

\begin{table}
    \centering
    \caption{Priors adopted for the HOD parameters used in the MCMC fitting. Mass parameters are expressed in log-space. The Gaussian prior on $\sigma_{\log_{10} M}$ is truncated at $\mu \pm 3\sigma$.}
    \label{tab:priors}
    \begin{tabular}{lcl}
    \hline
    Parameter & Prior type & Prior range \\
    \hline
    $\log_{10}(M_{\rm min}/{\rm M}_{\odot})$ & Uniform & $[10,\,15]$ \\
    $\log_{10}(M_1/{\rm M}_{\odot})$ & Uniform & $[\log_{10}(M_{\min}),\,15]$ \\
    $\log_{10}(M_0/{\rm M}_{\odot})$ & Uniform & $[8,\,\log_{10}(M_1)]$ \\
    $\sigma_{\log_{10} M}$ & $\mathcal{N}(\mu=0.3,\,\sigma=0.1)$ & $\mu \pm 3\sigma\;([0.0,\,0.6])$ \\
    $\alpha$ & Uniform & $[0.5,\,1.5]$ \\
    \hline
    \end{tabular}
\end{table}

We use the priors shown in Table \ref{tab:priors} for each HOD parameter, with the halo masses set in log-space, and we randomly initialise 30 walkers within these parameter bounds for 3000 steps, with a 10 per cent burn-in phase. As a convergence test, we repeated the fitting in each redshift bin using adaptively extended chains, requiring the post-burn-in chain length to exceed 50 times the integrated autocorrelation time, $\tau$, for every fitted parameter, and successive estimates of $\tau$ to agree to within 1 per cent. 
The resulting HOD parameters and $\chi^2$ values were effectively unchanged relative to the 3000-step results. 

As noted in Section~\ref{sec:RedshiftDistribution}, the form of the redshift distribution can influence the predicted level of clustering for an underlying HOD model. The main effect of a redshift distribution with wings that extend beyond the nominal boundaries of redshift bins is to lower the {\em observed} clustering signal for a given theoretical two-point correlation function due to the increase in the number of galaxies in the bin that are not scattered into the bin due to the uncertainty on the photometric redshifts. There is therefore a degeneracy between the HOD model parameters and the $N(z)$ used to translate the theoretical HOD model to the projected angular two-point correlation function. The parameter within the HOD model most affected by the wings in the $N(z)$ distribution is $\sigma_{\log_{10} M}$. Increasing $\sigma_{\log_{10} M}$ essentially increased the amount of scatter in the galaxy - halo mass relation, with a larger $\sigma_{\log_{10} M}$ resulting in a decorrelation of the data, or lowering of the amplitude in the two-point correlation function. Such a large scatter smooths the central occupation function so strongly that the minimum halo mass for hosting a central galaxy, $M_{\rm min}$, shifts upwards towards the satellite halo mass scale, $M_{1}$. 
To attempt to alleviate these effects, we therefore adopt a Gaussian prior for $\sigma_{\log_{10} M}$. The mean of the Gaussian prior is set to $\mu=0.3$, consistent with the level of scatter expected from the combined intrinsic variation in the stellar-to-halo mass relation and uncertainties in the measured stellar masses discussed by \cite{2011ApJ...738...45L}. By adopting a Gaussian prior with a $3\sigma$ limit, we still allow for substantial mass scatter, but avoid solutions in which the central and satellite halo mass scales become effectively indistinguishable. In this sense, the Gaussian broadening of the $N(z)$ distributions described in Section \ref{sec:RedshiftDistribution} and the prior on $\sigma_{\log_{10}M}$ play complementary roles; the broadened $N(z)$ accounts for the expected line-of-sight smearing from photometric-redshift uncertainties, while the Gaussian prior on $\sigma_{\log_{10}M}$ prevents the HOD fit from artificially using an excessively broad central-occupation transition to compensate for projection effects that should instead be captured by the redshift distribution.

The likelihood in each field is calculated using a $\chi^{2}$, which includes a number density term and a clustering term, respectively:
\begin{equation}
\begin{split}
    \chi^{2}_\text{{field}} &= \frac{(n_{\rm gal}^{\rm obs} - n_{\rm total}^{\rm mod})^2}{\sigma_{\rm total}^2} \\
           &\quad + \sum \left(\omega_{\rm mod}(\theta) - \omega_{\rm obs}(\theta)\right) 
                 C^{-1} \left(\omega_{\rm mod}(\theta) - \omega_{\rm obs}(\theta)\right)
\end{split}
\label{eq:chi-squared_field}
\end{equation}
where $n_{\rm gal}^{\rm obs}$ is the number of observed galaxies in the sample, $n_{\rm total}^{\rm mod}$ is the total number of galaxies in the model (see Section \ref{sec:HOD_model} for the derivation) and $\sigma_{\rm total}^{2}=\sigma_{\rm Poisson}^{2}+\sigma_{\text{cv}}^{2}$ is the error in the number counts originating from Poisson noise as $\sqrt{n_{\rm gal}}$ and from cosmic variance (see Section \ref{sec: cosmic variance}). The $n_{\rm gal}^{\rm obs}$ is specifically estimated by counting a galaxy as part of the redshift bin if its summed PDF weight within that $z$-bin is greater than 0.2. In this way, galaxies are included according to their probability of belonging to the bin, rather than purely using an assigned photometric redshift estimate.
For the clustering term, $\omega_{\rm mod}(\theta)$ and $\omega_{\rm obs}(\theta)$ are the modelled and observed angular correlation functions over angular scale $\theta$ and $C$ is the covariance matrix of the observed correlation function determined through jackknife resampling (see Section \ref{sec:Covariance Matrix}).

We compute the observed correlation function $\omega_{\rm obs}$ over uniform logarithmic angular separations $\theta=0.001\degr - 0.5\degr$ for the XMM-LSS and ECDF-S fields, and $\theta=0.001\degr - 0.2\degr$ for the smaller COSMOS field.

In addition to fitting the three fields independently, we combine them by applying the same HOD model to each field consistently across the different redshift and stellar mass samples fitted, coupled with the individual $N(z)$ for each field. 
We then jointly explore the likelihood space for the total $\chi^{2}_{\text{COMBINED}}$, summing over all fields to constrain the best-fitting HOD parameters. 
\begin{equation}
    \chi^{2}_{\text{COMBINED}} = \chi^{2}_{\text{COSMOS}} + \chi^{2}_\text{XMM-LSS} +
    \chi^{2}_\text{ECDF-S}
\end{equation}
This combined approach ensures consistency in the galaxy–halo connection and allows us to leverage information across multiple fields.

\section{Results}
\label{sec:Results}
For each parameter presented in our results, we report the median of the resulting posterior distribution obtained from the fitting described in Section \ref{sec:Fitting_process}. The quoted lower and upper uncertainties are given by the 16th and 84th percentiles of the sampled posterior.

\subsection{The two-point correlation function} \label{sec:obs_and_mod_tpcf_results}

\begin{figure*}
    \includegraphics[scale=0.34]{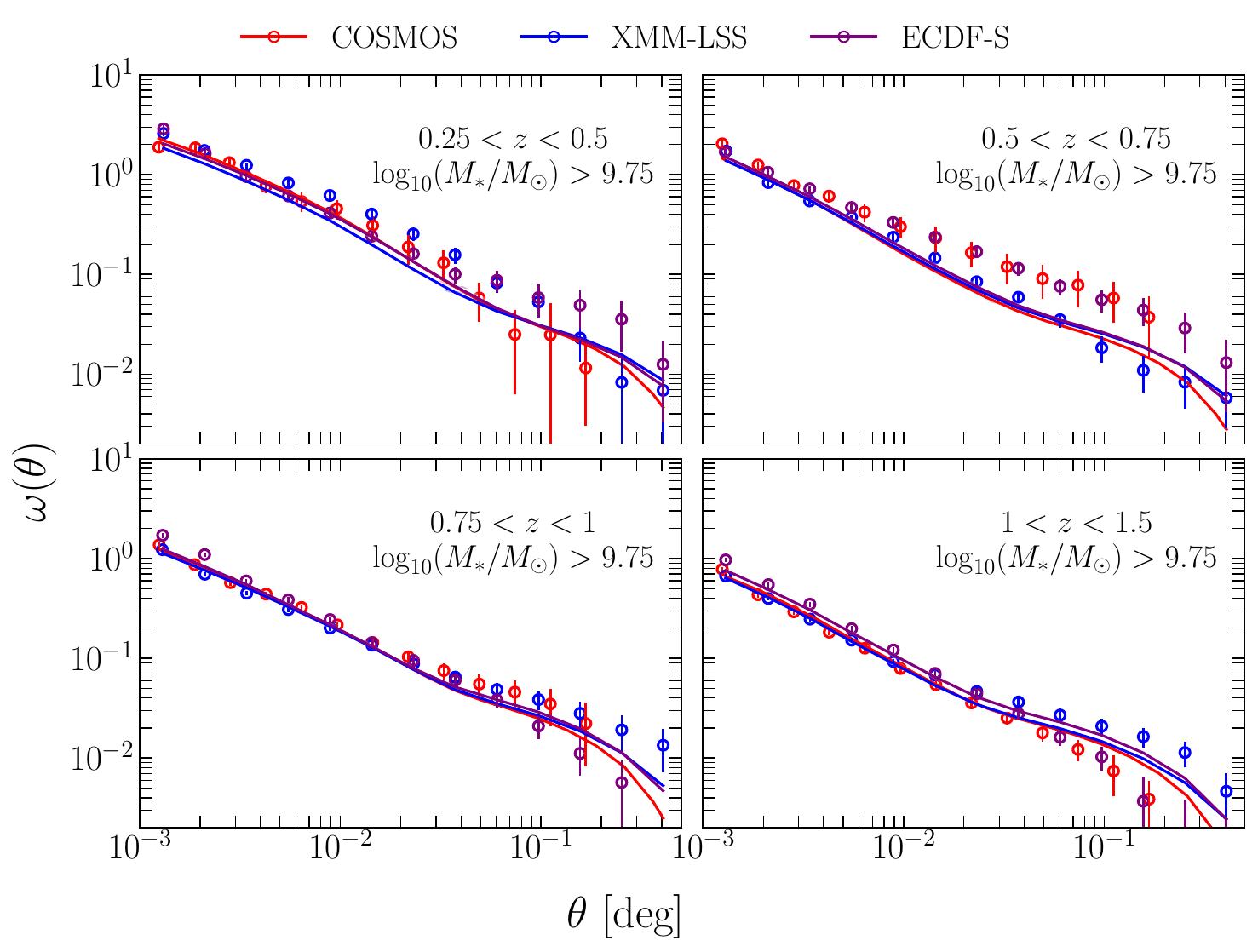}
    \includegraphics[scale=0.34]{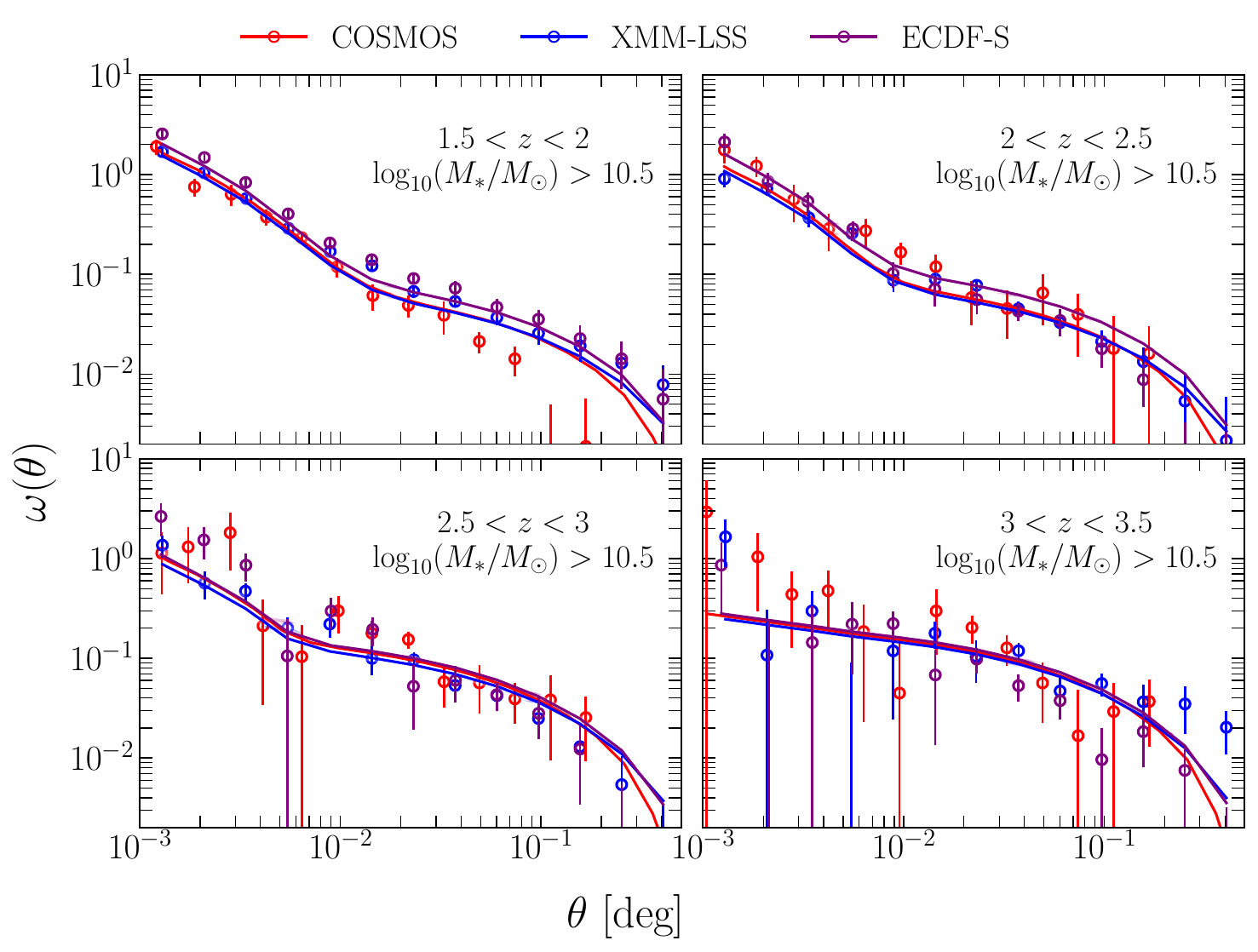}
    \caption{The observed angular two-point correlation functions for COSMOS (in red), XMM-LSS (in blue), and ECDF-S (in purple) fields are shown as circular points for galaxies with stellar mass threshold $\log_{10} (M_{\star}/$M$_{\odot})>9.75$ (left block of 4 panels) in four redshift bins: $0.25<z<0.5$ (top left), $0.5<z<0.75$ (top right), $0.75<z<1$ (bottom left) and $1<z<1.5$ (bottom right) and for $\log_{10} (M_{\star}/$M$_{\odot})>10.5$ (right block of four panels), for $1.5<z<2$ (top left), $2<z<2.5$ (top right), $2.5<z<3$ (bottom left) and $3<z<3.5$ (bottom right). The combined modelled correlation function is applied to each field and overplotted as solid lines in the corresponding field colours. The small differences among the model curves for the three fields arise from their different $N(z)$ distributions and integral constraint corrections.} \label{fig:tpcf_plots_mass-10.5}
\end{figure*}

\begin{figure*}
    \includegraphics[scale=0.34]{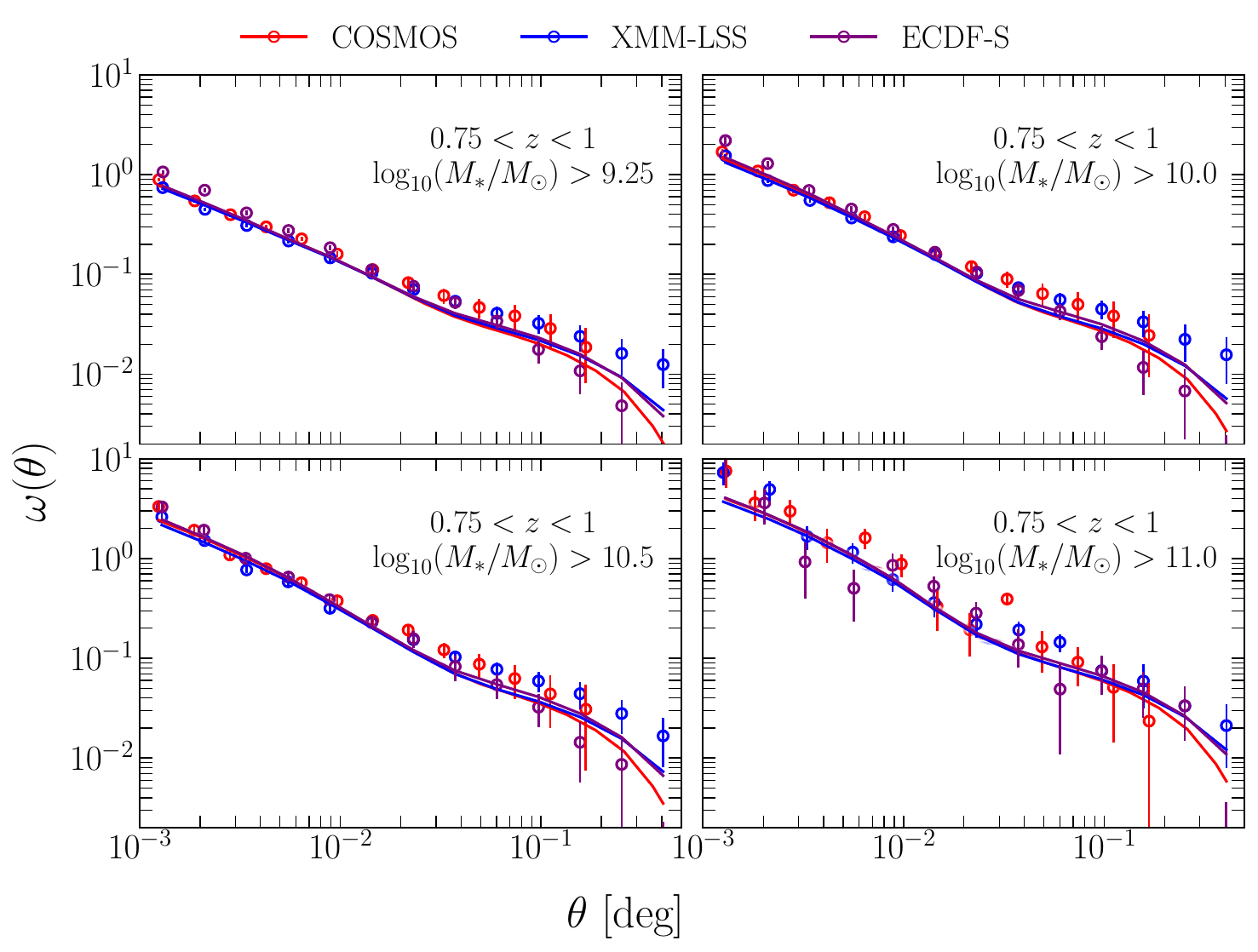}
    \centering
    \includegraphics[scale=0.34]{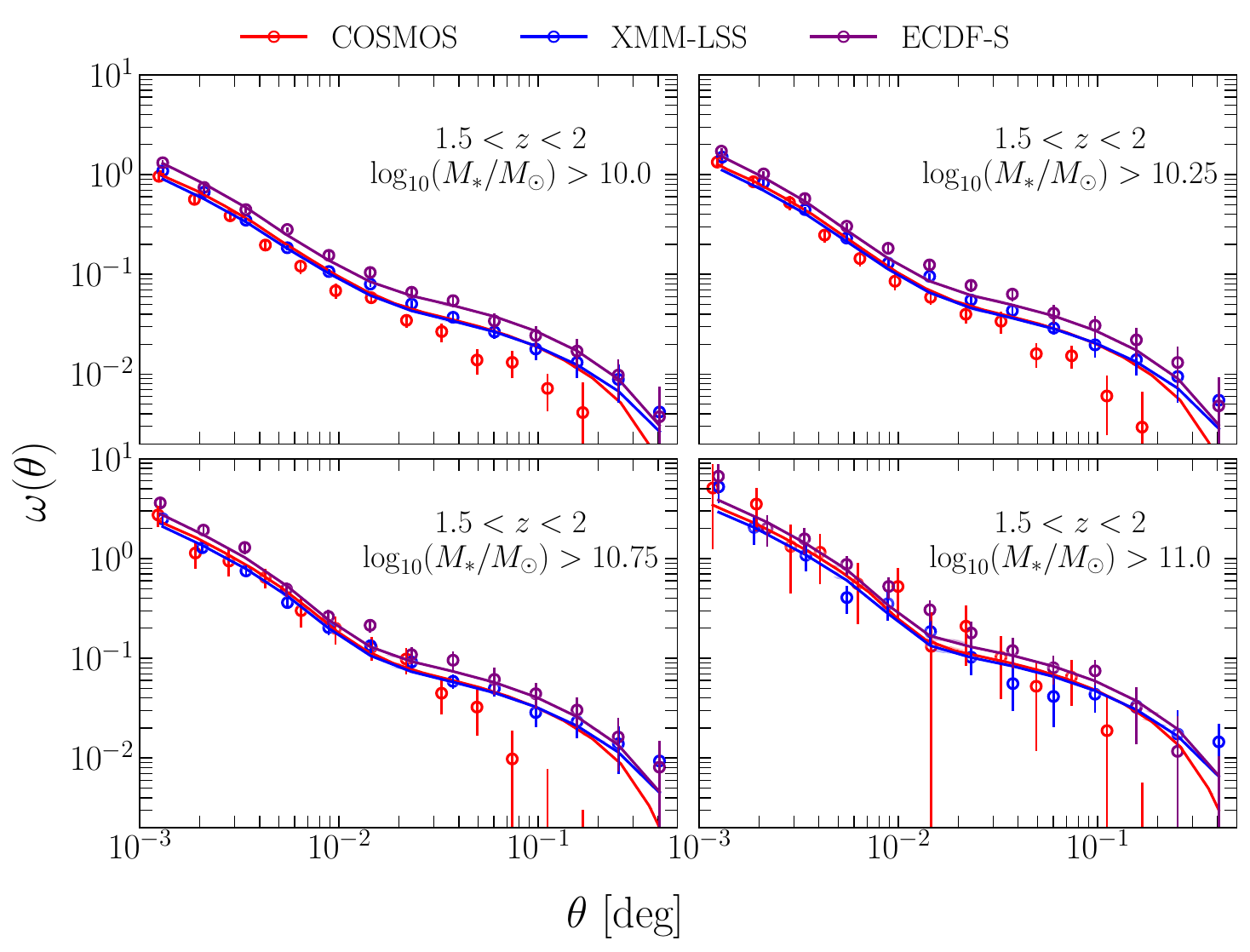}
    \caption{The observed angular two-point correlation functions for COSMOS (in red), XMM-LSS (in blue), and ECDF-S (in purple) fields are shown as circular points for galaxies at $0.75<z<1$  in four stellar mass thresholds (left block of four panels): $\log_{10} (M_{\star}/M_{\odot})>9.25$ (top left panel), $\log_{10} (M_{\star}/$M$_{\odot})>10.0$ (top right panel), $\log_{10} (M_{\star}/$M$_{\odot})>10.5$ (bottom left panel) and $\log_{10} (M_{\star}/$M$_{\odot})>11.0$ (bottom right panel), and for galaxies at $1.5<z<2$ in four stellar mass thresholds (right block of four panels): $\log_{10} (M_{\star}/$M$_{\odot})>10.0$ (top left panel), $\log_{10} (M_{\star}/$M$_{\odot})>10.25$ (top right panel), $\log_{10} (M_{\star}/$M$_{\odot})>10.75$ (bottom left panel) and $\log_{10} (M_{\star}/$M$_{\odot})>11.0$ (bottom right panel). The combined modelled correlation function is applied to each field and overplotted as solid lines in the corresponding field colours.The small differences among the model curves for the three fields arise from their different $N(z)$ distributions and integral constraint corrections.} \label{fig:tpcf_plots_1.5_z_2}
\end{figure*}

For each redshift and mass threshold sub-sample, the angular auto-correlation function was calculated using the method outlined in Section \ref{sec:TPCF} for the COSMOS, XMM-LSS, ECDF-S fields independently and then for the three fields COMBINED. 
Here we present the best-fitting modelled angular correlation functions that were obtained by combining data from the three fields for the redshift ranges and mass thresholds shown in Table \ref{tab:hod_params_combined_table}.

It is evident from the combined models in the examples in Figures~\ref{fig:tpcf_plots_mass-10.5} and  \ref{fig:tpcf_plots_1.5_z_2}, that certain redshift and mass bins exhibit a subtle change in slope, indicative of the transition from the 1-halo to the 2-halo term. This marks the shift from galaxy clustering within individual halos at small scales to clustering between separate halos at larger scales. This transition is especially noticeable for massive galaxies, whose larger halo masses support more satellite galaxies, thus boosting the amplitude of the 1-halo term. In the highest redshift bins, we find little evidence for a clear 1-halo term, and the model is not sufficiently constrained at small scales to make strong conclusions at $z>3$ on the satellite contributions.

There is a clear pattern of clustering becoming stronger as the stellar mass threshold increases across all scales, regardless of redshift range and field. This behaviour is due to the more massive galaxies being hosted by more massive dark matter halos, which exhibit stronger clustering due to their formation in the highest density peaks of the matter distribution. This is consistent with previous studies \citep[e.g.]{2006MNRAS.368...21L,2008A&A...478..299M,2018A&A...612A..42D,2018MNRAS.475.3730C}, which demonstrated that the degree of clustering strengthens as stellar mass increases.

It is important to note that although COSMOS offers greater depth compared to the other two fields, in this work, we have a greater area that can probe the higher-mass galaxy samples. Consequently, XMM-LSS and ECDF-S are more effective at accurately constraining the clustering of the more massive galaxies, particularly at higher redshift bins beyond $z > 2$ (where the COSMOS field alone has too small an area to accurately probe the clustering signal at such high masses), because of their larger volumes and greater capacity to contain more high-mass galaxies. This further highlights the importance of including these fields when studying the high-mass end of galaxy evolution.

\subsection{Evolution of HOD Model Parameters} \label{sec:HOD_params_results}

The best-fit HOD model parameters ($M_{\min}$, $M_0$, $M_1$, $\sigma_{\log_{10} M}$, $\alpha$) from the COMBINED case are shown in Table \ref{tab:hod_params_combined_table}, as functions of stellar mass and redshift, along with other sample statistics. 
The equivalent results for each independent field HOD model, COSMOS, XMM-LSS, and ECDF-S, can be found in the Appendix \ref{Appendix}.

\subsubsection{Halo Mass for a Central \texorpdfstring{$M_{\rm min}$}{}, and a Satellite \texorpdfstring{$M_{1}$}{}} \label{sec:Mmin and M1}

\begin{figure}  
    \centering
    \includegraphics[scale=0.38]{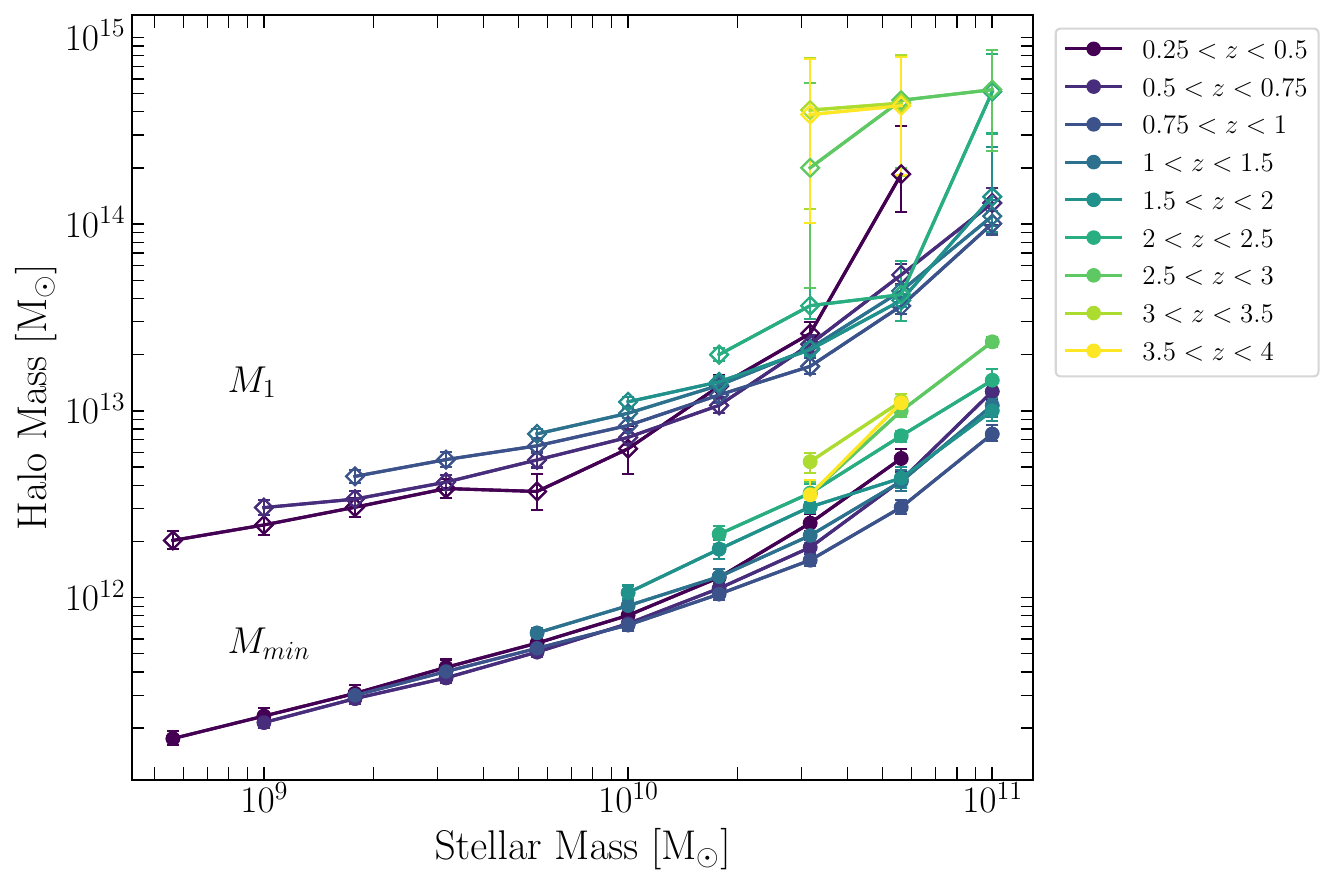}
    \caption{The evolution of the HOD parameters $M_{1}$ (typical halo mass for a halo to host one satellite galaxy) and $M_{\rm min}$ (minimum halo mass for a halo to host a central galaxy) as a function of stellar mass threshold for each redshift bin.}
    \label{fig:Mmin_M1_vs_M*}
\end{figure}

The HOD model parameters $M_{\rm min}$ and $M_{1}$ represent the halo masses needed to host a central galaxy and a satellite galaxy in a halo, respectively. Figure \ref{fig:Mmin_M1_vs_M*} shows the evolution of $M_{\rm min}$ and $M_{1}$ with respect to the stellar mass threshold for each redshift range. The $M_{\rm min}$ and $M_{1}$ parameters exhibit an increasing trend with stellar mass. This reflects the well-established observation that more massive galaxies occupy more massive dark matter halos. $M_{\rm min}$ ranges from $\sim 10^{11.3}$\,M$_{\odot}$ to $\sim 10^{13.4}$\,M$_{\odot}$, while $M_{1}$ ranges from $\sim 10^{12.3}$\,M$_{\odot}$ to $\sim 10^{14.7}$\,M$_{\odot}$. 
The halo masses to host a central and a satellite, $M_{\rm min}$ and $M_{1}$, show clear evidence of redshift dependence. Galaxies of a fixed stellar mass are occupying more massive halos with increasing redshift, as seen in Figure \ref{fig:Mmin_M1_vs_M*}.

\begin{figure}  
    \centering
    \includegraphics[scale=0.37]{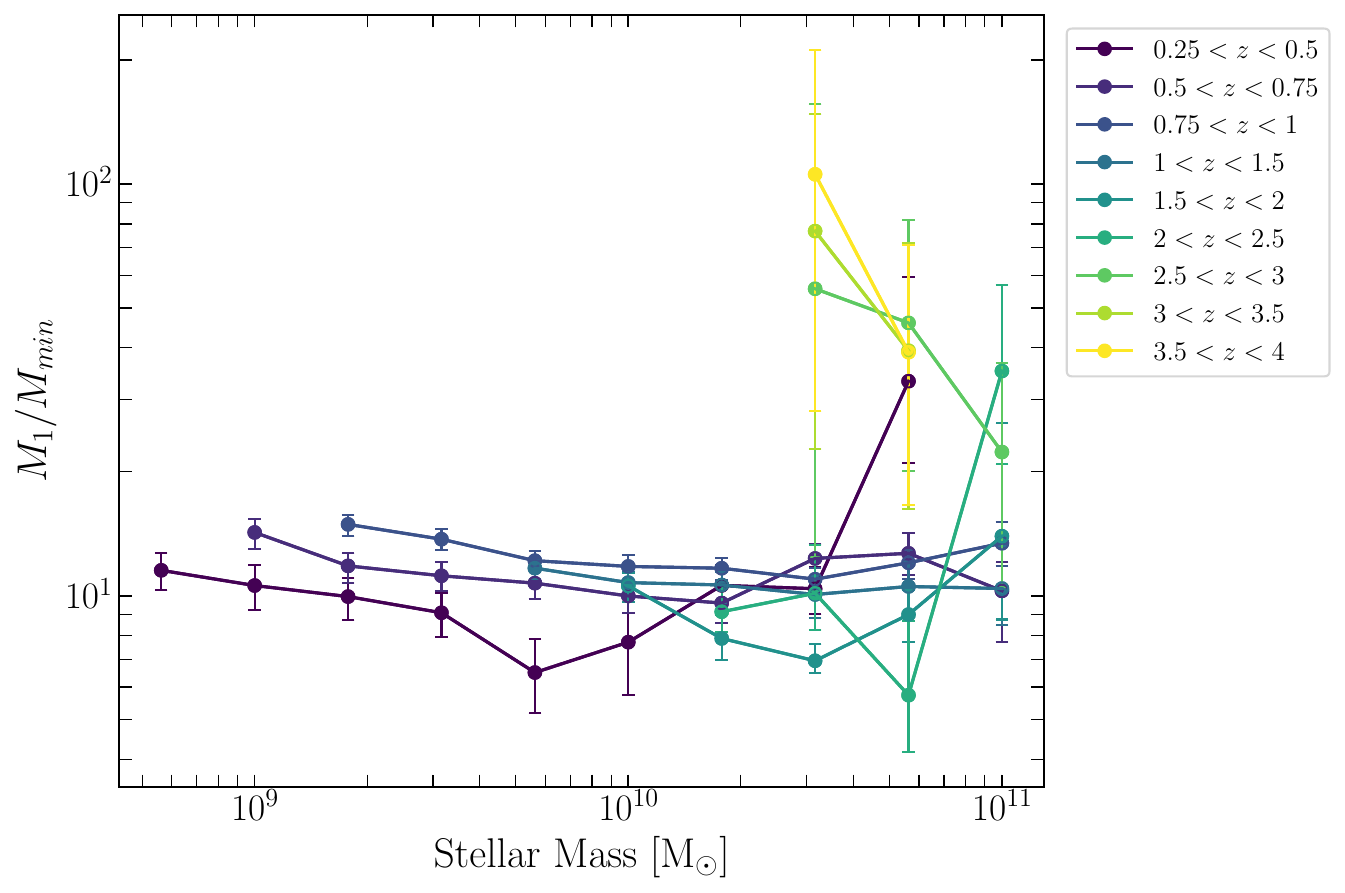}
    \caption{The evolution of the HOD parameters' ratio $M_{1}/M_{\rm min}$ as a function of stellar mass threshold for each redshift bin.}
    \label{fig:M1/Mmin_vs_M*}
\end{figure}

The ratio $M_{1}/M_{\rm min}$, shown in Figure~\ref{fig:M1/Mmin_vs_M*}, is indicative of the mass gap between the halo mass required to host a satellite galaxy and the minimum mass necessary to host a central galaxy. 
The local $M_{1}/M_{\rm min}$ ratio at $z\sim0.1$ from wide-field spec-$z$galaxy surveys is found to be $\sim20$. Specifically, studies from \cite{2005ApJ...630....1Z} with SDSS data and \cite{2013MNRAS.429.3604B} with 6dFGS data using a 3-parameter HOD model, found $M_{1}/M_{\rm min}$ to be $\sim23$ and $\sim22$ respectively. 
\cite{2015MNRAS.449..901M} at $0.5<z<2$ found a trend where $M_{1}/M_{\rm min}$ tends to be larger at both low ($\sim10^{9}-10^{9.75}$\,M$_{\odot}$) and high ($\sim10^{10.5}-10^{11}$\,M$_{\odot}$) stellar masses, with a ratio of $\sim20$, while the mass gap decreases to $\sim10$ at the intermediate masses. Our results in Figure \ref{fig:M1/Mmin_vs_M*} show a similar overall trend and are broadly consistent within the mass and redshift ranges in common to both studies; however, our values are enlarged in the regions where we extend to higher redshifts than \cite{2015MNRAS.449..901M}.

For galaxies with stellar masses $<10^{10.5}$\,M$_{\odot}$, and the galaxies with $z<2.5$, we find $M_{1}/M_{\rm min}\sim15-10$ (i.e. the halo mass required to host a satellite galaxy is $\sim15-10$ times the halo mass needed to host a central), with some outliers ranging down to 6. Over similar redshift and stellar masses, \cite{2016MNRAS.459.2618H} found $M_{1}/M_{\rm min}$ to be relatively constant with some scatter, ranging from approximately 10 to 30, with minimal redshift evolution and a potential increase at higher stellar masses. 

For galaxies with stellar masses $>10^{10.5}$\,M$_{\odot}$, we see an upturn in the mass gap ratio in Figure \ref{fig:M1/Mmin_vs_M*}. \cite{2015MNRAS.449..901M} notes this upturn in $M_{1}$ observed in the highest mass ranges $>10^{10.5}$\,M$_{\odot}$.

Apart from the increase in $M_{1}/M_{\rm min}$ with high stellar masses, we observe an increase in the ratio at redshifts $2.5<z<4$. As seen in both Figures \ref{fig:Mmin_M1_vs_M*} and \ref{fig:M1/Mmin_vs_M*}, the halo mass to host a satellite galaxy, $M_{1}$, increases more at $z>2.5$. This results in an increase in the ratio $M_{1}/M_{\rm min}$ by an order of magnitude. This can be interpreted such that, at high redshift and high stellar mass thresholds, significantly more massive halos are required to host a massive satellite galaxy compared to the halo mass to host a massive central galaxy. This is in agreement with \cite{2015MNRAS.449..901M} and \cite{2020ApJ...904..128I}, who suggest that high stellar mass satellites tend to form only in massive halos and that satellite formation is less efficient at high redshift.

\subsubsection{Minimum Satellite Halo Mass - \texorpdfstring{$M_{0}$}{}} \label{sec:M0}

The HOD parameter $M_{0}$ is commonly referred to as the minimum halo mass needed before any satellite galaxies appear in a dark matter halo. It is typically much smaller than the $M_{1}$ parameter, which is regarded as the halo mass at which halos begin to host one satellite on average. Consequently, the upper bounds for $M_{0}$ in our model have been set to $M_{1}$ during the fitting process.

It is well-known that the $M_{0}$ parameter is not as tightly constrained as the $M_{\rm min}$ and $M_{1}$ parameters. This arises from the HOD parameterisation, where the HOD model does not fit centrals and satellites independently (as in \citealt{2023MNRAS.525.4257C}) (see the satellite occupation in Equation \ref{eq:Nsat}, restricting satellites to halos with a central galaxy). This extra layer of complexity in our galaxy clustering analysis produces less stringent constraints on $M_{0}$. 

Other studies attempt to constrain $M_{0}$ by either setting it to $M_{0} = M_{\rm min}$ \citep{2018MNRAS.473.4318A} (i.e., satellite galaxies can begin to form when a central galaxy exists) or by using a fixed function such as $\log(M_{0}) = 0.76 \log(M_{1}) + 2.3$ (\citealt{2016ApJ...821..123H}; \citealt{2018MNRAS.477.3760H}; \citealt{2025arXiv250111674P}) following results from \citealt{2006ApJ...647..201C}.

The $M_{0}$ parameter does not affect the resulting galaxy clustering to the same degree as other HOD parameters (mainly $M_{\rm min}$, $M_{1}$, $\alpha$). In particular, we find that for our stellar mass and redshift ranges, $M_{0}$ spans from $\sim 10^{9}$\,M$_{\odot}$ to $\sim 10^{14}$\,M$_{\odot}$ and while it is exhibiting some increase with higher mass and redshift, it is not substantial considering the error bars.

\subsubsection{Stellar Mass Scatter - \texorpdfstring{$\sigma_{\log M}$}{}} \label{sec:sigma_logM}

The scatter in stellar mass at a fixed halo mass parameter, $\sigma_{\log M}$, essentially describes the scatter around the mean $\log M_{*} - \log M_{h}$ relation. This assumes that $M_{*}$ at a fixed $M_{h}$ follows a log-normal profile, and it characterises how tightly stellar mass correlates with halo mass. The total log–normal scatter $\sigma_{\log M}$, comprises of an intrinsic component, $\sigma^{i}_{\log M_{*}}$, and a stellar mass measurement error component, $\sigma^{m}_{\log M}$, arising from uncertainties in redshift, photometry, and stellar mass modelling, and it can be expressed as$(\sigma_{\log M})^{2}=(\sigma^{i}_{\log M})^{2}+(\sigma^{m}_{\log M})^{2}$ \citep{2011ApJ...738...45L}. While the measurement error can differ from study to study due to differences in data and stellar mass estimation methods, the intrinsic scatter should be consistent.
\citealt{2012ApJ...744..159L} determined that incorporating a stellar mass–dependent scatter in the measurement error did not improve their model fit over having a constant scatter, and since the scatter is effectively independent of halo mass, we therefore treat it as a free parameter in our model \citep{2013ApJ...778...93T}. 

We use Gaussian priors for $\sigma_{\log M}$, and we find that at $z<1$ the models constrain $\sigma_{\log M}$ around the Gaussian mean $\sigma_{\log M}\sim 0.3$, although we note that if wer adopt uniform priors at these redshifts then remains at $\sigma_{\log M}\sim 0.3$. For $z$-bins of $z>1.5$, we see that $\sigma_{\log M}$ prefers the upper $3\sigma$ limit. We find this is purely due to the uncertainty surrounding the actual $N(z)$ in each redshift bin, as discussed in Section~\ref{sec:RedshiftDistribution}. We therefore do not comment further on whether there is potential evolution in $\sigma_{\log M}$, due to this degeneracy with the $N(z)$ distribution.

\subsubsection{Satellite Slope - \texorpdfstring{$\alpha$}{}} \label{sec:alpha}

\begin{figure}  
    \centering
    \includegraphics[scale=0.35]{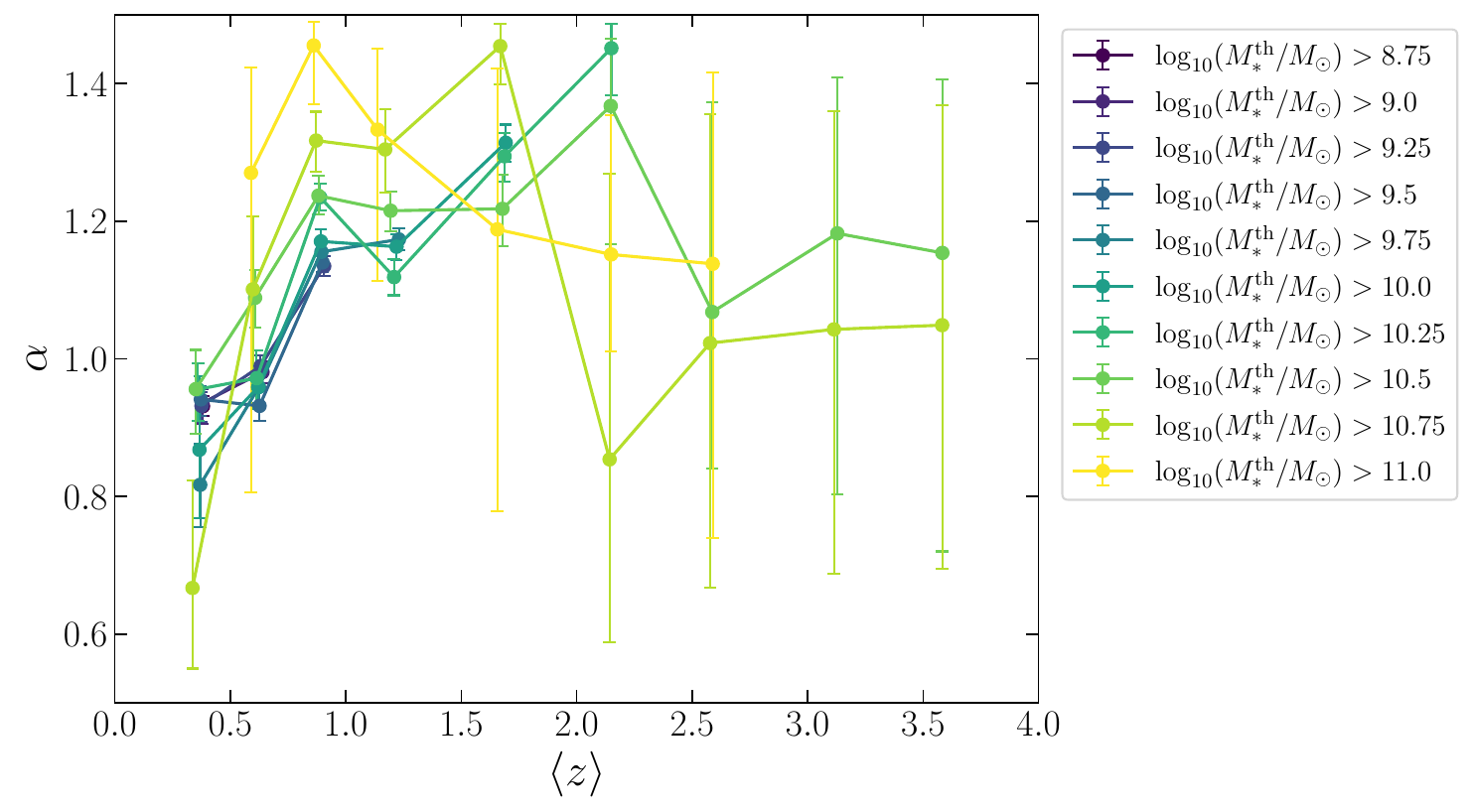}
    \caption{The evolution of the satellite slope $\alpha$ as a function of weighted mean redshift, for different stellar mass thresholds.}
    \label{fig:alpha_plot}
\end{figure}

The HOD parameter $\alpha$ describes the slope of the power-law that the number of satellite galaxies follow as they grow with mass. 
Physically, $\alpha \approx 1$ implies that the number of satellites scales linearly with halo mass, i.e., the subhalo abundance grows proportionally to host halo mass.  A wide range of studies find $\alpha$ to consistently vary around one, hence many tend to fix the slope to $\alpha=1$ in their HOD model (\citealt{2011ApJ...738...45L}; \citealt{2018MNRAS.477.3760H}; \citealt{2020JCAP...03..044N}). 

Studies that allow $\alpha$ to vary in their models also find values around one with negligible evolution with either redshift or stellar mass (\citealt{2015MNRAS.454.1161Z}; \citealt{2015ApJ...807..152S};  \citealt{2016MNRAS.459.2618H}). Having said that, studies working with luminosity-selected samples do report a slight increase in $\alpha$ with increasing luminosity thresholds, although given their errors these are not as tightly constrained (\citealt{2012A&A...542A...5C}; \citealt{2007ApJ...667..760Z}). 
Interestingly, \citealt{2018A&A...612A..42D} worked with both luminosity and stellar mass selected samples, and found an increase in $\alpha$ for their luminosity selected samples. However, for their stellar mass-selected samples, they find all their values to be $\alpha \sim 1$,  and report no substantial variation or evolution.

Our results strongly suggest that $\alpha$ does evolve with redshift (Figure~\ref{fig:alpha_plot}; top panel), at least out to $z\sim 1.5$. This is the redshift regime where we have the tightest constraints and also the evolution is reflected in all stellar mass bins (apart from the highest mass bin which has poorer statistics at low redshift). On the other hand we find much less evidence for dependence on stellar mass, where $\alpha$ appears to remain roughly constant as a function of stellar mass at a given redshift.

This evolution can be interpreted as the rate at which satellites are formed within haloes traced by the same central galaxy mass at low and high redshift. The is also related to the fact that at a given stellar mass the minimum halo mass to form a central increases with redshift (Figure~\ref{fig:Mmin_M1_vs_M*}) and the typical halo mass for the first satellite galaxy also increases, albeit less rapidly resulting the $M_1/M_{\rm min}$ ratio (Figure~\ref{fig:M1/Mmin_vs_M*}) that becomes larger with increasing redshift. It is within these haloes that the number of satellites has a much steeper slope with halo mass. This can be understood if the satellite build up in these more massive haloes at earlier times, traced by progressively relatively less massive centrals galaxies (and therefore relatively less massive satellites given the stellar mass thresholds in our analysis), is due to these haloes having formed earlier and where the centrals and satellites are forming more coevally than at low redshift where accretion of satellites onto haloes is more prevalent. Essentially the cores of these massive haloes are already in place, and the later time accretion of satellites, increases the number of satellites in line with the growth in the halo mass, i.e. large haloes are accreting smaller haloes which already have a central galaxy, which becomes a satellite of the larger halo. This process would naturally lead to the lowering of $\alpha$ towards lower redshifts for a given stellar mass threshold.

\subsection{Testing for Systematics}
We have also examined several additional scenarios and found the clustering results to be consistent. 

First, we measured the two-point correlation function (TPCF) using conservative $Ks$ magnitude limits of $24.7$ for COSMOS and $23.5$ for XMM-LSS and ECDF-S, which shifts all the samples from being near the $5\sigma$ detection depth to a more conservative $6.6\sigma$ depth. A conservative magnitude cut does not significantly impact the clustering pattern, but it reduces the number of galaxies in the sample. As a result, the error bars are larger in this case, but the overall clustering amplitude remains consistent.

Second, we calculated the TPCF using solely galaxies whose peak redshift $z_{\rm peak}$ in their photo-$z$PDF falls within the $z$-bin (see Section \ref{sec: Final Samples}) and ignored contributions from galaxies whose photo-$z$PDFs can partly reside inside the $z$-bin. The $z_{\rm peak}$-based TPCFs are broadly consistent with what we present in this paper. However, relying exclusively on the $z_{\rm peak}$ contribution leads to poorer constraints in the higher redshift bins (i.e., $z > 3$) for the high mass thresholds we are investigating, due to a  reduction in overall sources due to galaxies with significant wings in their photometric redshift probability distribution spanning these redshift bins, but  their peak redshift lying in other (often higher) redshift bins.

Thirdly, we conducted HOD fitting by adopting a model that includes a time-dependent smoothing parameter $\alpha_{\rm smooth}$ in the galaxy-galaxy power spectrum. This added parameter can potentially better model the transition between the 1-halo and 2-halo regimes, as described in \citep{2021MNRAS.502.1401M}. The modified power spectrum is given by $P_{gg}(k) = \left( P_{gg}^{1h}(k)^{\alpha_{\rm smooth}} + P_{gg}^{2h}(k)^{\alpha_{\rm smooth}} \right)^{1/\alpha_{\rm smooth}}$. We found that when using the transition parameter $\alpha_{\rm smooth}$ we obtained comparable results, and therefore decided to continue with the simpler 5-parameter HOD model described in Section~\ref{sec:HOD_model} for the purposes of this work.

\section{Derived quantities}

\subsection{Galaxy Bias} \label{sec:galaxy_bias_results}

\begin{figure}
    \centering
    \includegraphics[scale=0.4]{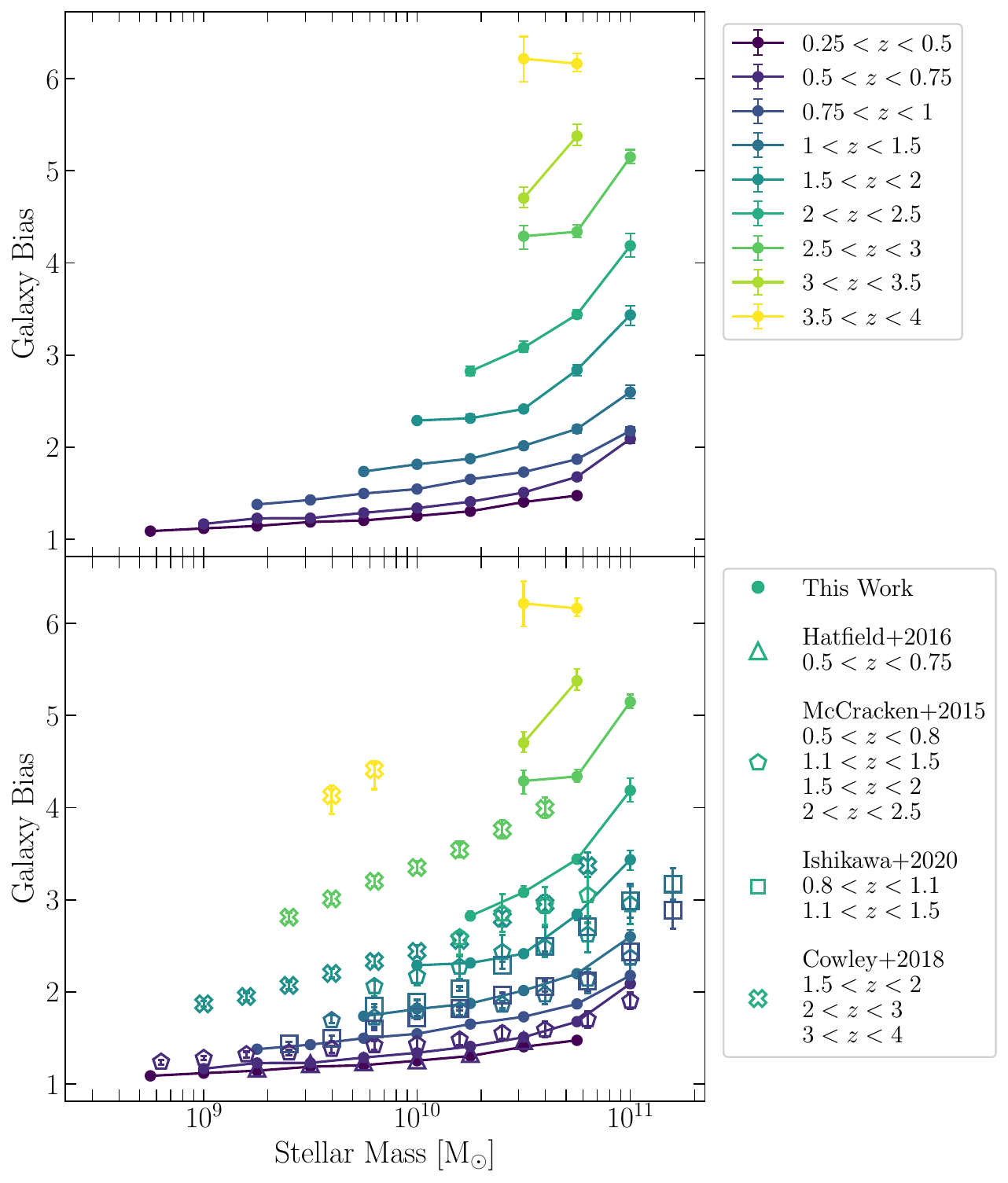}
    \caption{The evolution of galaxy bias as a function of stellar mass threshold and redshift bin, denoted in the top right legend. The top panel shows our galaxy bias results for each redshift bin. The bottom panel shows a comparison of our results (solid lines) with other papers, presented in the bottom legend, with their equivalent shapes.}
    \label{fig:Bias_vs_M*}
\end{figure}

The bias (see Equation~\ref{eq:galaxy_bias}) relates the galaxy overdensity to the dark matter overdensity and can equivalently be defined as the ratio of their correlation functions as:
\begin{equation}
b_g = \frac{\delta_g}{\delta_{\rm DM}}= \sqrt{\frac{\xi_g}{\xi_{\rm DM}}}
\end{equation}
where, $\delta_g$ and $\delta_{\rm DM}$ represent the galaxy and dark matter overdensities, while $\xi_g$ and $\xi_{\rm DM}$ represent the galaxy and dark matter spatial correlation functions. The bias therefore indicates whether galaxies are more or less strongly clustered than the dark matter distribution.

In the top panel of Figure \ref{fig:Bias_vs_M*}, we present our galaxy bias measurements from our best-fitting combined halo model, as a function of the stellar mass at each redshift bin.
The galaxy bias increases with higher galaxy mass, with high-mass galaxies showing a steeper increase than low-mass galaxies at the lowest redshifts, which can appear flatter. This indicates that more massive galaxies tend to form in larger dark matter halos, which are more strongly biased towards denser regions of the dark matter distribution.

Additionally, the galaxy bias decreases toward lower redshifts. At high redshift, haloes of a given mass correspond to rarer peaks in the density field, and hence the galaxies they host are therefore located in the most overdense regions and are highly biased relative to the mean matter distribution. As cosmic time progresses, and structure grows at lower redshifts, haloes of the same mass become more common and less strongly biased, and galaxies occupy a broader range of environments, causing the average galaxy bias to decrease.

The bottom panel of Figure \ref{fig:Bias_vs_M*} places our galaxy bias measurements in the context of previous work. We find excellent agreement with \cite{2016MNRAS.459.2618H} (at $0.5<z<0.75$) and with \cite{2015MNRAS.449..901M} (various $z$-bins in $z=0.5-2.5$). We report slightly higher bias values than \cite{2015MNRAS.449..901M} in the highest stellar masses, where we expect our measurements to be more reliable due to the larger area probed by our sample compared to the smaller just COSMOS field area used in \cite{2015MNRAS.449..901M}. 

The bias values of \cite{2020ApJ...904..128I} ($0.8<z<1.1$, $1.1<z<1.5$) are in good agreement with ours at lower stellar masses, while their bias estimates appear slightly larger at higher masses. This could also be attributed to differences in redshift binning, which, for $0.8<z<1.1$, are reflected in the inclusion of galaxies at $z\geq0.75$ that, on average, are more strongly biased than those in our lower and narrower $0.75<z<1$ bin.

At higher redshift, our bias results fall within agreement with \cite{2018ApJ...853...69C} $2<z<3$ and expectations from $3<z<4$ measurements, given that we split these into finer bins ($2<z<2.5$, $2.5<z<3$, $3<z<3.5$, $3.5<z<4$). Due to the small $\sim0.66$ deg$^{2}$ area, considered by \cite{2018ApJ...853...69C} in COSMOS, they can probe the lower-mass galaxies better than the higher-mass galaxies, which require sufficient volume to be representatively sampled (see discussion in Section \ref{sec: cosmic variance} and dark matter cosmic variance decreasing with increasing field area in Figure \ref{fig:sigma_dm_plot}). We therefore note agreement at lower stellar masses and assume that the trend continues from low to higher masses as such.

Overall, the trends we infer for the evolution of galaxy bias with stellar mass and redshift are consistent with previous studies.

\subsection{Cosmic Variance Evolution}
\label{sec: cosmic variance evolution}
\begin{figure*}
    \centering
    \includegraphics[scale=0.38]{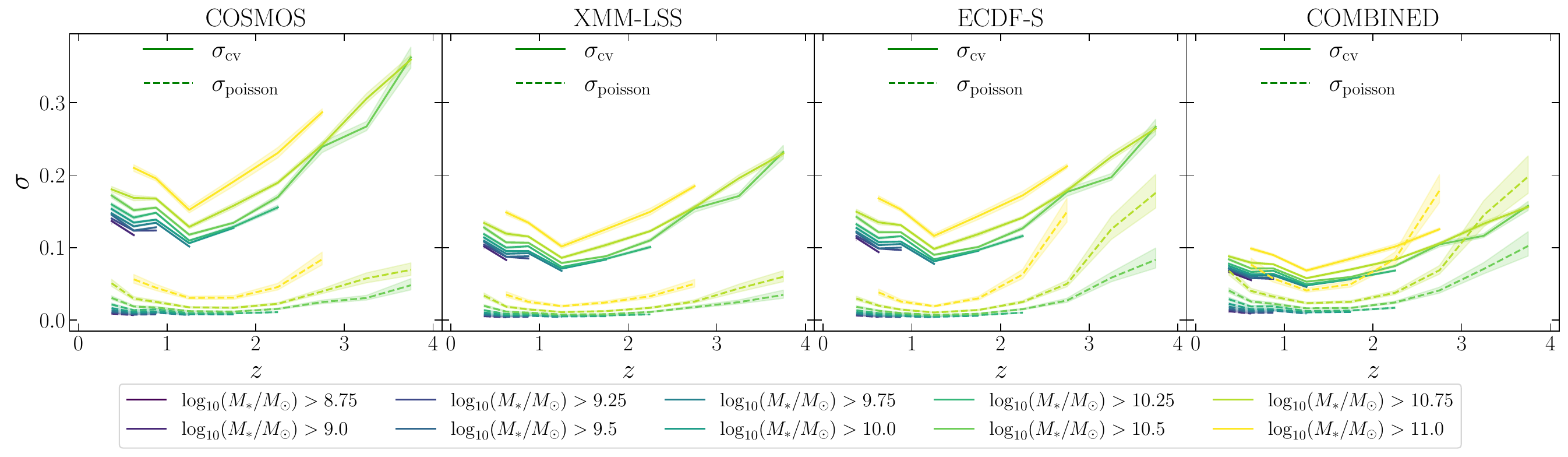}
    \caption{The fractional cosmic variance (solid lines) and Poisson uncertainties (dashed lines) as a function of redshift are shown for each stellar mass range across the three survey fields. The final plot on the right displays the combined uncertainties for the three fields, where the Poisson errors were combined in quadrature and the cosmic variances were combined using the volume-weighted equation \ref{eq:sigma_cv_combined} \citep{2011ApJ...731..113M}. Due to its smaller size, we see that COSMOS is more affected by cosmic variance. By including two other larger independent fields, we reduce cosmic variance by a factor $\sim2$ (right plot), making our sample more representative of the galaxy matter distribution.}
    \label{fig:sigma_cv_plot}
\end{figure*}
Figure \ref{fig:sigma_cv_plot}, presents the fractional cosmic variance compared to the Poisson uncertainties for each field, displaying how both change across redshift and stellar mass ranges. 
For all three fields (COSMOS, XMM-LSS, and ECDF-S), the $\sigma_{\mathrm{cv}}$ is the predominant source of uncertainty across the entire range of stellar mass bins and redshift, surpassing the Poisson uncertainties $\sigma_{\mathrm{poisson}}$. As the stellar mass threshold increases, both $\sigma_{\mathrm{cv}}$ and $\sigma_{\mathrm{poisson}}$ tend to rise. This is because while lower-mass galaxies are more randomly distributed, higher-mass galaxies tend to reside in the most massive, and therefore rarest halos. As a result, the higher-mass galaxies are stronger traces of the large-scale structure and more susceptible to its density fluctuations. The redshift evolution of $\sigma_{\mathrm{cv}}$ reflects both volume effects and structure formation. At low redshifts, the small comoving volume covered by each field results in higher sensitivity to local over- or underdensities, leading to larger $\sigma_{\mathrm{cv}}$. At intermediate redshifts ($z\sim1-2$), the larger volumes probed at those redshifts reduce this effect and help average over density fluctuations, leading to a minimum in $\sigma_{\mathrm{cv}}$. At the highest redshifts, $\sigma_{\mathrm{cv}}$ begins to gradually rise again due to increasing rarity and strong clustering of massive galaxies, along with less complete sampling of the galaxy population, making the sample sparser and thus more prone to variance. It is also important to note that the sudden drop at $z=1.5$ in the plots of Figure \ref{fig:sigma_cv_plot} is due to the increase of $\Delta z$ bin size from 0.25 to 0.5. This increase in $\Delta z$ increases the volume, leading to an abrupt decrease in cosmic variance. However, as the volume per $\Delta z$ gradually decreases, the cosmic variance begins to rise again.

In the COMBINED plot of Figure \ref{fig:sigma_cv_plot}, we also show the combined fractional $\sigma_{\mathrm{cv}_{\rm combined}}$, which we calculated using a comoving volume ($V$) weighting (also described in equation 7 in \citealt{2011ApJ...731..113M}).
\begin{equation}
    \sigma^{2}_{\mathrm{cv}_{\rm combined}}= \frac{\sum_{i} V_{i}^2 \sigma^{2}_{\mathrm{cv}_{i}}}{\left(\sum_{i} V_{i}\right)^2}
    \label{eq:sigma_cv_combined}
\end{equation}
The trends depicted in each individual field are broadly consistent, indicating that differences in field area do not drastically alter the qualitative behaviour of the uncertainties. Notably, when the cosmic variances from all three fields are combined, as shown by the final panels of Figure \ref{fig:sigma_cv_plot}, the impact of cosmic variance is significantly reduced. This is because the three fields provide a more representative sampling of the Universe. As a result, we show that when combining three independent fields, the cosmic variance uncertainty decreases by a factor of $\sim2$, depending slightly on redshift and stellar mass. This, in turn, reduces the associated uncertainties in galaxy clustering and galaxy stellar mass function.

\subsection{Satellite Fractions} \label{sec:sat_frac_results}

\begin{figure}
    \centering
    \includegraphics[scale=0.397]{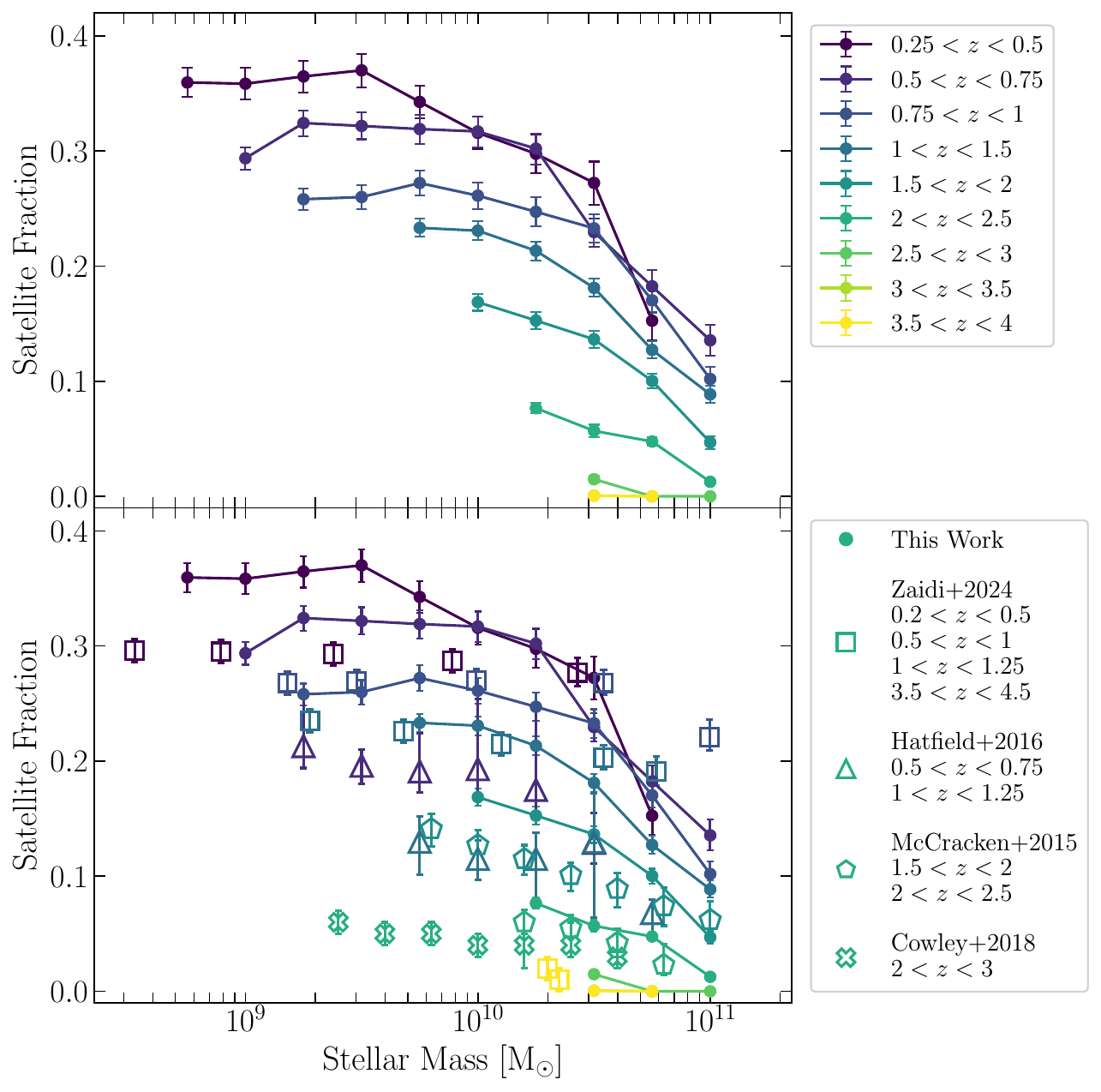}
    \caption{The evolution of the satellite fraction as a function of stellar mass threshold for each redshift bin, denoted in the top right legend. The top panel shows our satellite fraction results for each redshift bin. The bottom panel shows a comparison of our results (solid lines) with other studies, presented in the bottom legend, with their equivalent shapes.}
    \label{fig:FrSat_vs_M*}
\end{figure}

The change in the satellite population provides insights into the evolution of halo environments. It reflects the halo’s hierarchical assembly history and the impact of environmental effects, such as stripping, quenching, and merging, that influence the growth and survival of satellites. The fraction of satellites, and its evolution with stellar mass in each redshift bin, are shown in the top panel of Figure~\ref{fig:FrSat_vs_M*}. In all $z$-bins, the satellite fraction decreases toward higher stellar masses, with a noticeably sharper decline above $\log_{10}(M_{\star}^{th}/$M$_{\odot}) \sim 10.5$. 
Physically, this reflects the fact that the more massive galaxies are primarily the central galaxies in higher-mass halos and larger halos capable of hosting these massive galaxies as satellites are uncommon. 

We find that the fraction of satellites appears to plateau at $\log_{10}(M_{\star}^{th}/$M$_{\odot})< 10$. This can be understood by considering the number of low mass haloes which actually have a lower mass galaxy as their central galaxy and are also classed as satellites of larger haloes. As such the number of satellites and the number of centrals for these lower stellar mass galaxies is actually the same and the ratio over the whole population remains roughly constant.

The bottom panel of Figure \ref{fig:FrSat_vs_M*} compares our best-fit satellite fractions with similar studies. One can see that we place much better constraints on the satellite fraction at high stellar masses due to the increased volume of our survey over previous deep-field studies.

Our results at lower redshifts, $z<2$, resemble more the satellite fractions reported by \citealt{2024arXiv241104256Z} ($0.2<z<0.5$, $0.5<z<1$, $1<z<1.25$) rather than the lower values provided by \citealt{2016MNRAS.459.2618H} ($0.5<z<0.75$, $1<z<1.25$) and \citealt{2015MNRAS.449..901M} ($1.5<z<2$). 
Our results at $z>2$ are consistent with previous literature, including \citealt{2015MNRAS.449..901M}, \citealt{2018ApJ...853...69C} and \citealt{2024arXiv241104256Z}. The redshift colours assigned to the literature measurements in the bottom panel of Figure \ref{fig:FrSat_vs_M*} are chosen to approximately match our redshift bins, although the binning is not always identical. 

As expected, satellites are more abundant at lower redshifts because the massive halos that host the majority of satellite galaxies are still assembling at higher redshifts, whereas the satellite fraction declines toward higher redshift, consistent with expectations from hierarchical structure formation.

\subsection{Stellar-to-Halo Mass Relation} \label{sec:SHMR_results}

\begin{figure*}
    \centering
    \begin{tabular}{cc}
        \includegraphics[scale=0.34]{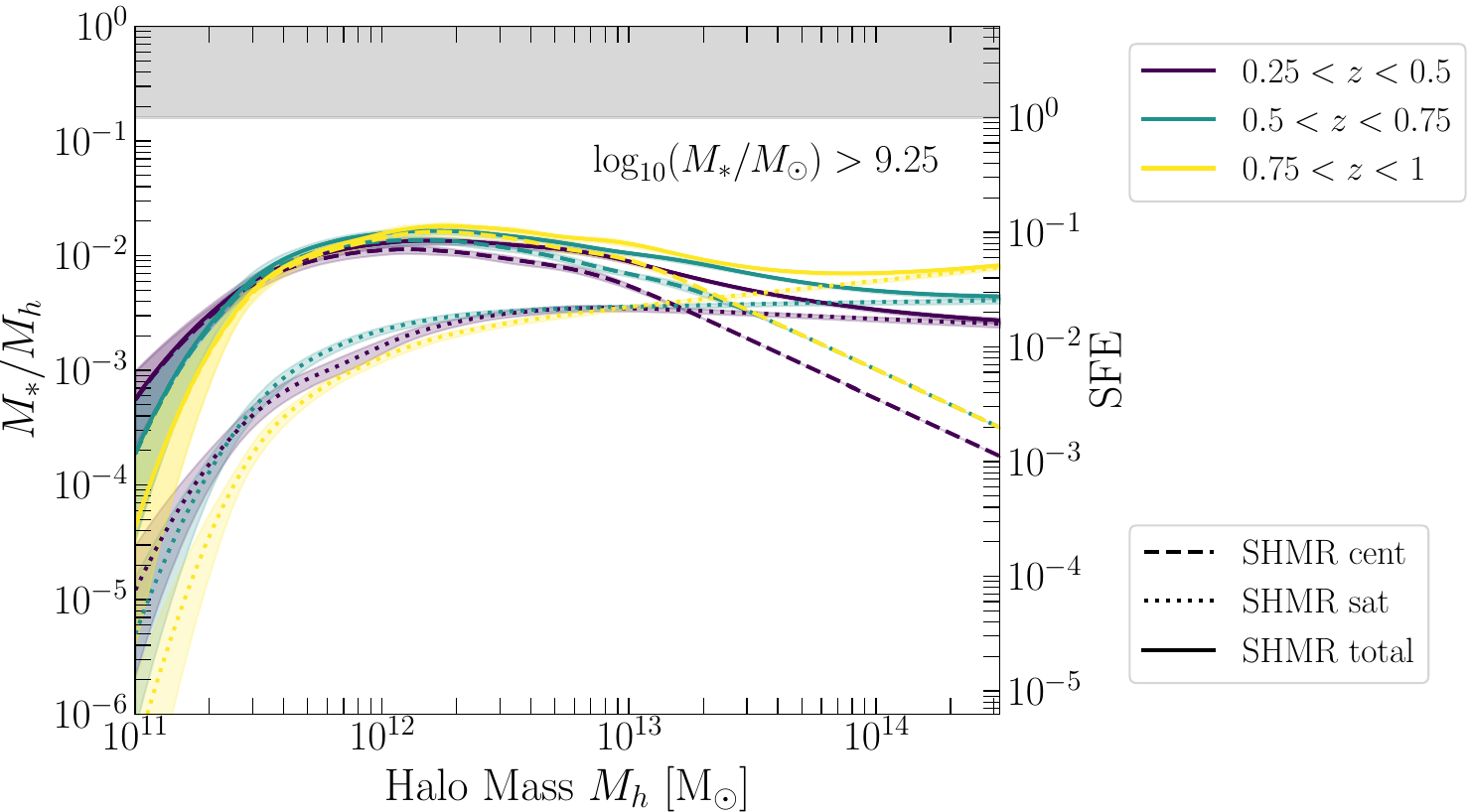} &
        \includegraphics[scale=0.34]{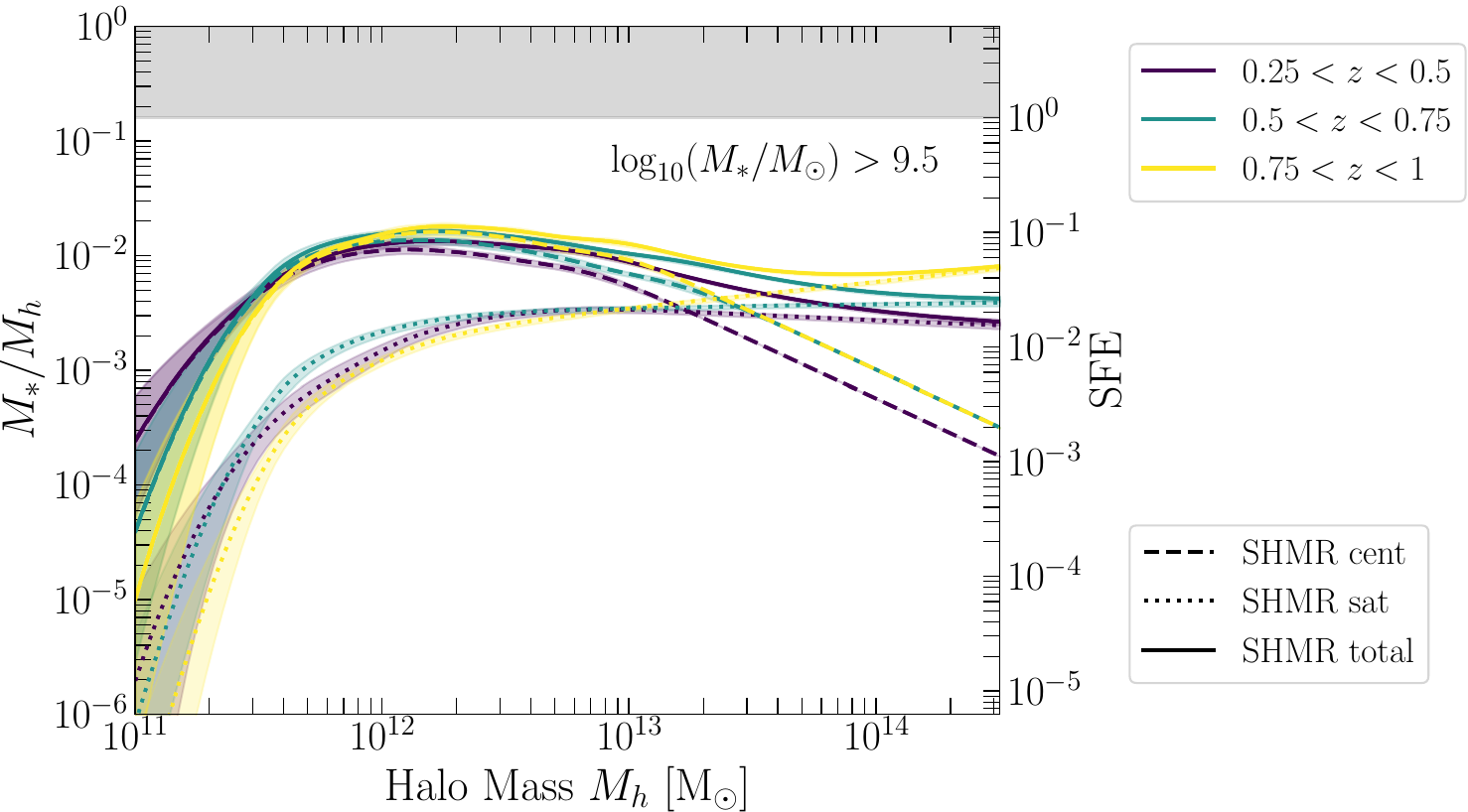} \\ 
        \includegraphics[scale=0.34]{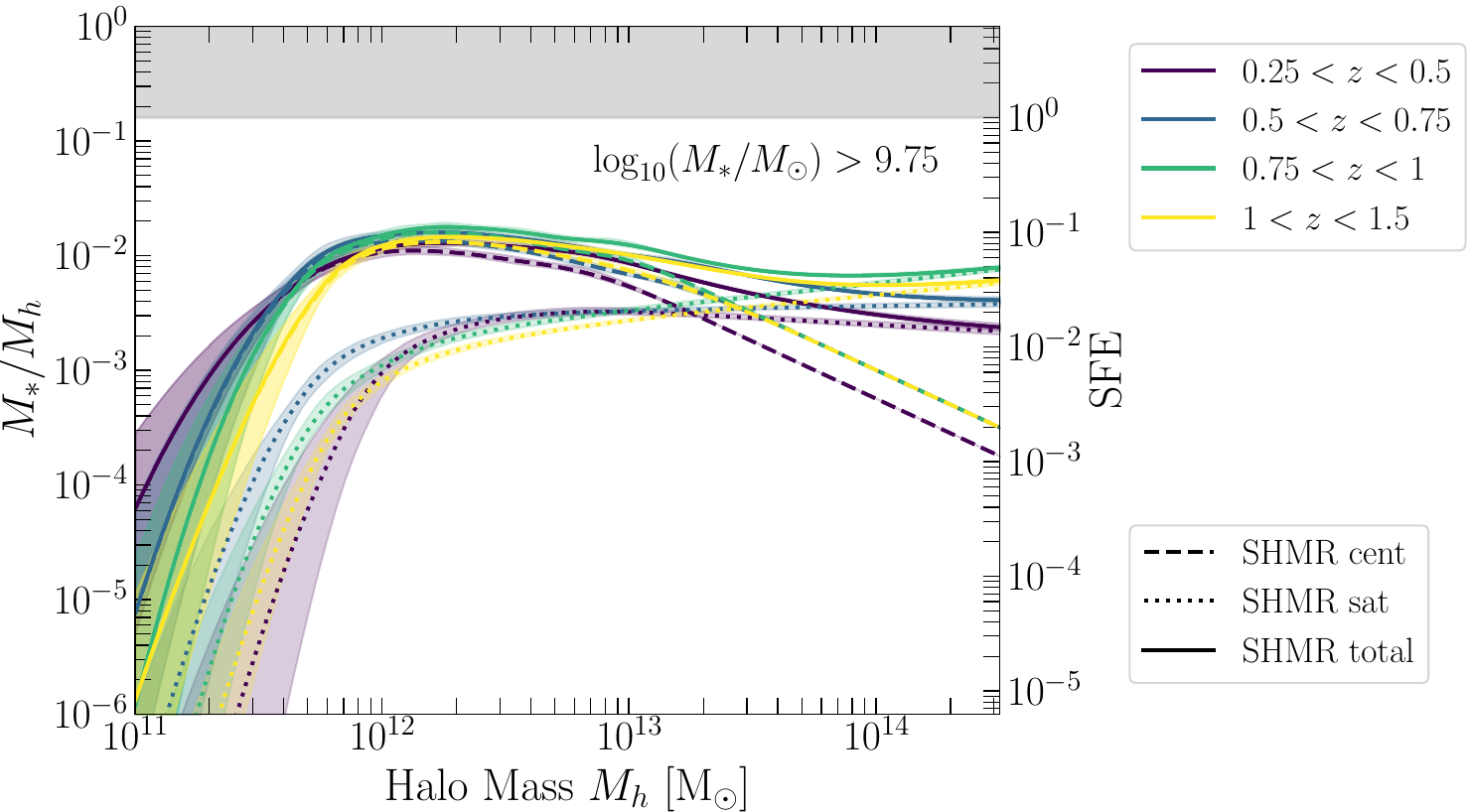} &
        \includegraphics[scale=0.34]{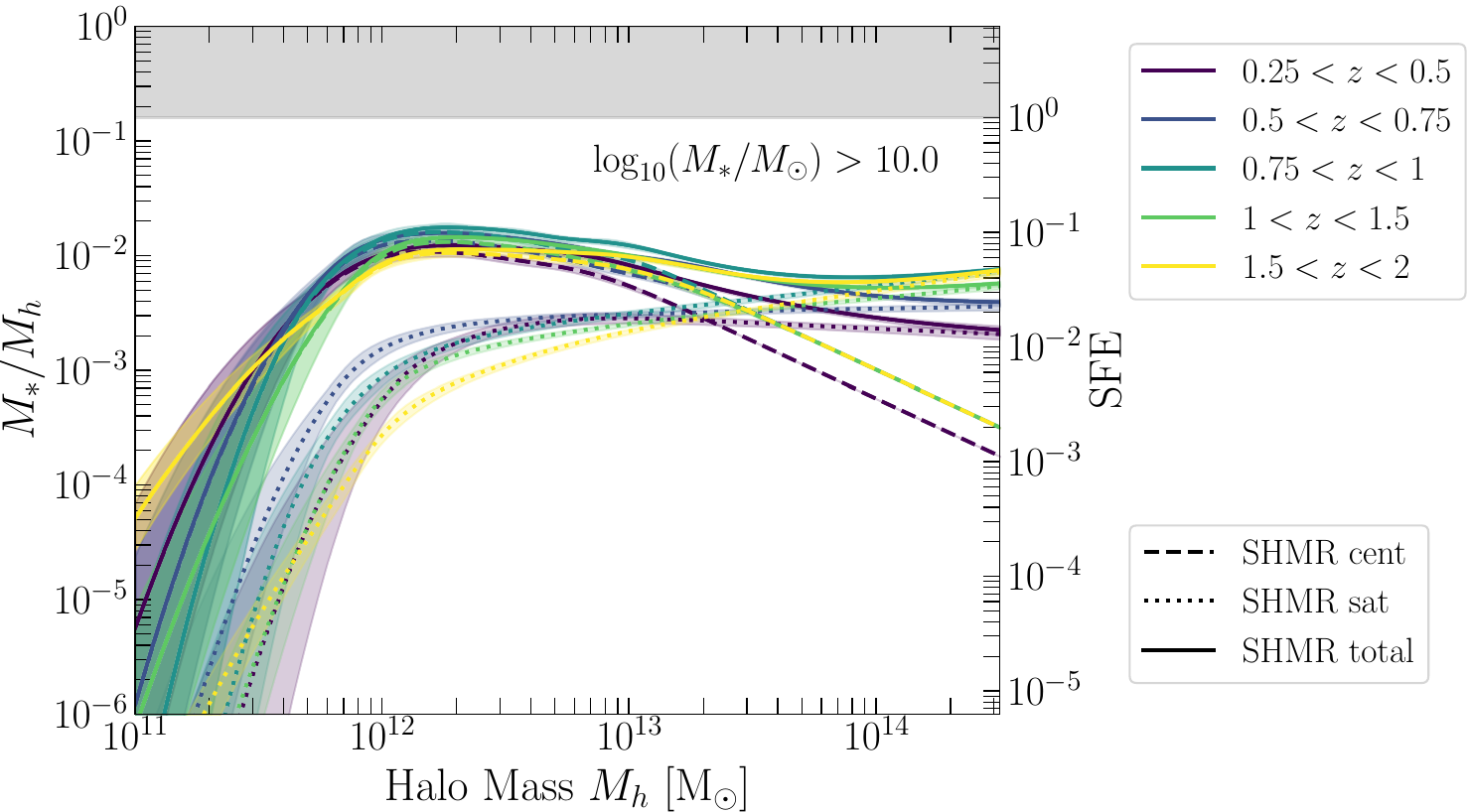} \\ 
        \includegraphics[scale=0.34]{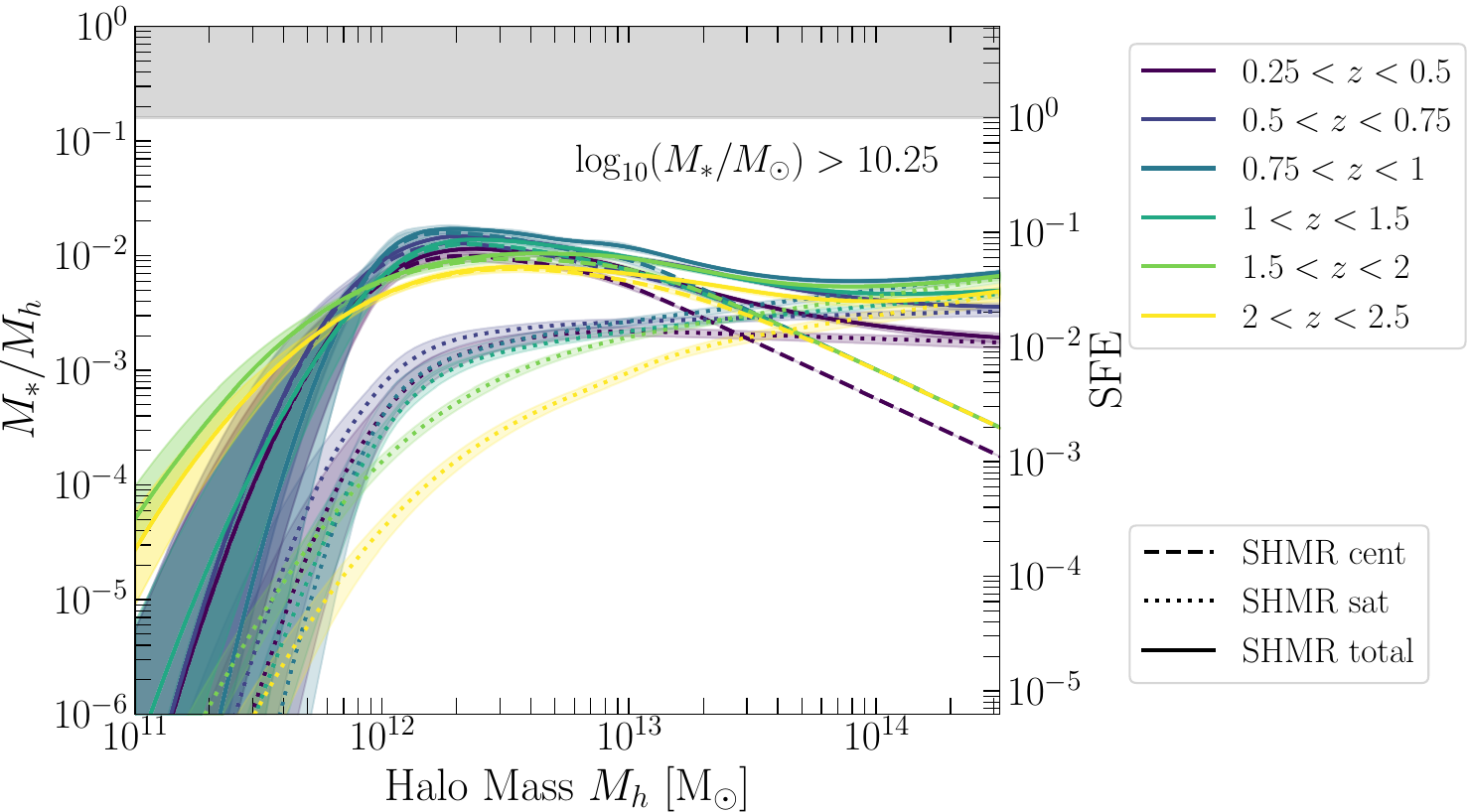} &
        \includegraphics[scale=0.34]{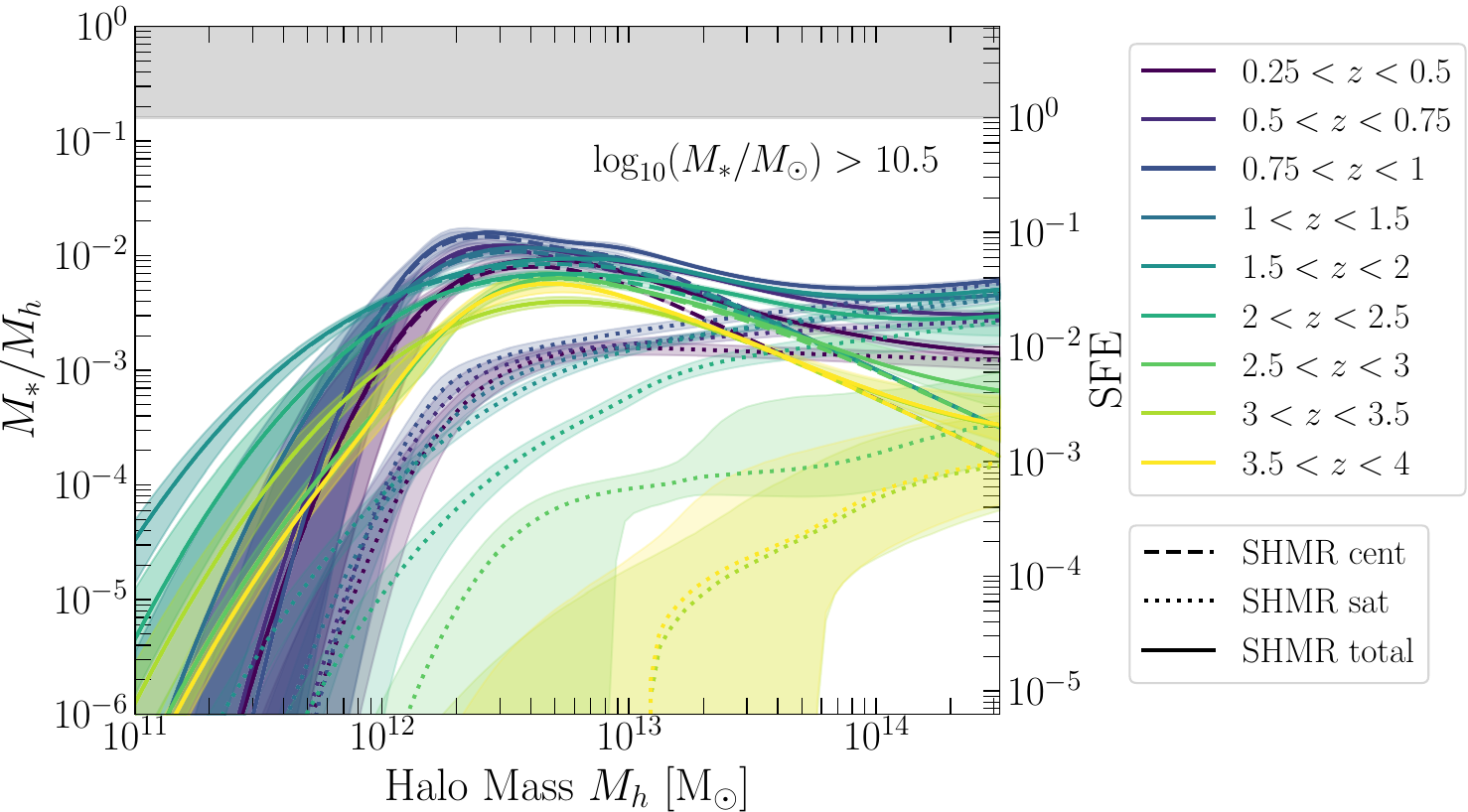} \\
    \end{tabular}
    \caption{The stellar-to-halo mass ratio ($\mathrm{SHMR}=M_{\star}/M_{h}$), for different redshifts ($z$-bins noted in each legend) and for galaxies above a certain stellar mass (i.e. each panel shows the SHMR of galaxies above a different stellar mass threshold, $\log_{10}(M_{\star}/M_\odot)> 9.25, 9.5, 9.75, 10, 10.25, 10.5$).
    The SMHR is separated into contributions from central galaxies (dashed lines), satellite galaxies (dotted lines) and total galaxies (solid lines). On the right y-axis, we present the corresponding integrated star formation efficiency $\mathrm{SFE} = (M_{\star}/M_{h})/f_{b}$. The grey shaded area in each plot highlights above $\mathrm{SFE}=1$ of where efficiencies would exceed the cosmic baryon fraction.}
    \label{fig:SHMR_plots}
\end{figure*}

\begin{figure}
    \centering
    \includegraphics[scale=0.41]{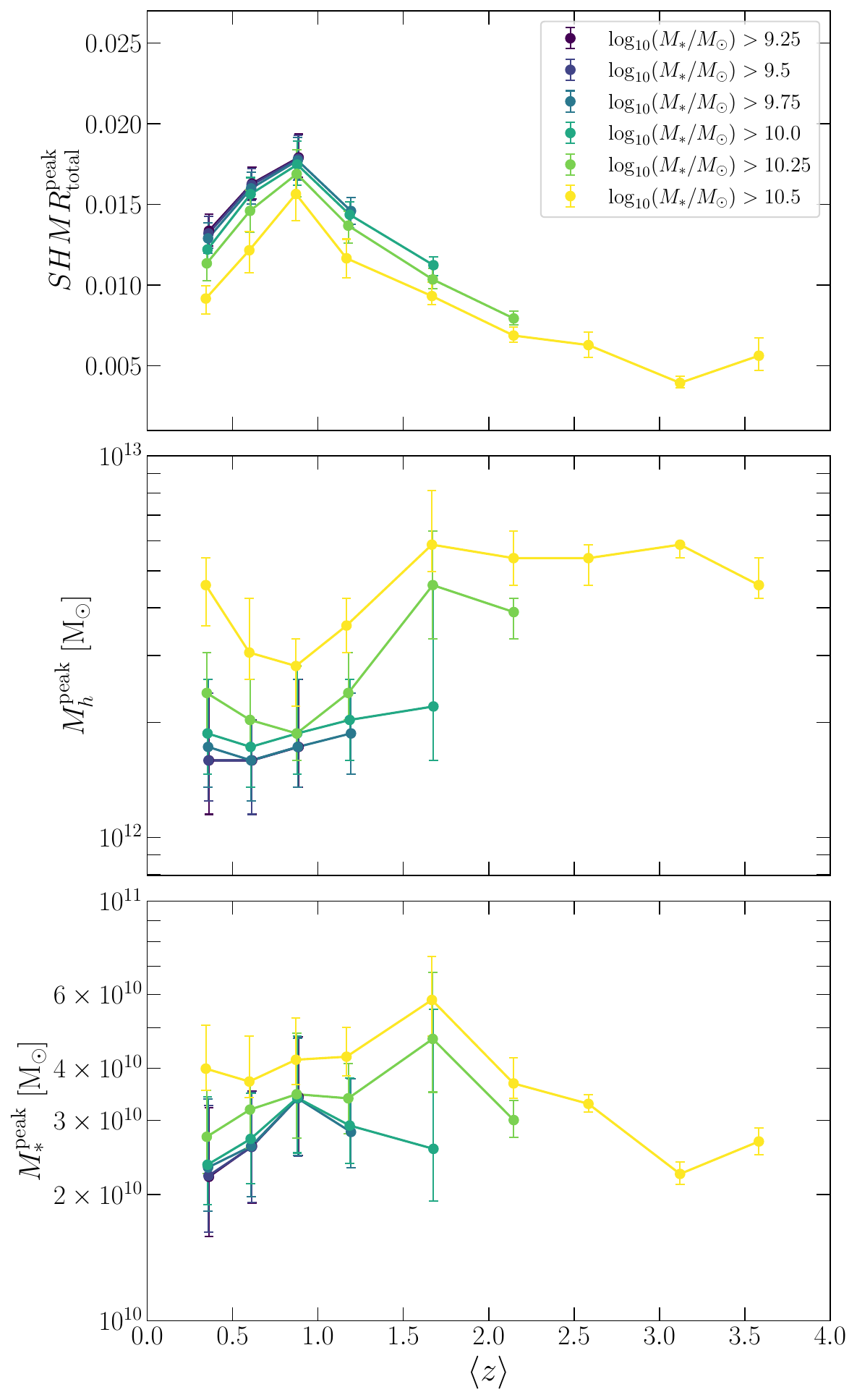}
    \caption{The peak $\mathrm{SHMR}$ of the total galaxies, $\mathrm{SHMR}^{\rm peak}_{\rm total}$ (top panel) and the corresponding peaks of halo mass, $M_{h}^{\rm peak}$ (central panel) and stellar mass, $M_{\star}^{\rm peak}$ (bottom panel), for different stellar mass thresholds, as a function of the weighted mean redshift.}
    \label{fig:Mh_and_M*_peaks_plots}
\end{figure}

The stellar-to-halo mass relation (SHMR) shows how efficient the dark matter halos of different stellar masses are at forming galaxies at different redshifts. Since the galaxy population is divided into centrals and satellites, the SHMR can be broken down into a central contribution (from halos hosting a central above a stellar-mass threshold)  and a satellite contribution (from halos hosting a satellite above a stellar-mass threshold). The total SHMR is thus the sum of these components, providing a direct way to quantify how stellar mass is distributed between central galaxies and satellites as a function of halo mass and redshift.

Following \cite{2025A&A...695A..20S}, the SHMR can be recast as an integrated star-formation efficiency, defined as $\mathrm{SFE} = (M_{\star}/M_{h})/f_{b}$, which expresses the stellar mass fraction relative to the baryonic reservoir expected in the halo. We assume that halos retain baryons in proportion to the universal baryonic fraction, and we adopt $f_{b}=\Omega_{b}/\Omega_{m}=0.16$. The SFE describes the efficiency with which available baryons in a halo of a given mass are converted into stars. In this framework, variations in $M_{\star}/M_{h}$  (and hence SFE) with halo mass and redshift reflect the impact of halo growth and the processes that promote or suppress the conversion of gas into stars, including gas cooling, merging, and stellar/AGN feedback.

We compute the SHMR for stellar-mass-threshold samples in multiple redshift bins, using the posterior HOD parameter chains from our clustering fits. For a given stellar-mass-threshold sample, the posterior provides the 5 HOD model parameters $M_{\min}$, $M_0$, $M_1$, $\sigma_{\log_{10} M}$, $\alpha$. These fully provide the central $\langle N_\mathrm{central}|M_{h} \rangle$ and satellite $\langle N_\mathrm{sat}|M_{h} \rangle$ distributions given by Equations \ref{eq:Ncent} and \ref{eq:Nsat}, which are essentially describing the number of galaxies above each mass threshold as a function of halo mass $\langle N (>M^{th}_{\star} )|M_{h} \rangle$. 
Since the mass threshold samples are cumulative, galaxies above a high stellar mass appear in all lower threshold samples. Hence, we convert the threshold occupations into differential stellar-mass intervals by subtracting adjacent thresholds:
$$\langle N(M^{th}_{\star, i}, M^{th}_{\star, i+1})|M_{h} \rangle = \langle N (>M^{th}_{\star, i} )|M_{h} \rangle - \langle N (>M^{th}_{\star, i+1} )|M_{h} \rangle $$
We assume that the galaxy stellar mass function declines steeply, so the $\left[ M^{th}_{\star, i}, M^{th}_{\star, i+1} \right]$ sample is expected to be dominated by galaxies near the lower stellar mass threshold, $M^{th}_{\star, i}$. 
The SHMR is then computed for the central and satellite populations separately, following the methodology of \citealt{2016MNRAS.459.2618H}, for stellar mass thresholds of $M^{th}_{\star, i} < M^{th}_{\star, i+1} < \ldots < M^{th}_{\star, k}$ as:
\begin{equation}
\begin{split}
    \mathrm{SHMR}(M_h) = \Bigg[
    &\sum_{i=1}^{k-1}
    \left(
    \langle N(>M^{\mathrm{th}}_{\star,i}) \mid M_h \rangle
    -
    \langle N(>M^{\mathrm{th}}_{\star,i+1}) \mid M_h \rangle
    \right)
    M^{\mathrm{th}}_{\star,i}
    \\
    &\quad
    + \langle N(>M^{\mathrm{th}}_{\star,k}) \mid M_h \rangle
    \, M^{\mathrm{th}}_{\star,k} 
    \Bigg] \frac{1}{M_h}
\end{split}
\end{equation}
where $M_{h}$ is a uniform halo mass grid of $\log_{10} M_{h} \in [11, 14.5]$. 

In each redshift bin, we build $\mathrm{SHMR}_\mathrm{cent} (M_{h})$ and $\mathrm{SHMR}_\mathrm{sat} (M_{h})$ for each halo mass $M_{h}$, which then sum up to $\mathrm{SHMR}_\mathrm{total} (M_{h}) = \mathrm{SHMR}_\mathrm{cent} (M_{h}) + \mathrm{SHMR}_\mathrm{sat} (M_{h})$. We evaluate the SHMR across all posterior samples, reporting the median along with the 16th and 84th percentiles as uncertainties at each $M_{h}$.

Because we compute SHMR by summing contributions from stellar-mass-threshold samples, our measurement should be interpreted as the SHMR of the observed, threshold-selected galaxy population, not a continuous $M_{\star}(M_{h})$ mapping. A limitation of this approach is that it cannot capture the full stellar mass budget in halos: it excludes galaxies below the surveys’ mass completeness limit, and it can omit some massive galaxies that are too rare to yield independent clustering constraints. However, this approach has the advantage of relying on galaxy populations directly probed by the data used, and it avoids introducing model-dependent extrapolations into regimes the data cannot directly constrain, making the calculation simple and conservative. 

We present a set of stellar-to-halo mass ratios ($\mathrm{SHMR}=M_{\star}/M_{h}$) in Figure \ref{fig:SHMR_plots} as functions of halo mass, across all available redshift bins, for different stellar mass thresholds, separated into contributions from the central, satellite and total galaxies. We provide a secondary y-axis with the integrated star-formation efficiency SFE, and the region above $\mathrm{SFE}=1$ in the plots is shaded to highlight efficiencies exceeding the cosmic baryon budget. All of our measurements remain safely below the $\mathrm{SFE}>1$ line, confirming that the stellar mass does not exceed the total baryons and that galaxies contain only a small fraction of available baryons, as expected. 

The central population in  Figure \ref{fig:SHMR_plots} is the dominant contribution in $\mathrm{SHMR}_\mathrm{total}$ at lower to intermediate halo masses, where most halos host at most one above-threshold galaxy. In contrast, the satellite population becomes increasingly important toward higher halo masses, beyond $M_{h}\sim10^{13}$\,M$_{\odot}$, where satellite occupations rise rapidly. The strong dependence of $M_{\star}/M_{h}$ on halo mass shows that different physical mechanisms regulate star formation most effectively on different halo mass scales. The total galaxy population shows the SHMR rising toward an intermediate peak halo mass of $M_{h}\approx 10^{12} $\,M$_{\odot}$ where galaxy formation is most efficient, and the balance between gas accretion, cooling, and feedback creates optimal conditions for star formation. The SHMR declines at both lower and higher halo masses, where stellar mass growth is suppressed.

\subsubsection{Evolution of $M_{h}^{\mathrm{peak}}$ and $M_{\star}^{\mathrm{peak}}$} \label{sec:Mh_peak and Mstar_peak}

We can go a step further by examining the evolution of the peak halo mass $M_{h}^{\mathrm{peak}}$, which identifies the halo masses most efficient at forming galaxies of a given stellar mass. We compute the peak halo mass $M_{h}^{\mathrm{peak}}$, by finding the $M_{h}$ where $\mathrm{SHMR}_\mathrm{total}$ is maximum, $\mathrm{SHMR}_\mathrm{total}^{\rm peak}$, for each posterior realisation and then taking the median and 16th-84th percentiles of the resulting $M_{h}^{\mathrm{peak}}$ distribution.

In Figure \ref{fig:Mh_and_M*_peaks_plots} (top panel), we show the evolution of the $\mathrm{SHMR}_\mathrm{total}^{\rm peak}$ with the weighted mean redshift. Across all stellar mass thresholds, the maximum SHMR increases towards $z\sim1$, where the inferred galaxy-formation efficiency is highest, and then declines at higher redshifts. This trend is consistent with the picture discussed previously, where $M_{\star}/M_{h}$ depends on the balance between the build-up of stellar mass and the growth of the host halo. At higher redshifts, halos are still growing rapidly, while the stellar mass has not yet built up as much relative to the halo mass. By $z\sim1$, halos close to the most efficient mass scale have formed a larger fraction of their stars, producing the peak in the SHMR. The lower values at below $z\sim 1$ suggest that halo growth continues at a higher rate than the stellar mass build-up, at least for galaxies with $\log_{10}(M_{\star}/{\rm M}_\odot) > 9.25$. The lack of build up in stellar mass suggests that below $z< 1$ star formation in haloes traced by these galaxies is essentially shutting down, having used up all the available gas, which is reflected in the universal decline in the star-formation rate density of the Universe \citep[e.g.][for a review]{MadauDickinson2014}. The main cause of this shut down in star formation is often attributed to AGN-driven feedback \citep{Bower2006,Croton2006,BestHeckman2012} in the form of mechanical feedback from low-power radio jets, and the relative ubiquity of such sources in the most massive haloes hosting the most massive galaxies  at $z<1$ supports this \citep[e.g.][]{2018MNRAS.477.3760H, Magliocchetti2022, Hamlett2026}.

This maximum in the peak SHMR is also at the redshift where we see the power-law slope of the satellite galaxies begin to decline (Figure~\ref{fig:alpha_plot}) for all mass thresholds, yet the posteriors shown in Fig.~\ref{fig:Corner_Plots_HOD-5-parameters_lower-zbins} show that there is not a significant degeneracy between minimum halo mass for a central and the power-law slope for the satellite occupation. This suggests that the flattening of the $\alpha$ parameter is closely related to the evolution in the peak of the SHMR, yet the satellites themselves contribute very little to the position of the peak of the SHMR (Figure~\ref{fig:SHMR_plots}). We interpret this as haloes continue to grow through hierarchically accreting lower mass haloes, but where these haloes are also largely devoid of the gas needed to fuel further star formation and stellar mass growth. Indeed, the fact that we see a decrease in the peak SHMR and the peak halo mass for all stellar mass threshold samples supports this.

Moreover, the cause for the maximum at $z\sim 0.8$ appears to be the result of the relative difference in the evolution in stellar mass and halo mass for the different stellar-mass threshold samples. In Figure~\ref{fig:Mh_and_M*_peaks_plots} (middle panel) we show the evolution of the peak halo mass $M_{h}^{\mathrm{peak}}$ with redshift (i.e. weighted mean z) for galaxies above different stellar masses thresholds. For the low-stellar mass thresholds $\log_{10}(M_{\star}/{\rm M}_\odot) \lesssim 10$ there appears to be very little evolution in the peak halo mass, with a marginal slight rise from $z\sim 0\rightarrow 1$. On the other hand, the peak halo mass for the more massive stellar-mass threshold samples $(\log_{10}(M_{\star}/{\rm M}_\odot) \gtrsim  10)$ appears to show a clear minimum at $z\sim 0.8$ with subsequent rise towards $z=0$ and $z\sim 1.5$, followed by a plateau to higher redshifts. This is most obvious in the highest stellar mass threshold bin $(\log_{10}(M_{\star}/{\rm M}_\odot) \gtrsim  10.5)$ where our wide and deep data set allows us to probe to $z\sim 3.5$, but is also apparent out to $z\sim 2.5$ for the $\log_{10}(M_{\star}/{\rm M}_\odot) \gtrsim  10.25$ threshold sample.

In the bottom panel of Figure~\ref{fig:Mh_and_M*_peaks_plots} we show the corresponding evolution of the peak stellar mass. We compute the peak stellar mass in a similar way to the peak halo mass, where for each posterior realisation, $M_{\star}^{\mathrm{peak}}$ is found at the halo mass where the total SHMR peaks via $M_{\star}^{\mathrm{peak}}=\mathrm{SHMR}_\mathrm{tot}^{\mathrm{peak}}(M_{h}^{\mathrm{peak}})\times M_{h}^{\mathrm{peak}}$. We then take the median and 16th-84th percentile range of this $M_{\star}^{\mathrm{peak}}$ distribution.  
We find that the peak stellar mass generally increases from low to intermediate redshifts for all stellar mass thresholds.
However, for the lower stellar mass thresholds ($\log_{10} (M_{\star}/$M$_{\odot})<10$), $M_{\star}^{\mathrm{peak}}$ reaches a maximum at $z\sim1$, with $M_{\star}^{\mathrm{peak}}\approx 3\times10^{10} $\,M$_{\odot}$. On the other hand, for the higher stellar masses thresholds $\log_{10} (M_{\star}/$M$_{\odot})>10$, the peak occurs at a significantly higher redshift, around $z\sim1.75$, with $M_{\star}^{\mathrm{peak}}\approx 5.5\times10^{10}$\,M$_{\odot}$. In both cases, $M_{\star}^{\mathrm{peak}}$ then declines towards higher redshifts.

This suggests that the stellar mass scale at which galaxies are most efficiently formed also evolves with redshift. Therefore, we observe that both halos and galaxies convert baryons into stars more efficiently at intermediate redshifts ($1<z<2$), and their efficiency falls at both lower and higher redshifts. Similar to the evolution of $M_{h}^{\mathrm{peak}}$, $M_{\star}^{\mathrm{peak}}$ also depends on the stellar mass threshold. The lower-mass samples show little evolution in $M_{\star}^{\mathrm{peak}}$, while the higher-mass samples show a shift towards larger $M_{\star}^{\mathrm{peak}}$, indicating that the characteristic stellar mass of efficient galaxy formation increases for more massive galaxy populations. This is a manifestation of cosmic downsizing \citep{Cowie1996}, where the most massive galaxies form efficiently in the most massive halos at $1.5 < z <2$. 

\subsection{AGN feedback as an explanation of the evolution in the SHMR}
The cause of this transition at $z \sim 0.8$ for the most massive galaxies is likely related to the onset of radio-mode feedback from low-accretion rate active galactic nuclei. Feedback from radio jets has been a cornerstone of the vast majority of simulations of galaxy formation and evolution for the past two decades \citep[e.g.][]{Bower2006, Croton2006, HorizonAGN,SIMBA,Schaye2026}, and is invoked to stop the most massive galaxies in the simulations becoming more massive than observations find. However, observationally the evidence is less secure, with only the bubbles detected within the X-ray halos of galaxy clusters providing robust evidence for this effect \citep[e.g.][]{2012ARA&A..50..455F}. On the other hand, it is clear that these low-luminosity radio sources, that often do not show evidence for ongoing accretion at other wavelengths \citep[e.g.][]{HardcastleCroston2007, BestHeckman2012}, are abundant in the most massive galaxies \citep{WilliamsRottgering2015, Kondapally2025} at $z< 1$ \citep[e.g.][]{McAlpine2013,Smolcic2017,Kondapally2023,Thykkathu2026} and they also preferentially occupy the most massive dark matter halos \citep[e.g.][]{Lindsay2014,Hale2018,Magliocchetti2022,Hale2025}.

In particular, \cite{Hamlett2026} recently showed using data from the MeerKAT International Tiered Extragalactic Exploration \citep[MIGHTEE;][]{MIGHTEE} Survey, which covers the same fields as used in this work, that radio galaxies reside in a factor of $\sim 1.5$ more massive halos that similarly massive galaxies without evidence of radio jet activity. This suggests that the ongoing build up in peak halo mass that we observe at $z<0.8$, coupled with the decrease in the SHMR, could be regulated by this radio-mode feedback activity. The intermittent radio mode feedback ensures that the central massive galaxies do not become overly massive, while the dark matter halo continues to grow hierarchically, resulting in the observed decrease in the SHMR and the increase in peak halo mass for the most massive galaxies.

Further work on linking the radio emission from active galaxies to the underlying evolution of the SHMR, peak-halo and peak-stellar mass would help confirm whether indeed this is the case, following for example the prescription set out in \cite{2017MNRAS.472.3570H}.

\section{Conclusions}
\label{sec:Conclusions}
We have used deep optical and near-infrared data from UltraVISTA and VIDEO, together with complementary optical data across the COSMOS, XMM-LSS and ECDF-S fields to study the connection between galaxies and their host dark matter haloes over $0.25 < z < 4$. We constructed stellar-mass-selected galaxy samples, and computed the angular two-point correlation function in each field across a wide range of redshift and stellar mass thresholds, with the photometric-redshift PDFs used to weight the contribution of each galaxy to each redshift bin. We then fitted the observed clustering with a combined-field five-parameter HOD model of \cite{2005ApJ...633..791Z}, allowing us to infer the evolution of the halo occupation, galaxy bias, satellite fraction and stellar-to-halo mass relation.

The key strength of this work is the combination of three independent extragalactic survey fields covering an area of 9.3 degrees$^2$. This allows us to retain the depth required for high-redshift galaxy samples while substantially increasing the volume available for understanding the link between massive galaxies and their dark-matter halos.
This combined-field approach provides a substantially larger effective area than previous deep-field clustering studies, at comparable redshifts, mainly based on COSMOS-sized fields. Since massive galaxies are rare, this larger effective area is essential for allowing us to better sample the high-mass galaxy population, reduce the impact of cosmic variance.

Our findings are summarised as follows:
\begin{itemize}
    \item The clustering amplitude increases with stellar mass threshold at all redshifts, consistent with more massive galaxies residing in more massive and more strongly clustered dark matter haloes. The larger volume provided by XMM-LSS and ECDF-S is essential for constraining the clustering of the most massive galaxies at higher redshifts, than COSMOS alone.

\item The HOD halo-mass parameters, $M_{\min}$ and $M_{1}$, both increase with stellar mass threshold, as expected if progressively more massive central and satellite galaxies occupy progressively more massive haloes. We also find clear evidence for redshift evolution, with galaxies of fixed stellar mass occupying more massive haloes at higher redshift. 

\item We find an upturn in the ratio $M_{1}/M_{\min}$ at the highest stellar masses and at $z>2.5$, suggesting that substantially more massive haloes are required to host massive satellite galaxies compared to central galaxies at early times.

\item The satellite slope $\alpha$ shows evidence for redshift evolution, decreasing towards lower redshift at $z\lesssim1.5$. This is consistent with a picture in which massive haloes continue to assemble through the accretion of smaller haloes whose central galaxies become satellites.

\item By combining the three independent fields of COSMOS, XMM-LSS and ECDF-S, we reduced the cosmic variance contribution by approximately a factor of two compared to the individual fields, better sampling the large-scale structure and demonstrating the importance of wide, independent fields for clustering measurements, especially on the high-mass galaxy end.

\item The galaxy bias increases with higher galaxy mass, as more massive galaxies tend to form in larger, more biased dark matter halos. Galaxy bias also decreases toward lower redshifts, as galaxies shift from forming in overdense, biased regions at high-$z$ to more typical environments at low-z, thereby better tracing the dark matter field.

\item The satellite fraction decreases with increasing stellar mass threshold, indicating that the most massive galaxies are predominantly central galaxies in massive haloes. The satellite fraction is also higher at lower redshift, likely because the massive halos that host the majority of satellite galaxies are still assembling at higher redshifts. 

\item The SHMR of the total (centrals and satellites) population reaches a maximum at an intermediate halo mass scale of $M_{h}\approx 10^{12}$\,M$_{\odot}$, corresponding to the halo mass where galaxy formation is most efficient. The peak SHMR is at a maximum at $z\sim1$  declining both towards higher redshift and towards lower redshift, highlighting that $z\sim 1$ is the epoch at which massive galaxy formation starts becoming less efficient.

\item The evolution of the peak SHMR seems to be driven by different behaviour for low- and high-stellar-mass threshold samples. At lower stellar masses, the decline in the peak SHMR below $z\sim1$ is mainly associated with a decrease in $M_{\star}^{\rm peak}$ at approximately constant $M_{h}^{\rm peak}$, implying that lower-mass galaxies are tracing similar halo masses at later times. At higher stellar masses, the decline in the peak SHMR  below $z\sim1$ is instead driven by an increase in $M_{h}^{\rm peak}$ while $M_{\star}^{\rm peak}$ remains approximately constant. This indicates that massive haloes continue to grow below $z\sim1$, while the stellar mass of their central galaxies does not grow at the same rate.

\item Together, the evolution in $M_{h}^{\rm peak}$, $M_{\star}^{\rm peak}$, the increasing satellite fraction and the evolution of the satellite slope, $\alpha$, suggest a picture in which the most massive haloes continue to assemble hierarchically at late times, while the stellar mass growth of their central galaxies is suppressed. This may reflect a combination of passive halo growth through satellite accretion and feedback processes, such as AGN feedback, which regulate the growth of massive central galaxies. 

\end{itemize}

Overall, our results show that the massive end of the galaxy-halo connection contains important information about the growth of massive haloes, especially below $z\sim1$ and the regulation of massive galaxies. Our combined COSMOS, XMM-LSS and ECDF-S field analysis with the larger area allows us to probe this regime of the SHMR with improved constraints compared to smaller-area deep-field studies, and provides a foundation for future work linking halo growth, satellite accretion and AGN feedback in the same fields.

\section*{Acknowledgements}

We thank the anonymous referee for providing comments that improved the paper.
NS, MJJ, RGV and TY acknowledge funding from a UKRI Frontiers Research Grant [EP/X026639/1].
MJJ also acknowledges support from the Hintze
Family Charitable Foundation through the Oxford Hintze Centre for Astrophysical Surveys. 
JH acknowledges funding from the Science and Technology Facilities Council (STFC) [grant code ST/Y509474/1].
CLH acknowledges support from the Science and Technology Facilities Council (STFC) through grant ST/Y000951/1 and from the Oxford Hintze Centre for Astrophysical Surveys which is funded through generous support from the Hintze Family Charitable Foundation.

\section*{Data Availability}

All imaging data was obtained from original sources in the public domain. The photometric redshift catalogues used in this study can be provided on request.



\bibliographystyle{mnras}
\bibliography{example} 




\appendix 

\section{}\label{Appendix}

We provide the best-fit HOD parameters for the combined analysis of the three fields in Table~\ref{tab:hod_params_combined_table}. For each redshift bin and stellar mass threshold, we list the best-fitting values of the five HOD parameters, together with the main derived quantities used in the interpretation of the results. These include the modelled galaxy number, satellite fraction, galaxy bias, Poisson uncertainty, cosmic variance uncertainty and reduced $\chi^2$. This table therefore provides a complete summary of both the fitted combined-field HOD model and the derived galaxy-halo connection properties and forms the basis for the trends discussed in the main text.

\begin{figure*}
    \includegraphics[scale=0.39]{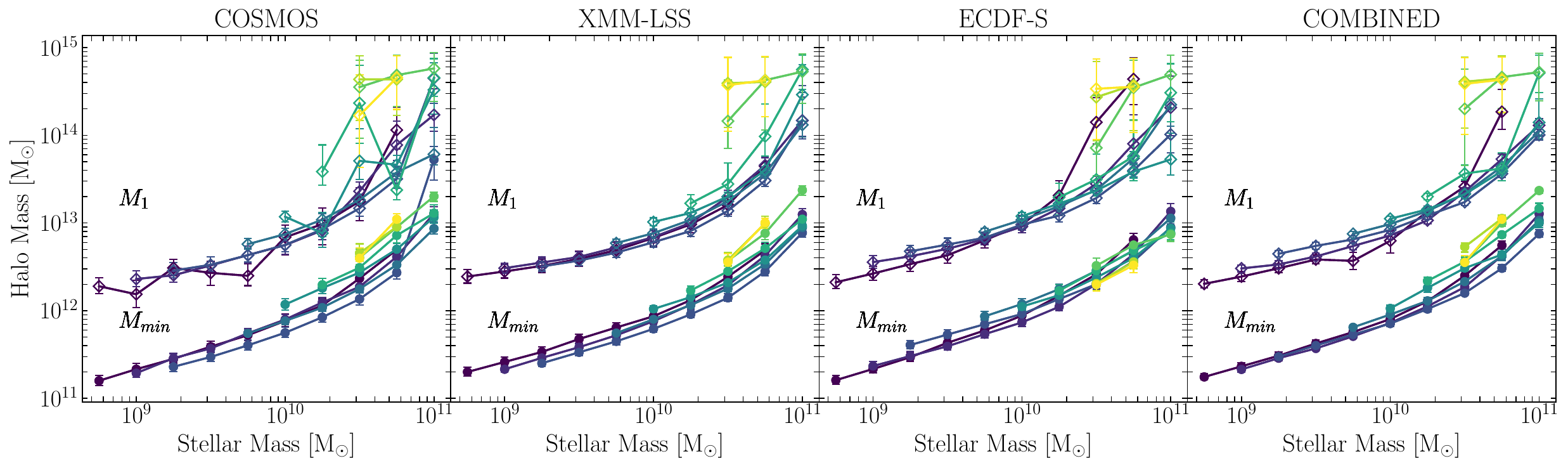}
    \includegraphics[scale=0.39]{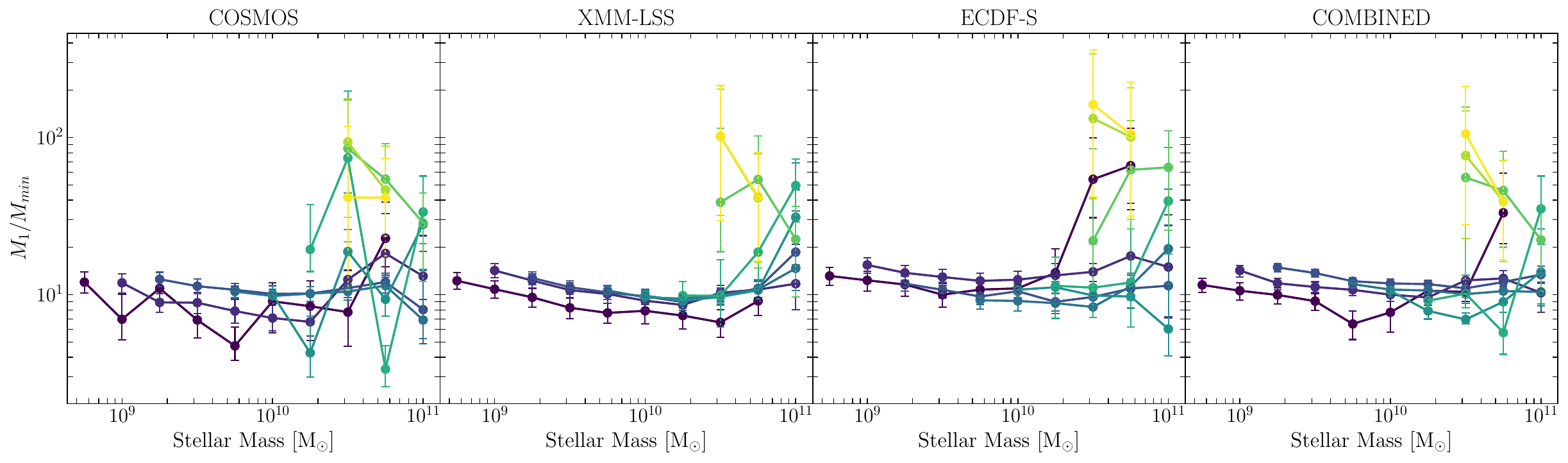}
    \includegraphics[scale=0.39]{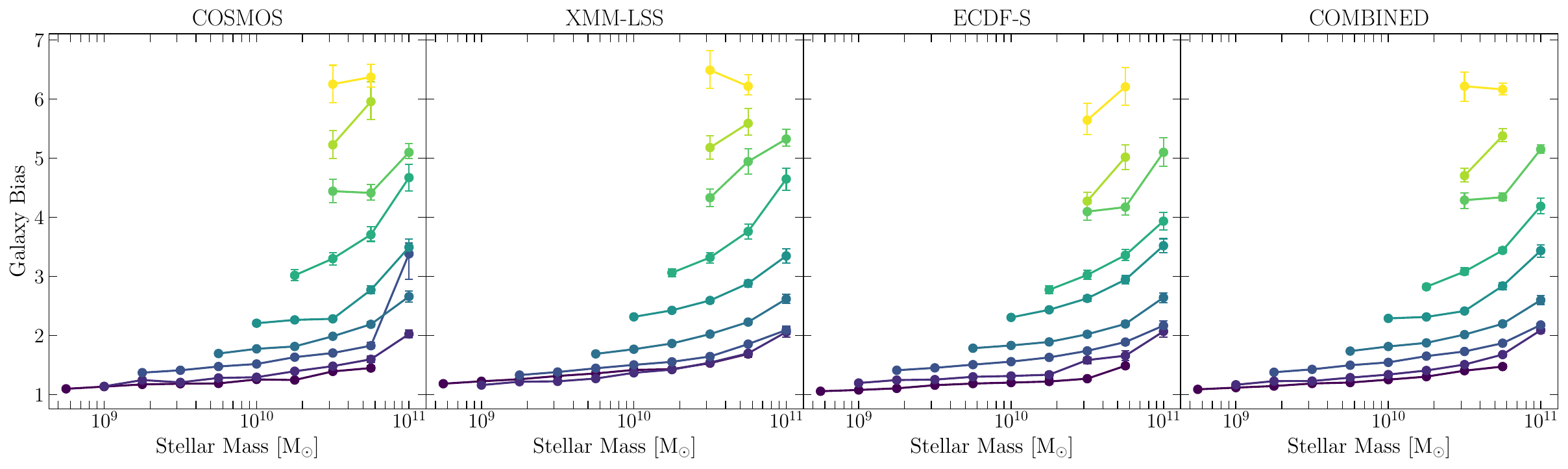}
    \includegraphics[scale=0.39]{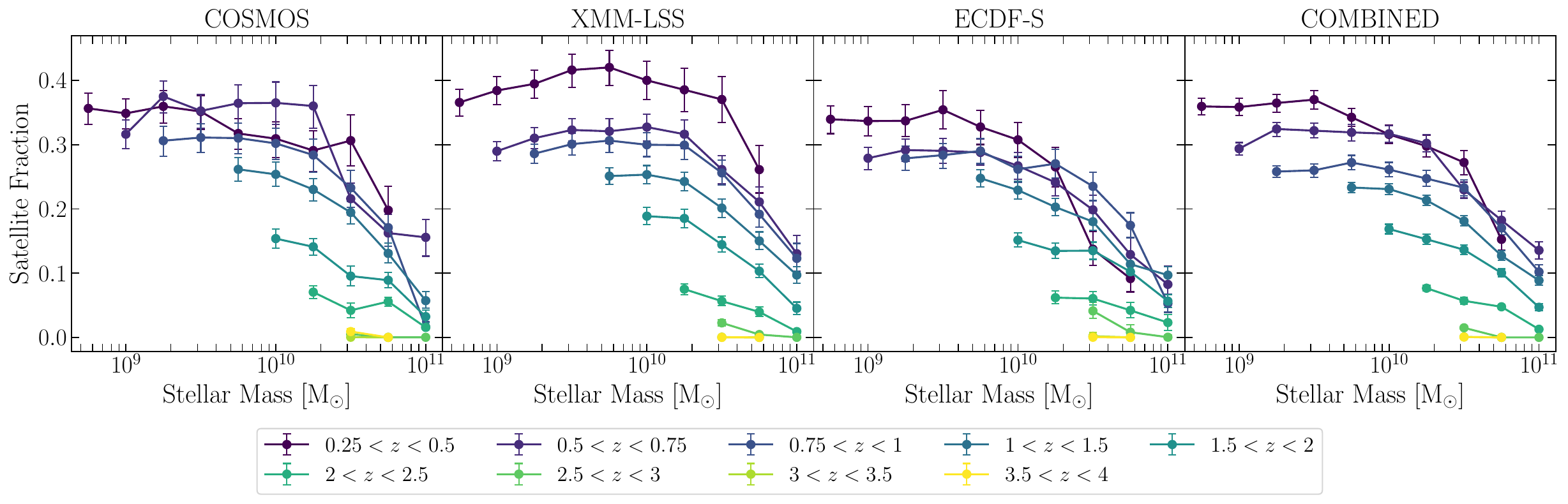}
    \caption{Comparison of the HOD-derived quantities measured separately in COSMOS, XMM-LSS and ECDF-S, and for the COMBINED-field analysis. The quantities shown are $M_{\min}$, $M_{1}$, $M_{1}/M_{\min}$, galaxy bias $b_g$ and satellite fraction $f_{\rm sat}$, plotted as a function of stellar mass threshold in each redshift bin. These comparisons show that the smaller area of COSMOS limits the constraints at the highest stellar masses, while the larger XMM-LSS and ECDF-S fields provide improved leverage on the rare massive galaxy population. The combined-field analysis provides the most reliable constraints used in the main results.}
    \label{fig:Comparison_Plots}
\end{figure*}

We also show a comparison between the individual-field and combined-field HOD results in Figure~\ref{fig:Comparison_Plots}. The comparison plots of Figure~\ref{fig:Comparison_Plots} highlight the importance of using multiple independent fields and demonstrate the role that survey area plays in constraining HOD parameters. Although COSMOS provides the deepest data, its smaller area limits its ability to constrain the rare high-mass galaxy population, particularly at the highest stellar mass thresholds and towards high redshift. The larger VIDEO fields, XMM-LSS and ECDF-S, provide greater volume and therefore stronger constraints on the massive galaxy regime. The combined-field result therefore represents our most robust estimate of the galaxy-halo connection, especially at the high-mass end.

\begin{figure*}
    \centering
    \begin{tabular}{cc}
        \includegraphics[scale=0.2]{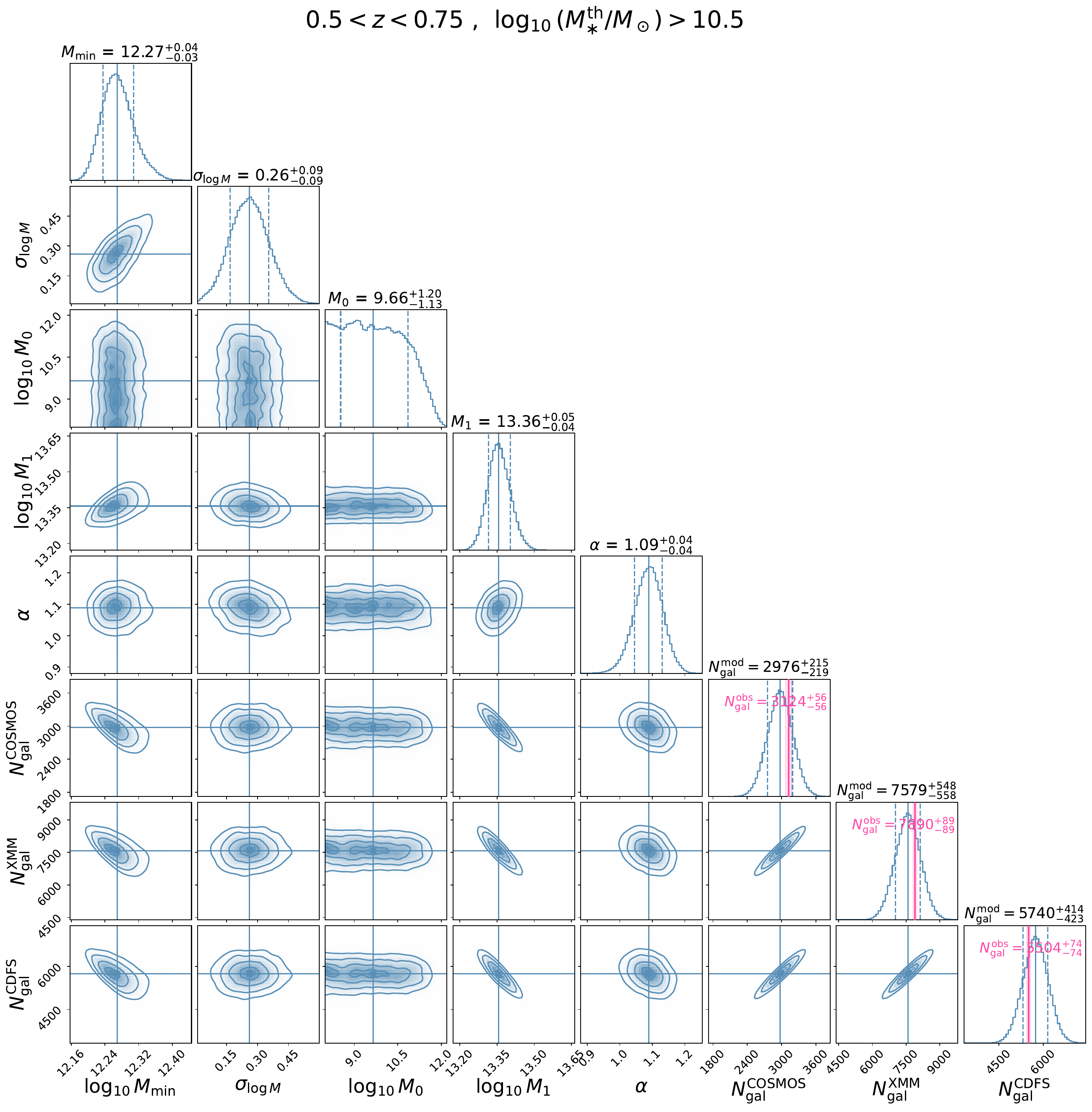} &
        \includegraphics[scale=0.2]{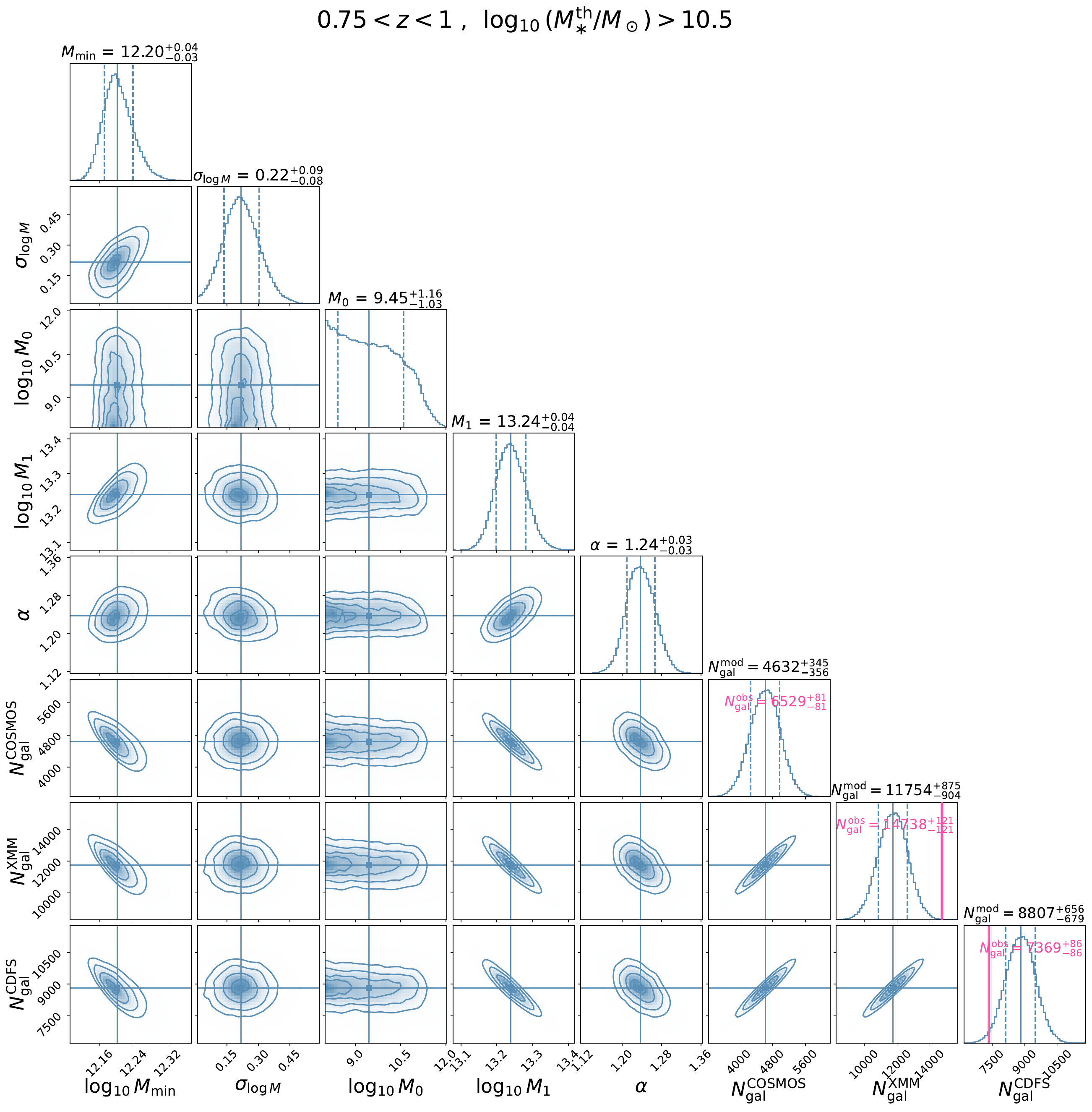} 
        \vspace{4mm}  \\
        \includegraphics[scale=0.2]{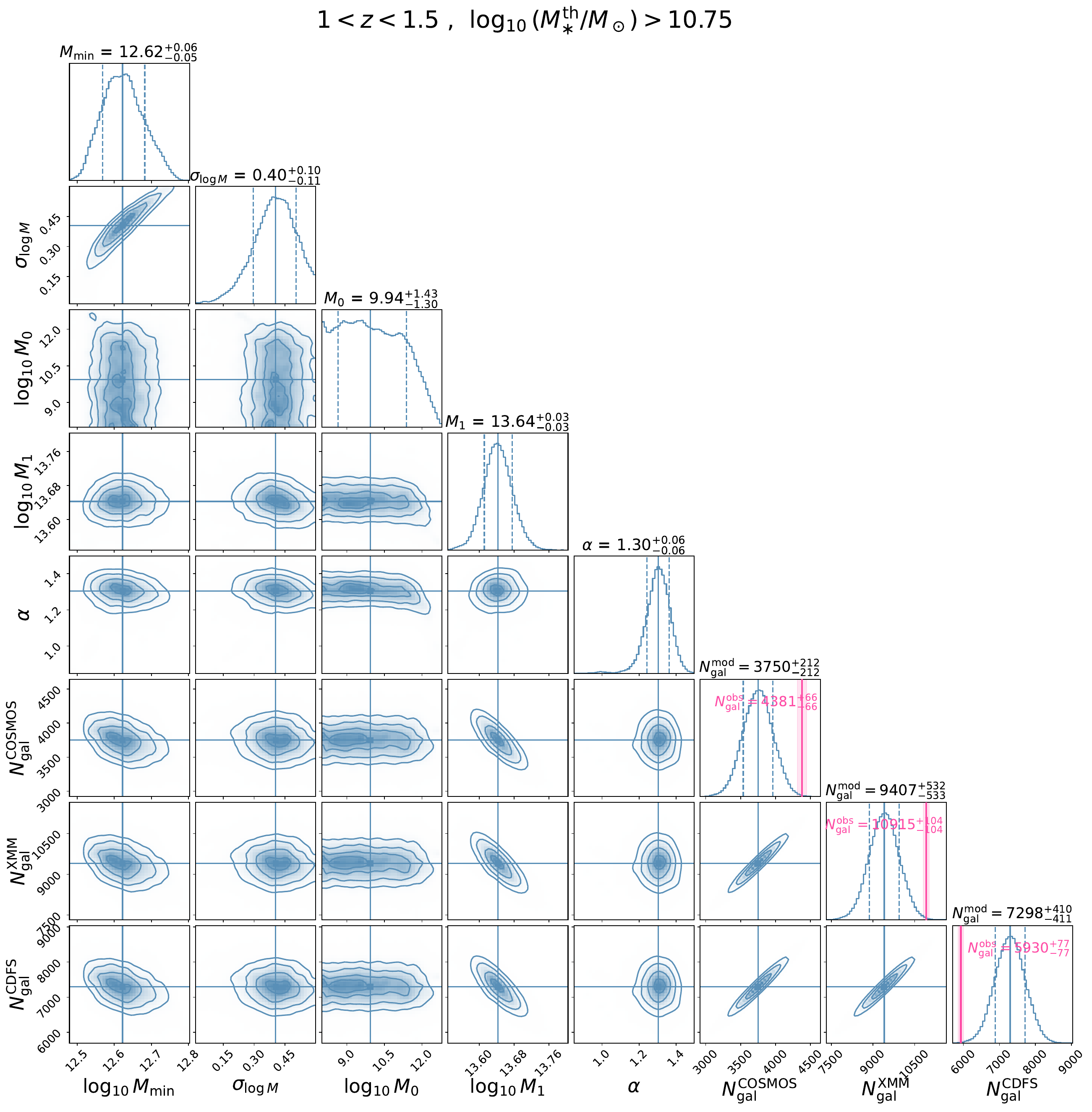} &
        \includegraphics[scale=0.2]{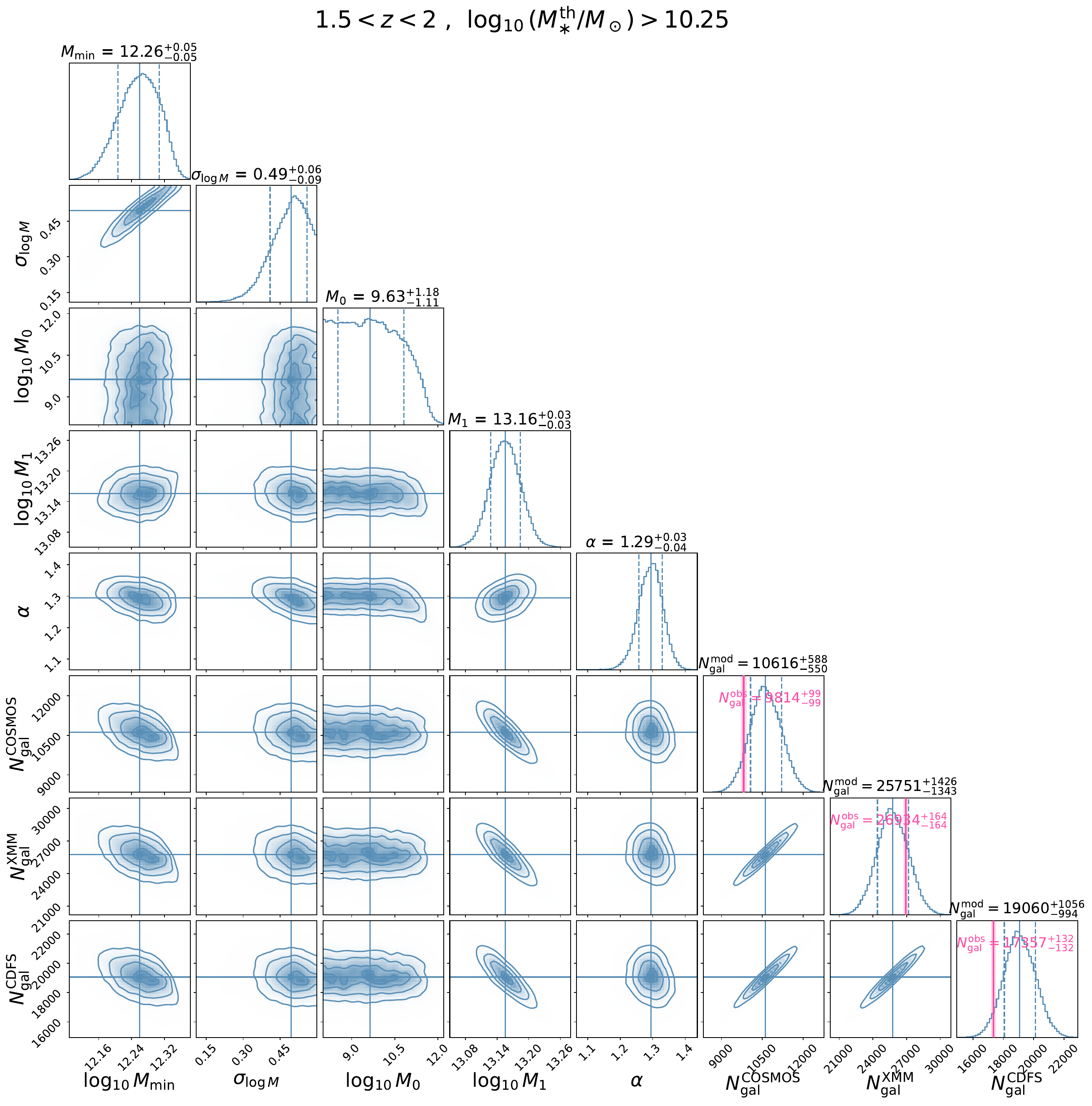}        
    \end{tabular}
    \caption{Corner plots showing the posterior distributions of the HOD parameters, $M_{\min}$, $M_0$, $M_1$, $\sigma_{\log_{10}M}$ and $\alpha$, for the COMBINED-field model in redshift bins; $0.5<z<0.75$, $0.75<z<1$, $1<z<1.5$ and $1.5<z<2$. The plots illustrate the constraints on each parameter and the degeneracies between parameters in the HOD model, for selected stellar mass threshold $\log_{10} (M_*^{\mathrm{th}}/M_\odot)$ samples. We additionally include the posterior distributions of model-predicted number of galaxies $N_\mathrm{gal}^\mathrm{mod}$ for each field, while the observed number of galaxies $N_\mathrm{gal}^\mathrm{obs}$ obtained from the summed photo-$z$ PDFs, are shown in pink for comparison.}
    \label{fig:Corner_Plots_HOD-5-parameters_lower-zbins}
\end{figure*}

\begin{figure*}
    \centering
    \begin{tabular}{cc}
        \includegraphics[scale=0.2]{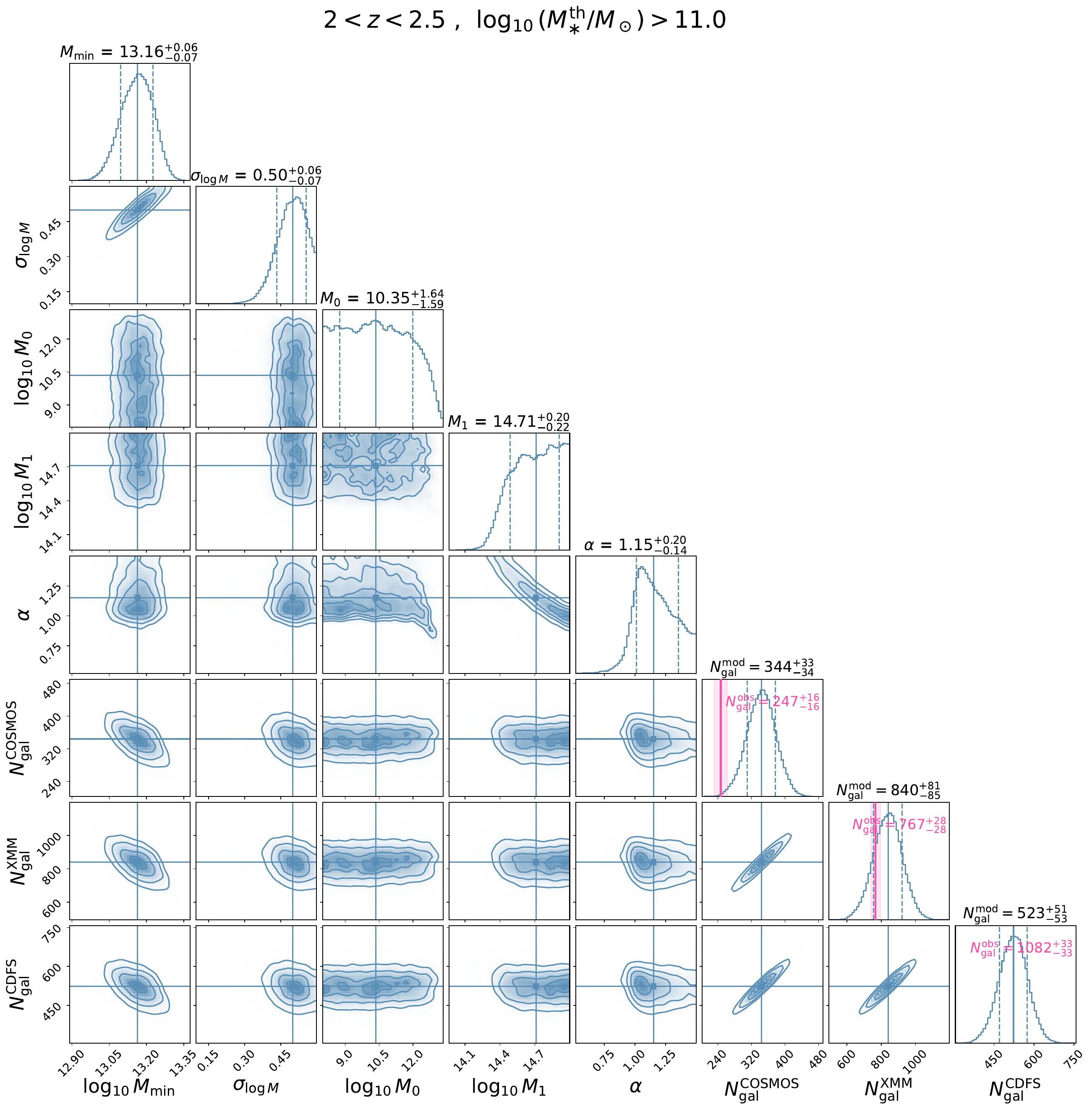} &
        \includegraphics[scale=0.2]{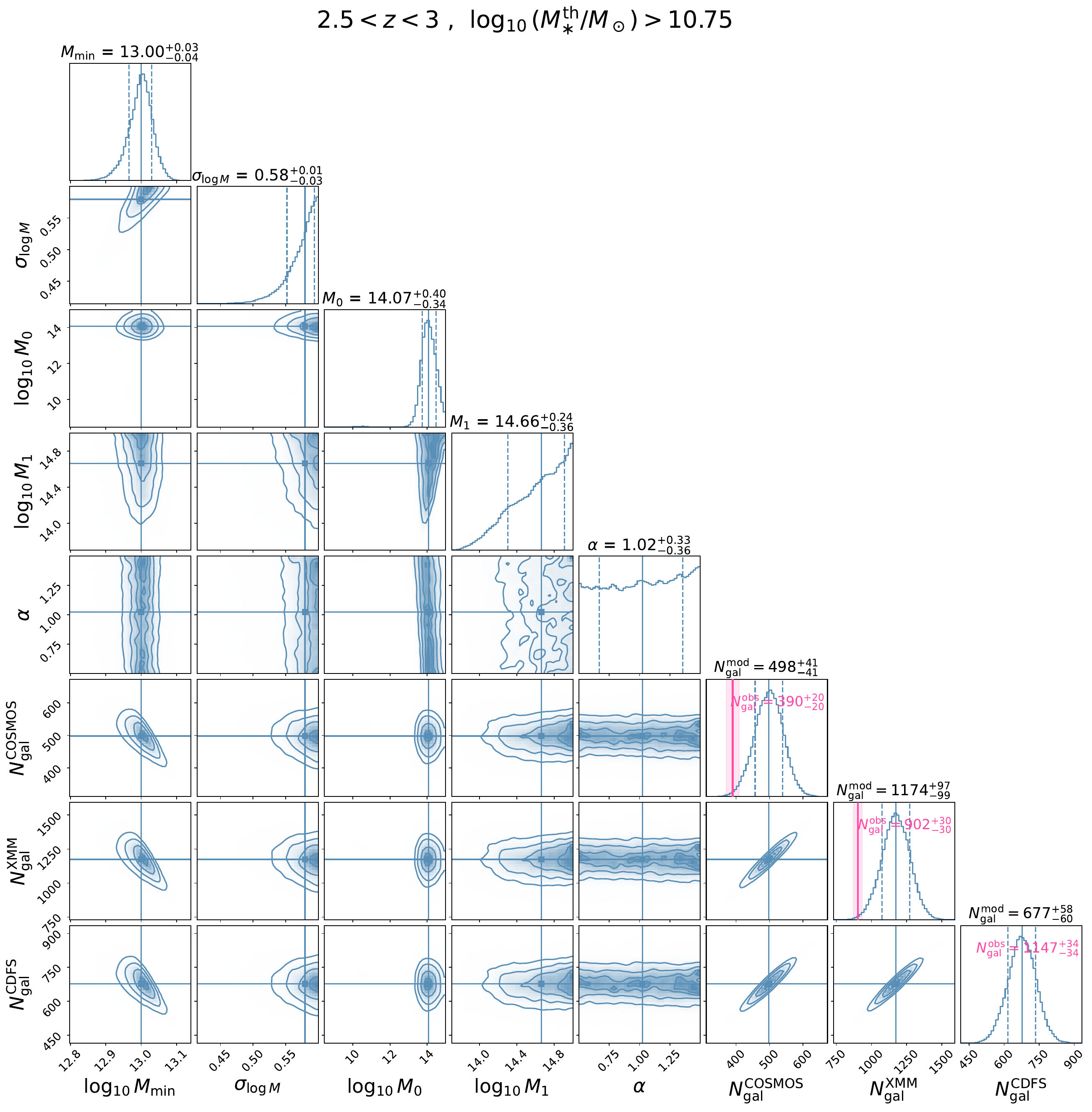} 
        \vspace{4mm}  \\
        \includegraphics[scale=0.2]{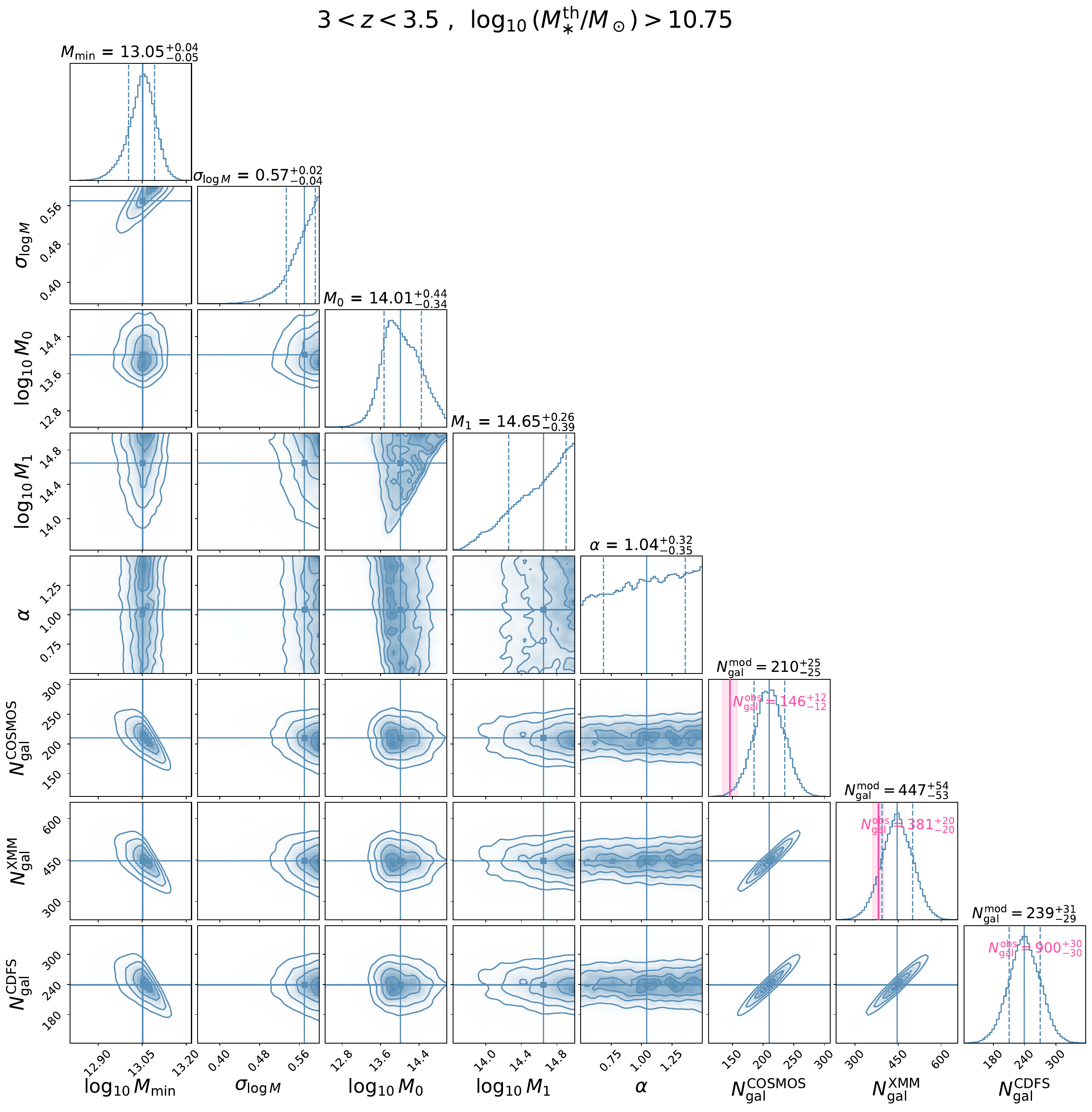} &
        \includegraphics[scale=0.2]{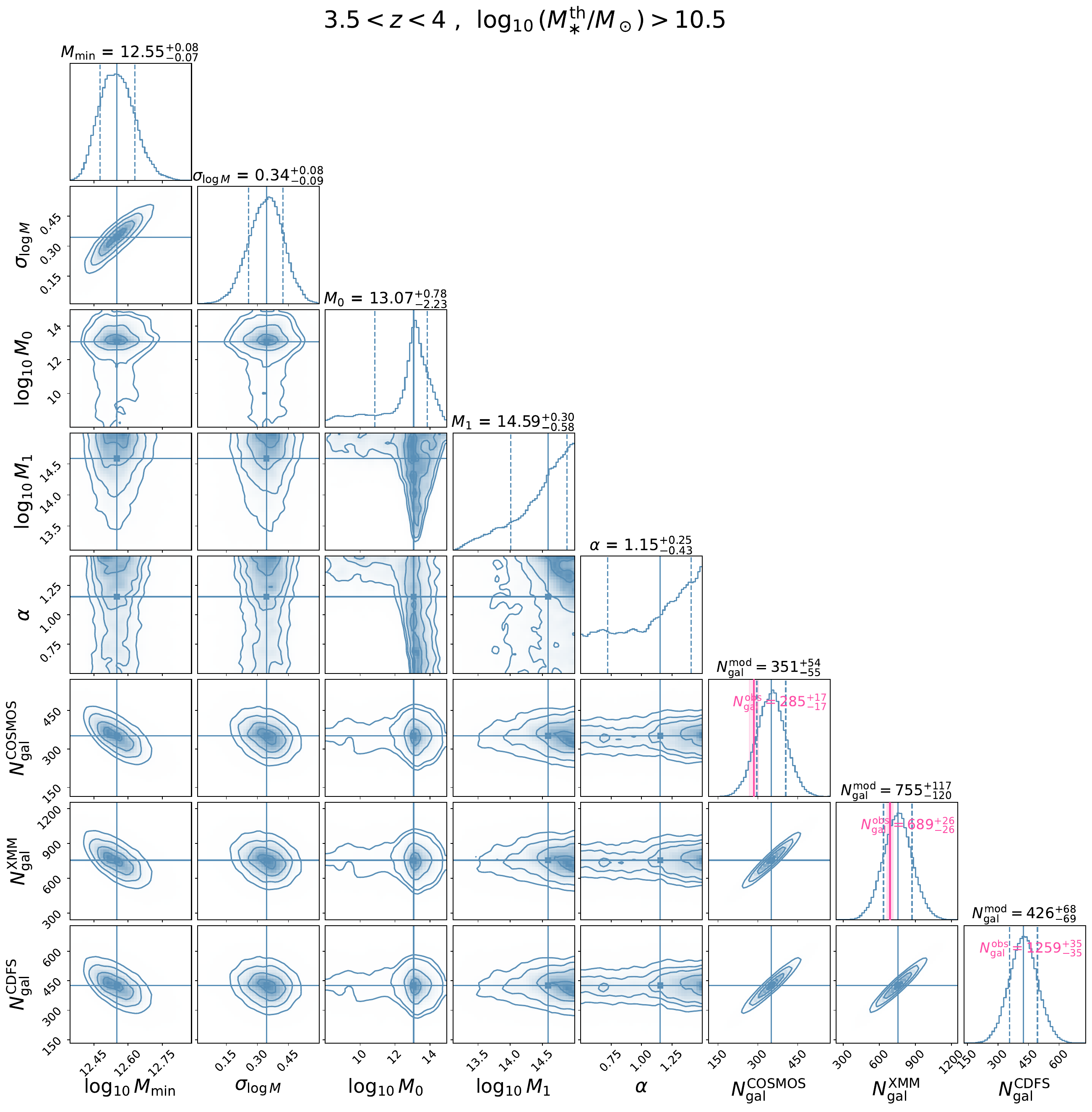}        
    \end{tabular}
    \caption{Corner plots showing the posterior distributions of the HOD parameters, $M_{\min}$, $M_0$, $M_1$, $\sigma_{\log_{10}M}$ and $\alpha$, for the COMBINED-field model in redshift bins; $2<z<2.5$, $2.5<z<3$, $3<z<3.5$ and $3.5<z<4$. The plots illustrate the constraints on each parameter and the degeneracies between parameters in the HOD model, for selected stellar mass threshold $\log_{10} (M_*^{\mathrm{th}}/M_\odot)$ samples. We additionally include the posterior distributions of model-predicted number of galaxies $N_\mathrm{gal}^\mathrm{mod}$ for each field, while the observed number of galaxies $N_\mathrm{gal}^\mathrm{obs}$ obtained from the summed photo-$z$ PDFs, are shown in pink for comparison.}
    \label{fig:Corner_Plots_HOD-5-parameters_higher-zbins}
\end{figure*}

At the end, we include the corner plots for the COMBINED-field HOD model in different redshift bins in Figures \ref{fig:Corner_Plots_HOD-5-parameters_lower-zbins} and \ref{fig:Corner_Plots_HOD-5-parameters_higher-zbins}. These corner plots show the posterior distributions of the fitted HOD parameters and highlight parameter degeneracies, for selected stellar mass threshold samples.

\begin{landscape}
\begin{table}
\centering
\caption{The best-fit parameter values from the HOD COMBINED model ($M_{\min}$, $M_0$, $M_1$, $\sigma_{\log_{10} M}$, $\alpha$) are shown for each $z$-bin and stellar mass threshold $M_*^{\mathrm{th}}$. For each sample, we provide the mean stellar mass $M_*^{\mathrm{mean}}$ and the number of observed galaxies $N_\mathrm{gal}^\mathrm{obs}$ (calculated from the sum of the photo-$z$ PDFs in each sample), as well as other model outputs; the number of modelled galaxies $N_\mathrm{gal}^\mathrm{mod}$, the satellite fraction $f_\mathrm{sat}$, the galaxy bias $b_g$, the Poisson uncertainty $\sigma_\mathrm{Poisson}$ measured from $N_\mathrm{gal}^\mathrm{mod}$, the cosmic variance uncertainty $\sigma_\mathrm{CV}$ (see Section \ref{sec: cosmic variance}), and the $\chi^2_\mathrm{reduced}$, which is the best-fit $\chi^2$ divided by the degrees of freedom in our model. The stellar mass columns $M_{*}^{th}$ and $M_*^{\mathrm{mean}}$ are expressed in units of $\log_{10}(M_{*}/M_\odot)$, while the HOD mass parameters, $M_{\min}$, $M_0$ and $M_1$ are expressed in units of $\log_{10}(M_{h}/M_\odot)$.}
\label{tab:hod_params_combined_table}
\begin{tabular}{|c|c|c|c|c|c|c|c|c|c|c|c|c|c|c|}
\hline
\textbf{$z$-bin} & \textbf{$M_*^{\mathrm{th}}$} & \textbf{$M_*^{\mathrm{mean}}$} & \textbf{$N_\mathrm{gal}^\mathrm{obs}$} & \textbf{$N_\mathrm{gal}^\mathrm{mod}$} & \textbf{$M_{\min}$} & \textbf{$M_0$} & \textbf{$M_1$} & \textbf{$\sigma_{\log_{10} M}$} & \textbf{$\alpha$} & \textbf{$f_\mathrm{sat}$} & \textbf{$b_g$} & \textbf{$\sigma_\mathrm{Poisson}$} & \textbf{$\sigma_\mathrm{CV}$} & \textbf{$\chi^2_\mathrm{reduced}$} \\ \hline
$0.25 - 0.5$ & 8.75 & 9.59 & 92163 & $91976_{-5875}^{+5994}$ & $11.25_{-0.03}^{+0.04}$ & $9.39_{-0.92}^{+0.94}$ & $12.31_{-0.05}^{+0.05}$ & $0.30_{-0.10}^{+0.10}$ & $0.93_{-0.01}^{+0.01}$ & $0.36_{-0.01}^{+0.01}$ & $1.09_{-0.01}^{+0.01}$ & $0.011$ & $0.065$ & 3.1 \\ 
 & 9.00 & 9.76 & 73089 & $72441_{-4913}^{+4993}$ & $11.37_{-0.04}^{+0.04}$ & $10.14_{-1.45}^{+0.91}$ & $12.39_{-0.06}^{+0.05}$ & $0.32_{-0.10}^{+0.10}$ & $0.94_{-0.03}^{+0.02}$ & $0.36_{-0.01}^{+0.01}$ & $1.12_{-0.01}^{+0.01}$ & $0.012$ & $0.067$ & 3.5 \\ 
 & 9.25 & 9.93 & 56608 & $56620_{-3874}^{+3559}$ & $11.49_{-0.04}^{+0.04}$ & $9.70_{-1.15}^{+1.15}$ & $12.48_{-0.05}^{+0.05}$ & $0.32_{-0.10}^{+0.10}$ & $0.93_{-0.02}^{+0.02}$ & $0.36_{-0.01}^{+0.01}$ & $1.14_{-0.01}^{+0.01}$ & $0.014$ & $0.068$ & 2.7 \\ 
 & 9.50 & 10.10 & 42749 & $42832_{-2839}^{+2948}$ & $11.63_{-0.04}^{+0.04}$ & $9.91_{-1.29}^{+1.11}$ & $12.58_{-0.05}^{+0.05}$ & $0.33_{-0.09}^{+0.10}$ & $0.94_{-0.02}^{+0.02}$ & $0.37_{-0.01}^{+0.01}$ & $1.19_{-0.01}^{+0.01}$ & $0.016$ & $0.071$ & 3.5 \\ 
 & 9.75 & 10.26 & 31340 & $31006_{-2265}^{+2190}$ & $11.76_{-0.04}^{+0.05}$ & $11.86_{-0.20}^{+0.15}$ & $12.57_{-0.10}^{+0.10}$ & $0.32_{-0.10}^{+0.10}$ & $0.82_{-0.06}^{+0.06}$ & $0.34_{-0.01}^{+0.01}$ & $1.20_{-0.01}^{+0.01}$ & $0.019$ & $0.072$ & 4.7 \\ 
 & 10.00 & 10.42 & 21816 & $21377_{-1582}^{+1608}$ & $11.90_{-0.04}^{+0.05}$ & $11.94_{-0.63}^{+0.22}$ & $12.80_{-0.14}^{+0.12}$ & $0.31_{-0.10}^{+0.10}$ & $0.87_{-0.10}^{+0.11}$ & $0.32_{-0.01}^{+0.01}$ & $1.25_{-0.02}^{+0.02}$ & $0.022$ & $0.075$ & 3.2 \\ 
 & 10.25 & 10.59 & 13767 & $13258_{-1038}^{+1000}$ & $12.11_{-0.04}^{+0.05}$ & $10.14_{-1.43}^{+1.30}$ & $13.13_{-0.06}^{+0.06}$ & $0.29_{-0.10}^{+0.10}$ & $0.96_{-0.05}^{+0.04}$ & $0.30_{-0.02}^{+0.02}$ & $1.30_{-0.02}^{+0.02}$ & $0.029$ & $0.078$ & 2.4 \\ 
 & 10.50 & 10.78 & 7251 & $6811_{-563}^{+550}$ & $12.40_{-0.04}^{+0.05}$ & $9.88_{-1.29}^{+1.40}$ & $13.41_{-0.06}^{+0.06}$ & $0.31_{-0.10}^{+0.10}$ & $0.96_{-0.07}^{+0.06}$ & $0.27_{-0.02}^{+0.02}$ & $1.40_{-0.03}^{+0.02}$ & $0.041$ & $0.084$ & 3.2 \\ 
 & 10.75 & 10.98 & 2941 & $2572_{-243}^{+238}$ & $12.75_{-0.04}^{+0.05}$ & $10.28_{-1.58}^{+1.53}$ & $14.27_{-0.20}^{+0.26}$ & $0.28_{-0.10}^{+0.10}$ & $0.67_{-0.12}^{+0.16}$ & $0.15_{-0.02}^{+0.02}$ & $1.47_{-0.03}^{+0.03}$ & $0.068$ & $0.088$ & 2.3 \\ 
\hline
$0.5 - 0.75$ & 9.00 & 9.78 & 148473 & $142877_{-8134}^{+7804}$ & $11.33_{-0.03}^{+0.04}$ & $9.13_{-0.76}^{+0.86}$ & $12.48_{-0.04}^{+0.04}$ & $0.27_{-0.09}^{+0.09}$ & $0.98_{-0.02}^{+0.02}$ & $0.29_{-0.01}^{+0.01}$ & $1.17_{-0.01}^{+0.01}$ & $0.009$ & $0.055$ & 3.3 \\ 
 & 9.25 & 9.95 & 115956 & $112356_{-6543}^{+6499}$ & $11.46_{-0.03}^{+0.03}$ & $9.11_{-0.74}^{+0.83}$ & $12.53_{-0.04}^{+0.04}$ & $0.25_{-0.10}^{+0.09}$ & $0.99_{-0.02}^{+0.02}$ & $0.32_{-0.01}^{+0.01}$ & $1.23_{-0.01}^{+0.01}$ & $0.010$ & $0.058$ & 4.0 \\ 
 & 9.50 & 10.12 & 87817 & $86817_{-5009}^{+4907}$ & $11.57_{-0.03}^{+0.03}$ & $9.16_{-0.77}^{+0.78}$ & $12.62_{-0.04}^{+0.04}$ & $0.23_{-0.09}^{+0.09}$ & $0.93_{-0.02}^{+0.02}$ & $0.32_{-0.01}^{+0.01}$ & $1.23_{-0.01}^{+0.01}$ & $0.011$ & $0.058$ & 4.4 \\ 
 & 9.75 & 10.29 & 64626 & $63989_{-3891}^{+3740}$ & $11.71_{-0.03}^{+0.03}$ & $9.23_{-0.85}^{+0.97}$ & $12.74_{-0.04}^{+0.04}$ & $0.23_{-0.09}^{+0.09}$ & $0.96_{-0.02}^{+0.02}$ & $0.32_{-0.01}^{+0.01}$ & $1.29_{-0.01}^{+0.01}$ & $0.013$ & $0.061$ & 3.4 \\ 
 & 10.00 & 10.44 & 45934 & $45959_{-2868}^{+2830}$ & $11.86_{-0.03}^{+0.03}$ & $9.30_{-0.89}^{+0.95}$ & $12.86_{-0.04}^{+0.04}$ & $0.25_{-0.09}^{+0.09}$ & $0.96_{-0.03}^{+0.03}$ & $0.32_{-0.01}^{+0.01}$ & $1.34_{-0.02}^{+0.02}$ & $0.015$ & $0.063$ & 3.4 \\ 
 & 10.25 & 10.60 & 30036 & $30069_{-1948}^{+1908}$ & $12.05_{-0.03}^{+0.04}$ & $9.52_{-1.02}^{+1.04}$ & $13.03_{-0.04}^{+0.04}$ & $0.28_{-0.09}^{+0.10}$ & $0.97_{-0.04}^{+0.04}$ & $0.30_{-0.01}^{+0.01}$ & $1.41_{-0.02}^{+0.02}$ & $0.019$ & $0.066$ & 3.2 \\ 
 & 10.50 & 10.77 & 16518 & $16294_{-1199}^{+1178}$ & $12.27_{-0.03}^{+0.04}$ & $9.66_{-1.13}^{+1.20}$ & $13.36_{-0.04}^{+0.05}$ & $0.26_{-0.09}^{+0.09}$ & $1.09_{-0.04}^{+0.04}$ & $0.23_{-0.01}^{+0.01}$ & $1.51_{-0.02}^{+0.02}$ & $0.026$ & $0.071$ & 2.5 \\ 
 & 10.75 & 10.95 & 6825 & $6682_{-545}^{+534}$ & $12.63_{-0.04}^{+0.04}$ & $10.05_{-1.36}^{+1.45}$ & $13.73_{-0.05}^{+0.06}$ & $0.29_{-0.10}^{+0.10}$ & $1.10_{-0.12}^{+0.11}$ & $0.18_{-0.01}^{+0.01}$ & $1.68_{-0.04}^{+0.03}$ & $0.040$ & $0.079$ & 1.8 \\ 
 & 11.00 & 11.16 & 1892 & $1848_{-196}^{+193}$ & $13.10_{-0.05}^{+0.05}$ & $11.85_{-2.67}^{+1.42}$ & $14.11_{-0.12}^{+0.08}$ & $0.30_{-0.10}^{+0.10}$ & $1.27_{-0.46}^{+0.15}$ & $0.14_{-0.01}^{+0.01}$ & $2.09_{-0.05}^{+0.04}$ & $0.077$ & $0.099$ & 1.7 \\ 
\hline
$0.75 - 1$ & 9.25 & 9.98 & 167295 & $137941_{-8331}^{+8760}$ & $11.48_{-0.03}^{+0.03}$ & $9.14_{-0.77}^{+0.83}$ & $12.65_{-0.04}^{+0.04}$ & $0.20_{-0.08}^{+0.08}$ & $1.13_{-0.01}^{+0.01}$ & $0.26_{-0.01}^{+0.01}$ & $1.38_{-0.01}^{+0.01}$ & $0.010$ & $0.057$ & 4.9 \\ 
 & 9.50 & 10.15 & 129436 & $104863_{-7301}^{+7051}$ & $11.60_{-0.03}^{+0.03}$ & $9.18_{-0.79}^{+0.87}$ & $12.74_{-0.04}^{+0.04}$ & $0.21_{-0.08}^{+0.08}$ & $1.14_{-0.01}^{+0.02}$ & $0.26_{-0.01}^{+0.01}$ & $1.43_{-0.01}^{+0.01}$ & $0.012$ & $0.059$ & 6.4 \\ 
 & 9.75 & 10.31 & 97333 & $80099_{-5392}^{+5552}$ & $11.73_{-0.03}^{+0.03}$ & $9.17_{-0.82}^{+0.86}$ & $12.81_{-0.04}^{+0.04}$ & $0.21_{-0.08}^{+0.08}$ & $1.16_{-0.01}^{+0.02}$ & $0.27_{-0.01}^{+0.01}$ & $1.50_{-0.01}^{+0.01}$ & $0.013$ & $0.062$ & 4.8 \\ 
 & 10.00 & 10.47 & 70619 & $60011_{-4164}^{+4141}$ & $11.85_{-0.03}^{+0.04}$ & $9.40_{-0.89}^{+0.95}$ & $12.92_{-0.04}^{+0.04}$ & $0.23_{-0.08}^{+0.09}$ & $1.17_{-0.02}^{+0.02}$ & $0.26_{-0.01}^{+0.01}$ & $1.54_{-0.01}^{+0.01}$ & $0.015$ & $0.064$ & 3.4 \\ 
 & 10.25 & 10.63 & 48056 & $38932_{-3043}^{+3188}$ & $12.02_{-0.03}^{+0.03}$ & $9.33_{-0.93}^{+1.00}$ & $13.09_{-0.04}^{+0.04}$ & $0.19_{-0.08}^{+0.08}$ & $1.23_{-0.02}^{+0.02}$ & $0.25_{-0.01}^{+0.01}$ & $1.65_{-0.01}^{+0.01}$ & $0.019$ & $0.068$ & 5.0 \\ 
 & 10.50 & 10.79 & 28636 & $25193_{-1939}^{+1876}$ & $12.20_{-0.03}^{+0.04}$ & $9.45_{-1.03}^{+1.16}$ & $13.24_{-0.04}^{+0.04}$ & $0.22_{-0.08}^{+0.09}$ & $1.24_{-0.03}^{+0.03}$ & $0.23_{-0.01}^{+0.01}$ & $1.73_{-0.01}^{+0.02}$ & $0.022$ & $0.071$ & 3.5 \\ 
 & 10.75 & 10.96 & 13118 & $11786_{-977}^{+953}$ & $12.48_{-0.03}^{+0.04}$ & $9.94_{-1.37}^{+1.34}$ & $13.56_{-0.04}^{+0.04}$ & $0.25_{-0.08}^{+0.09}$ & $1.32_{-0.04}^{+0.04}$ & $0.17_{-0.01}^{+0.01}$ & $1.87_{-0.02}^{+0.02}$ & $0.032$ & $0.077$ & 2.6 \\ 
 & 11.00 & 11.16 & 4042 & $3752_{-364}^{+354}$ & $12.88_{-0.04}^{+0.05}$ & $10.52_{-1.74}^{+1.71}$ & $14.00_{-0.06}^{+0.07}$ & $0.22_{-0.09}^{+0.09}$ & $1.46_{-0.09}^{+0.03}$ & $0.10_{-0.01}^{+0.01}$ & $2.18_{-0.04}^{+0.04}$ & $0.057$ & $0.090$ & 4.9 \\ 
\hline
$1 - 1.5$ & 9.75 & 10.29 & 171332 & $151484_{-7097}^{+7156}$ & $11.81_{-0.02}^{+0.03}$ & $9.18_{-0.79}^{+0.92}$ & $12.88_{-0.03}^{+0.03}$ & $0.20_{-0.08}^{+0.08}$ & $1.17_{-0.02}^{+0.02}$ & $0.23_{-0.01}^{+0.01}$ & $1.74_{-0.01}^{+0.01}$ & $0.009$ & $0.046$ & 4.0 \\ 
 & 10.00 & 10.47 & 118381 & $107360_{-4942}^{+5066}$ & $11.96_{-0.02}^{+0.03}$ & $9.33_{-0.91}^{+0.96}$ & $12.99_{-0.03}^{+0.03}$ & $0.22_{-0.08}^{+0.08}$ & $1.16_{-0.02}^{+0.02}$ & $0.23_{-0.01}^{+0.01}$ & $1.81_{-0.01}^{+0.01}$ & $0.011$ & $0.048$ & 4.1 \\ 
 & 10.25 & 10.64 & 77806 & $73539_{-3514}^{+3384}$ & $12.11_{-0.03}^{+0.04}$ & $9.31_{-0.90}^{+0.98}$ & $13.14_{-0.03}^{+0.03}$ & $0.26_{-0.09}^{+0.10}$ & $1.12_{-0.03}^{+0.03}$ & $0.21_{-0.01}^{+0.01}$ & $1.88_{-0.02}^{+0.02}$ & $0.013$ & $0.049$ & 5.6 \\ 
 & 10.50 & 10.80 & 45898 & $43377_{-2207}^{+2223}$ & $12.33_{-0.04}^{+0.05}$ & $9.67_{-1.12}^{+1.09}$ & $13.33_{-0.03}^{+0.03}$ & $0.34_{-0.10}^{+0.10}$ & $1.22_{-0.03}^{+0.03}$ & $0.18_{-0.01}^{+0.01}$ & $2.01_{-0.03}^{+0.02}$ & $0.016$ & $0.053$ & 3.1 \\ 
 & 10.75 & 10.97 & 21226 & $20456_{-1157}^{+1152}$ & $12.62_{-0.05}^{+0.06}$ & $9.94_{-1.30}^{+1.43}$ & $13.64_{-0.03}^{+0.03}$ & $0.40_{-0.11}^{+0.10}$ & $1.30_{-0.06}^{+0.06}$ & $0.13_{-0.01}^{+0.01}$ & $2.20_{-0.04}^{+0.04}$ & $0.023$ & $0.058$ & 2.7 \\ 
 & 11.00 & 11.17 & 6473 & $6429_{-439}^{+422}$ & $13.03_{-0.06}^{+0.07}$ & $10.46_{-1.67}^{+1.72}$ & $14.04_{-0.05}^{+0.09}$ & $0.43_{-0.10}^{+0.09}$ & $1.33_{-0.22}^{+0.12}$ & $0.09_{-0.01}^{+0.01}$ & $2.60_{-0.07}^{+0.08}$ & $0.041$ & $0.068$ & 2.3 \\ 
\hline

\end{tabular}
\end{table}
\end{landscape}

\begin{landscape}
\begin{table}
\centering
\contcaption{The best-fit HOD parameter outputs and sample statistics for the COMBINED halo model case.}

\begin{tabular}{|c|c|c|c|c|c|c|c|c|c|c|c|c|c|c|}
\hline
\textbf{$z$-bin} & \textbf{$M_*^{\mathrm{th}}$} & \textbf{$M_*^{\mathrm{mean}}$} & \textbf{$N_\mathrm{gal}^\mathrm{obs}$} & \textbf{$N_\mathrm{gal}^\mathrm{mod}$} & \textbf{$M_{\min}$} & \textbf{$M_0$} & \textbf{$M_1$} & \textbf{$\sigma_{\log_{10} M}$} & \textbf{$\alpha$} & \textbf{$f_\mathrm{sat}$} & \textbf{$b_g$} & \textbf{$\sigma_\mathrm{Poisson}$} & \textbf{$\sigma_\mathrm{CV}$} & \textbf{$\chi^2_\mathrm{reduced}$} \\ \hline
$1.5 - 2$ & 10.00 & 10.47 & 80378 & $79569_{-4414}^{+4263}$ & $12.03_{-0.03}^{+0.04}$ & $9.44_{-0.98}^{+1.09}$ & $13.05_{-0.03}^{+0.03}$ & $0.27_{-0.09}^{+0.08}$ & $1.31_{-0.03}^{+0.03}$ & $0.17_{-0.01}^{+0.01}$ & $2.29_{-0.03}^{+0.02}$ & $0.011$ & $0.056$ & 3.8 \\ 
 & 10.25 & 10.63 & 54105 & $55429_{-2891}^{+3069}$ & $12.26_{-0.05}^{+0.05}$ & $9.63_{-1.11}^{+1.18}$ & $13.16_{-0.03}^{+0.03}$ & $0.49_{-0.09}^{+0.06}$ & $1.29_{-0.04}^{+0.03}$ & $0.15_{-0.01}^{+0.01}$ & $2.31_{-0.03}^{+0.04}$ & $0.013$ & $0.057$ & 4.1 \\ 
 & 10.50 & 10.79 & 31918 & $33396_{-1861}^{+1727}$ & $12.49_{-0.03}^{+0.03}$ & $9.74_{-1.18}^{+1.22}$ & $13.33_{-0.03}^{+0.03}$ & $0.57_{-0.04}^{+0.02}$ & $1.22_{-0.05}^{+0.05}$ & $0.14_{-0.01}^{+0.01}$ & $2.41_{-0.02}^{+0.03}$ & $0.017$ & $0.060$ & 3.6 \\ 
 & 10.75 & 10.96 & 14508 & $14767_{-976}^{+992}$ & $12.64_{-0.06}^{+0.06}$ & $10.23_{-1.50}^{+1.42}$ & $13.59_{-0.03}^{+0.04}$ & $0.40_{-0.10}^{+0.09}$ & $1.45_{-0.06}^{+0.03}$ & $0.10_{-0.01}^{+0.01}$ & $2.84_{-0.06}^{+0.06}$ & $0.025$ & $0.070$ & 2.3 \\ 
 & 11.00 & 11.15 & 4233 & $3970_{-333}^{+341}$ & $13.00_{-0.06}^{+0.07}$ & $12.16_{-2.68}^{+0.98}$ & $14.15_{-0.19}^{+0.27}$ & $0.36_{-0.10}^{+0.09}$ & $1.19_{-0.41}^{+0.23}$ & $0.05_{-0.01}^{+0.01}$ & $3.44_{-0.11}^{+0.10}$ & $0.050$ & $0.085$ & 1.6 \\ 
\hline
$2 - 2.5$ & 10.25 & 10.58 & 31699 & $32412_{-2001}^{+2022}$ & $12.34_{-0.05}^{+0.04}$ & $10.01_{-1.38}^{+1.43}$ & $13.30_{-0.03}^{+0.04}$ & $0.52_{-0.07}^{+0.05}$ & $1.45_{-0.07}^{+0.04}$ & $0.08_{-0.00}^{+0.00}$ & $2.82_{-0.05}^{+0.05}$ & $0.017$ & $0.068$ & 3.2 \\ 
 & 10.50 & 10.77 & 16076 & $16093_{-1109}^{+1161}$ & $12.56_{-0.06}^{+0.05}$ & $10.40_{-1.65}^{+1.57}$ & $13.56_{-0.07}^{+0.10}$ & $0.52_{-0.07}^{+0.05}$ & $1.37_{-0.20}^{+0.10}$ & $0.06_{-0.01}^{+0.01}$ & $3.08_{-0.05}^{+0.07}$ & $0.024$ & $0.075$ & 2.5 \\ 
 & 10.75 & 10.97 & 6720 & $6699_{-481}^{+486}$ & $12.87_{-0.03}^{+0.03}$ & $12.86_{-0.26}^{+0.15}$ & $13.62_{-0.14}^{+0.18}$ & $0.58_{-0.03}^{+0.02}$ & $0.85_{-0.27}^{+0.42}$ & $0.05_{-0.00}^{+0.00}$ & $3.44_{-0.04}^{+0.05}$ & $0.038$ & $0.083$ & 2.6 \\ 
 & 11.00 & 11.18 & 2096 & $1707_{-172}^{+165}$ & $13.16_{-0.07}^{+0.06}$ & $10.35_{-1.59}^{+1.64}$ & $14.71_{-0.22}^{+0.20}$ & $0.50_{-0.07}^{+0.06}$ & $1.15_{-0.14}^{+0.20}$ & $0.01_{-0.00}^{+0.00}$ & $4.19_{-0.12}^{+0.13}$ & $0.084$ & $0.101$ & 2.2 \\ 
\hline
$2.5 - 3$ & 10.50 & 10.78 & 6189 & $6047_{-594}^{+603}$ & $12.56_{-0.05}^{+0.07}$ & $10.91_{-1.92}^{+2.01}$ & $14.30_{-0.64}^{+0.46}$ & $0.33_{-0.09}^{+0.09}$ & $1.07_{-0.23}^{+0.31}$ & $0.01_{-0.00}^{+0.00}$ & $4.29_{-0.14}^{+0.12}$ & $0.041$ & $0.104$ & 2.8 \\ 
 & 10.75 & 11.00 & 2439 & $2349_{-200}^{+196}$ & $13.00_{-0.04}^{+0.03}$ & $14.07_{-0.34}^{+0.40}$ & $14.66_{-0.36}^{+0.24}$ & $0.58_{-0.03}^{+0.01}$ & $1.02_{-0.36}^{+0.33}$ & $0.00_{-0.00}^{+0.00}$ & $4.34_{-0.06}^{+0.08}$ & $0.069$ & $0.106$ & 3.5 \\ 
 & 11.00 & 11.24 & 772 & $538_{-57}^{+55}$ & $13.37_{-0.03}^{+0.03}$ & $14.12_{-0.44}^{+0.39}$ & $14.72_{-0.33}^{+0.21}$ & $0.59_{-0.02}^{+0.01}$ & $1.14_{-0.40}^{+0.28}$ & $0.00_{-0.00}^{+0.00}$ & $5.15_{-0.07}^{+0.08}$ & $0.179$ & $0.125$ & 4.8 \\ 
\hline
$3 - 3.5$ & 10.50 & 10.74 & 4531 & $3104_{-337}^{+322}$ & $12.73_{-0.06}^{+0.05}$ & $13.07_{-2.87}^{+0.83}$ & $14.61_{-0.53}^{+0.28}$ & $0.55_{-0.05}^{+0.04}$ & $1.18_{-0.38}^{+0.23}$ & $0.00_{-0.00}^{+0.00}$ & $4.70_{-0.10}^{+0.12}$ & $0.071$ & $0.116$ & 2.3 \\ 
 & 10.75 & 10.99 & 1427 & $896_{-106}^{+110}$ & $13.05_{-0.05}^{+0.04}$ & $14.01_{-0.34}^{+0.44}$ & $14.65_{-0.39}^{+0.26}$ & $0.57_{-0.04}^{+0.02}$ & $1.04_{-0.35}^{+0.32}$ & $0.00_{-0.00}^{+0.00}$ & $5.38_{-0.10}^{+0.13}$ & $0.145$ & $0.133$ & 3.9 \\ 
\hline
$3.5 - 4$ & 10.50 & 10.75 & 2233 & $1532_{-244}^{+239}$ & $12.55_{-0.07}^{+0.08}$ & $13.07_{-2.23}^{+0.78}$ & $14.59_{-0.58}^{+0.30}$ & $0.34_{-0.09}^{+0.08}$ & $1.15_{-0.43}^{+0.25}$ & $0.00_{-0.00}^{+0.00}$ & $6.22_{-0.25}^{+0.24}$ & $0.102$ & $0.158$ & 1.5 \\ 
 & 10.75 & 11.01 & 707 & $472_{-58}^{+61}$ & $13.04_{-0.04}^{+0.03}$ & $14.01_{-0.28}^{+0.41}$ & $14.63_{-0.37}^{+0.26}$ & $0.59_{-0.02}^{+0.01}$ & $1.05_{-0.35}^{+0.32}$ & $0.00_{-0.00}^{+0.00}$ & $6.16_{-0.09}^{+0.11}$ & $0.198$ & $0.156$ & 5.5 \\ 
\hline

\end{tabular}
\end{table}
\end{landscape}

\bsp	
\label{lastpage}
\end{document}